\documentclass[review]{elsarticle}

\usepackage{bm}
\usepackage{lineno,hyperref}
\usepackage{mathtools}
\usepackage{graphicx}
\usepackage{multirow}
 \usepackage{textcomp}
 \usepackage{tikz}
 \usetikzlibrary{arrows,positioning}
\usepackage{makecell} 
\setcellgapes{5pt}

\usepackage{amsmath,amsfonts,amssymb}

\DeclareMathOperator*{\argmin}{arg\,min}

\usepackage{bm}
\usepackage{lineno,hyperref}
\usepackage{mathtools}
\usepackage{graphicx}
\usepackage{multirow}
\newcommand{\norm}[1]{\left\lVert #1 \right\rVert}
\usepackage{algorithm}
\usepackage{algpseudocode}
\usepackage{algpascal}
\usepackage{textcomp}
\usepackage{tikz}
\usepackage{smartdiagram}
\usetikzlibrary{arrows,positioning}
\usepackage{makecell} 
\setcellgapes{5pt}

\smartdiagramset{border color=none,
	set color list={blue!50!cyan,green!60!lime,orange!50!red,red!80!black},
	back arrow disabled=true, module minimum width = 4}

\smartdiagramset{module minimum height=1cm,
	module minimum width=0.8cm,
	module x sep=2}

\usepackage{lineno,hyperref}
\modulolinenumbers[5]

\usepackage{xcolor}

\usepackage{lineno,hyperref}
\modulolinenumbers[5]

\journal{Statistical Papers}

\begin{document}

\begin{frontmatter}

\title{Quantile-based fuzzy $C$-means clustering of multivariate time series: Robust techniques}


\author[mymainaddress]{\'Angel L\'opez-Oriona\corref{mycorrespondingauthor} (ORCID 0000-0003-1456-7342)}
\ead{oriona38@hotmail.com}

\author[dursoaddress]{Pierpaolo D'Urso (ORCID 0000-0002-7406-6411)}
\ead{pierpaolo.durso@uniroma1.it}

\author[mymainaddress,mysecondaryaddress]{Jos\'e A. Vilar (ORCID 0000-0001-5494-171X)}
\cortext[mycorrespondingauthor]{Corresponding author}
\ead{jose.vilarf@udc.es}

\author[mymainaddress]{Borja Lafuente-Rego (ORCID 0000-0002-2443-3084)}
\ead{borja.lafuente@udc.es}

\address[mymainaddress]{Research Group MODES, Research Center for Information and Communication Technologies (CITIC), University of A Coru\~na, 15071 A Coru\~na, Spain.}
\address[mysecondaryaddress]{Technological Institute for Industrial Mathematics (ITMATI), Spain.}
\address[dursoaddress]{Department of Social Sciences and Economics, Sapienza University of Rome, P. le Aldo Moro 5, Roma, Italy.}

\begin{abstract}
Three robust methods for clustering multivariate time series from the point of view of generating processes are proposed. The procedures are robust versions of a fuzzy $C$-means model based on: (i) estimates of the quantile cross-spectral density and (ii) the classical principal component analysis. Robustness to the presence of outliers is achieved by using the so-called metric, noise and trimmed approaches. The metric approach incorporates in the objective function a distance measure aimed at neutralizing the effect of the outliers, the noise approach builds an artificial cluster expected to contain the outlying series and the trimmed approach eliminates the most atypical series in the dataset. All the proposed techniques inherit the nice properties of the quantile cross-spectral density, as being able to uncover general types of dependence. Results from a broad simulation study including multivariate linear, nonlinear and GARCH processes indicate that the algorithms are substantially effective in coping with the presence of outlying series (i.e., series exhibiting a dependence structure different from that of the majority), clearly poutperforming alternative procedures. The usefulness of the suggested methods is highlighted by means of two specific applications regarding financial and environmental series.
\end{abstract}


\end{frontmatter}


\section{Introduction}\label{sectionintroduction}

In the last years there has been an unprecedented growth in complexity, speed and volume of data. In particular, time series data have become ubiquitous in our days, arising frequently in a broad variety of fields including medicine, computer science, finance, environmental sciences, machine learning, marketing and neuroscience, among many others. Typically, time series involve a huge number of records, present dynamic behavior patterns which might change over time, and one frequently has to deal with realizations of different length. Due to this complex nature, standard techniques to perform data mining tasks as classification, clustering or anomaly detection often produce unsatisfactory results. Complexity is still greater by treating with high dimensional time series, where the interdependence structure and large dimensionality are serious obstacles to develop efficient procedures. Univariate time series (UTS) were the main focus of intensive research until recently, but multivariate time series (MTS) have received lately a great deal of attention due to the advance of technology and storage capabilities of everyday devices. Well-known examples of MTS are multi-lead ECG signals of patients or records containing several economic indicators of a given country over time, but many other examples can be easily obtained from different fields.

Among time series data mining tasks, clustering is a central problem. In fact, identifying groups of similar series is basic for many applications in order to detect a few representative patterns, forecast future performances, quantify affinity, recognize dynamic changes and structural breaks\ldots\,However, unlike traditional databases, similarity search in time series data is a complex issue that cannot be addressed with conventional methods. For instance, it is not straightforward to define a proper distance between time series objects since such a distance should take into account the underlying dynamic patterns. This problem accentuates when coping with MTS data, since a proper dissimilarity in this framework must also consider the interdependence relationship between the different dimensions. In addition, it is not uncommon for MTS databases to contain some outlying MTS, i.e. series showing a very different behaviour from the rest of the MTS in the dataset. In this setting, a suitable distance but lacking robustness to anomalous series is likely to fail when utilised to perform clustering. Another challenging issue when treating with MTS data is the computational complexity of the clustering procedure. The high dimension of MTS samples implies a fast increase of the dataset size thus requiring more storage space and computing resources. In fact, clustering procedures specifically designed to deal with UTS data can be inefficient or even infeasible with MTS data. In sum, the complex nature of the dependence structure and large dimensionality of MTS make particularly challenging to develop effective and robust clustering procedures. 

A range of procedures have been proposed for clustering of time series during the last two decades, either considering the ``hard'' or the ``soft'' paradigm. The former techniques locate each data object in exactly one cluster, whereas the latter strategies provide a more versatile approach, constructing a partition where each object can belong to several groups with specific membership degrees. While a lot of works have addressed the topic of clustering of UTS \citep{lafuente2016clustering, lafuente2018robust, d2009autocorrelation, d2012wavelets, d2013clustering, d2017autoregressive, d2015time, d2014robust, maharaj2011fuzzy, piccolo1990distance, d2016garch, alonso2019clustering, maharaj1996significance, vilar2018quantile, izakian2015fuzzy, caiado2006periodogram}, the literature on MTS clustering is considerably more scarce \citep{kakizawa1998discrimination, d2014wavelet, lopez2021quantile, oriona2021c, he2018unsupervised, maharaj1999comparison, d2005fuzzy, d2018robust}. Comprehensive reviews on the topic can be seen in \cite{aghabozorgi2015time, maharaj2019time}. 

Robust clustering procedures for time series have also arisen in recent years. In the univariate framework, \cite{d2013noise} introduced a robust procedure based on the traditional fuzzy $C$-medoids algorithm, where the effect of outlying series is neutralised by introducing the so-called noise cluster, expected to contain the outlier elements. This work is extended in \cite{d2015time} by considering a suitable robust metric for time series, the exponential distance measure, and further in \cite{d2017autoregressive}, where a trimming-based rule is considered to trimming away the most anomalous series. \cite{d2016garch} proposed different fuzzy robust clustering approaches for UTS based on GARCH models. Specifically, the called metric, noise and trimmed approaches are considered to achieve robustness. \cite{lafuente2018robust} developed three robust fuzzy clustering strategies relying on quantile autocovariances. In \cite{rivera2017robust}, a robust clustering for stationary series is introduced by considering estimates of the spectral densities as functional data and then applying a robust clustering algorithm.

Some papers have also proposed robust clustering strategies for MTS. \cite{d2005fuzzy} introduced three fuzzy $C$-means clustering approaches  for  multivariate time trajectories considering the called positional dissimilarity, velocity dissimilarity and a generalization of both. \cite{d2018robust} proposed four fuzzy methods for grouping MTS based on the fuzzy $C$-medoids approach and making use of the exponential transformation for dissimilarity measures. The procedures differentiate from one another by the kind discrepancy they take into account. Three methods consider instantaneous-based, longitudinal-based, and both types of features, respectively, whereas the fourth algorithm employs the dynamic time warping (DTW) distance for MTS. DTW is also used in \cite{d2019trimmed} to suggest a novel trimmed fuzzy approach to cluster multivariate financial time series, which is successfully applied to 40 MTS representing companies in the Italian Stock Exchange. 

It is worth highlighting that the previous works regarding the multivariate framework are mainly aimed at capturing dissimilarity in shape between two MTS, that is, they intend to discriminate between different geometric profiles. To the best of our knowledge, no robust clustering method for MTS has been suggested in the literature when the outlyingness is characterised by an abnormal dependence pattern.  

Based on previous references, one can deduce that the majority of robust clustering strategies for time series consider a fuzzy approach and are based on at least one of the following three methodological strategies:

\begin{enumerate}
	\item \textbf{Metric approach}. It is based on incorporating in the objective function of the clustering algorithm distances with high robustness against outliers.
	\item \textbf{Noise approach}. A term representing an artificial cluster is included in the objective function. This cluster is expected to contain the outlying series in the dataset with a high membership. 
	\item \textbf{Trimmed approach}. The clustering technique is applied to the time series remaining after a fixed fraction of the most distant series is removed.
\end{enumerate}

The aim of this paper is to develop clustering procedures for MTS capable of effectively neutralizing the effect of outlying series. It is assumed that the target is on grouping MTS according to the underlying dependence structures, i.e., on identifying groups of MTS generated by the same multivariate stochastic process. This problem often arises when dealing with sets of MTS consisting of environmental, financial, EEG and fMRI data,which may exhibit complex serial dependence structures. According to this clustering principle, a given MTS is considered to be an outlier if it has been generated from a stochastic process different from those of the majority of the series in the dataset. This definition of outlier has already been considered in some works \citep{lafuente2018robust, blazquez2020review}. The proposed clustering strategies are based on a distance measure relying on the quantile cross-spectral density (QCD) and principal component analysis (PCA). This dissimilarity has been used in our previous work \citep{oriona2021c} to perform fuzzy clustering of MTS through the standard fuzzy $C$-means and fuzzy $C$-medoids clustering algorithms. The proposed approaches showed very good results in grouping MTS simulated from a wide range of generating processes, clearly outperforming other alternative metrics suggested in the literature. However, the procedures presented there lack the capability to correctly identify anomalous MTS, thus being prone to errors when a given dataset contains some abnormal series. To be able of properly managing outlying data, we propose three  robust extensions of the fuzzy $C$-means model based on QCD (QCD-FCMn) introduced in \cite{oriona2021c} by considering the metric, noise and trimmed approaches. We use the fuzzy $C$-means algorithm because it produced slightly better results than the $C$-medoids algorithm in \cite{oriona2021c}. This way, each method achieves robustness in different manner and all of them take advantage of the high capability of the QCD-based metric to discriminate between independent realizations from a broad range of stationary processes.

To analyze the behaviour of the suggested robust approaches, an extensive simulation study is carried out including linear, nonlinear and the so-called BEKK models, one well-known specification of multivariate GARCH processes. The simulation experiments consist of some scenarios containing two well-defined clusters and contaminated with one or two outlying series. Both the ability of the methods to correctly locate the non-abnormal MTS and to properly handle the anomalous series are evaluated. Due to the lack of robust alternative procedures in the literature, the dissimilarity proposed in \cite{d2012wavelets} and a natural extension of that introduced in \cite{d2009autocorrelation} are also analysed for comparison purposes. Furthermore, to ensure a fair comparison, robust versions of the fuzzy algorithms based on these measures considering the metric, robust and noise approaches are also considered. The usefulness and effectiveness of the proposed robust fuzzy models is illustrated considering two study cases with real data.


The remainder of this paper is structured as follows. 
Section~\ref{sectionqcd} is devoted to introduce the concept of quantile cross-spectral density (QCD) and construct a dissimilarity measure between a pair of MTS ($d_{QCD}$) based on proper estimates of the corresponding QCD. In Section~\ref{sectionrobustclustering}, three novel robust fuzzy clustering approaches relying on both the traditional fuzzy $C$-means algorithm and the proposed dissimilarity are developed. A direct modification of $d_{QCD}$ based on the PCA transformation is defined following \cite{oriona2021c}. The dissimilarity based on QCD and PCA is used to perform robust fuzzy clustering in Section \ref{sectionsimulationstudy}. Lastly, Section \ref{sectionapplications} contains applications to real datasets and some concluding remarks are summarized in Section \ref{sectionconcludingremarks}. 

\section{A distance measure based on the quantile cross-spectral density}\label{sectionqcd}

Consider a set of $s$ multivariate time series $\mathcal{S} = \left\{ \bm{X}_t^{(1)}, \ldots, \bm{X}_t^{(s)} \right\}$, where the $j$-th element $\bm{X}_t^{(j)}= \left\{ \bm{X}_{1}^{(j)},\ldots ,\bm{X}_{T_j}^{(j)}\right\}$ is a $T_j$-length partial realization from any $d$-variate real-valued strictly stationary stochastic process $(\bm{X}_t)_{t\in\mathbb{Z}}$. We wish to perform clustering on $\mathcal{S}$ in such a way that the series generated from the same stochastic process are grouped together. We propose to use a partitional algorithm starting from a pairwise dissimilarity matrix based on comparing estimated quantile cross-spectral densities. In this section, the quantile cross-spectral density notion is presented and then used to define a distance between MTS.

\subsection{The quantile cross-spectral density}\label{subsectionqcd}

Let  $\{\bm{X}_t, \, t\in\mathbb{Z}\} = \{(X_{t,1},\ldots,X_{t,d}), \, t\in\mathbb{Z}\}$ be a $d$-variate real-valued strictly stationary stochastic process. Denote by $F_j$ the marginal distribution function of $X_{t,j}$, $j=1,\ldots,d$, and by $q_j(\tau)=F_j^{-1}(\tau)$, $\tau \in [0,1]$, the corresponding quantile function. Fixed $l \in \mathbb{Z}$ and an arbitrary couple of quantile levels $(\tau,\tau^{\prime}) \in [0,1]^2$, consider the cross-covariance of the indicator functions $ I\left\{ X_{t,j_1} \leq  q_{j_1}(\tau) \right\}$ and $I\left\{ X_{t+l, j_2} \leq  q_{j_2} (\tau^{\prime}) \right\}$ given by
\begin{equation}		\label{qac}
\gamma_{j_1,j_2}(l,\tau,\tau^{\prime}) = \mbox{Cov} 
\left(  I \left\{ X_{t, j_1} \leq q_{j_1}(\tau) \right\}, I \left\{ X_{t+l, j_2} \leq q_{j_2}(\tau^{\prime}) \right\} \right),
\end{equation}
for $1\leq j_1,j_2 \leq d$. Taking $j_1=j_2=j$, the function $\gamma_{j,j}(l,\tau,\tau^{\prime})$, with $(\tau,\tau^{\prime}) \in [0,1]^2$, so-called quantile autocovariance function (QAF) of lag $l$, generalizes the traditional autocovariance function. While autocovariances measure linear dependence between different lags evaluating covariability with respect to the average, quantile autocovariances examine how a part of the range of variation of $X_{j}$ helps to predict whether the series will be below quantiles in a future time. This way, QAF entirely describes  the joint distribution of $( X_{t,j}, X_{t+l,j} )$, enabling us to capture serial features that standard autocovariances cannot detect. Note that $\gamma_{j_1,j_2}(l,\tau,\tau^{\prime})$ always exists since no assumptions about moments are required. Furthermore, QAF also takes advantage of the local distributional properties inherent to the quantile methods, including robustness against heavy tails, dependence in the extremes and changes in the conditional shapes (skewness, kurtosis). Motivated by these nice properties, a dissimilarity between UTS based on comparing estimated quantile autocovariances over a common range of quantiles was proposed by \cite{lafuente2016clustering} to perform UTS clustering with very satisfactory results. 

In the case of the multivariate process $\{\bm{X}_t, \, t\in\mathbb{Z}\}$, we can consider the $d \times d$ matrix
\begin{equation}  \label{gammaMTS}
\bm{\Gamma} (l,\tau,\tau^{\prime}) = 
\left( \gamma_{j_1,j_2}(l,\tau,\tau^{\prime}) \right)_{1\leq j_1,j_2\leq d},
\end{equation}
which jointly provides information about both the cross-dependence (when $j_1\neq j_2$) and the serial dependence (because the lag $l$ is considered). To obtain a much richer picture of the underlying dependence structure, $\bm{\Gamma} (l,\tau,\tau^{\prime})$ can be computed over a range of prefixed values of $L$ lags, $\mathcal{L}=\{l_1,\ldots,l_L\}$, and $r$ quantile levels, $\mathcal{T} = \{\tau_1,\ldots,\tau_r\}$, thus having available the set of matrices
\begin{equation}  \label{gammaProcess}
\bm{\Gamma}_{\bm{X}_t} \left(\mathcal{L},\mathcal{T}\right) = 
\left\{ \bm{\Gamma} (l,\tau,\tau^{\prime}), \,\,  l \in \mathcal{L}, \, \, \tau,\tau^{\prime} \in \mathcal{T} 
\right\}. 
\end{equation}

In the same way as the spectral density is the representation in the frequency domain of the autocovariance function, the spectral counterpart for the cross-covariances $\gamma_{j_1,j_2}(l,\tau,\tau^{\prime})$ can be introduced. Under suitable summability conditions (mixing conditions), the Fourier transform of the cross-covariances is well-defined and  the \textit{quantile cross-spectral density} is given by 
\begin{equation} \label{qcrossdens.j1j2}
{\mathfrak f}_{j_1,j_2} (\omega, \tau,\tau^{\prime}) = (1/2\pi) \sum_{l=-\infty}^{\infty} 
\gamma_{j_1,j_2}(l,\tau,\tau^{\prime}) e^{-il\omega},
\end{equation}
for $1\leq j_1,j_2 \leq d$, $\omega \in \mathbb{R}$ and $\tau,\tau^{\prime} \in [0,1]$. Note that ${\mathfrak f}_{j_1,j_2}(\omega, \tau,\tau^{\prime})$ is complex-valued so that it can be represented in terms of its real and imaginary parts, which will be denoted by $\Re({\mathfrak f}_{j_1,j_2}(\omega, \tau,\tau^{\prime}))$ and $\Im({\mathfrak f}_{j_1,j_2}(\omega, \tau,\tau^{\prime}))$, respectively. The quantity $\Re({\mathfrak f}_{j_1,j_2}(\omega, \tau,\tau^{\prime}))$ is known as quantile cospectrum of $(X_{t, j_1})_{t\in\mathbb{Z}}$ and $(X_{t, j_2})_{t\in\mathbb{Z}}$, whereas the quantity -$\Im({\mathfrak f}_{j_1,j_2}(\omega, \tau,\tau^{\prime}))$ is called quantile quadrature spectrum of $(X_{t, j_1})_{t\in\mathbb{Z}}$ and $(X_{t, j_2})_{t\in\mathbb{Z}}$. 

For fixed quantile levels $(\tau,\tau^{\prime})$, the quantile cross-spectral density is the cross-spectral density of the bivariate process
\begin{equation}
( I \{ X_{t, j_1} \leq q_{j_1}(\tau) \}, I \{ X_{t, j_2} \leq q_{j_2}(\tau^{\prime}) \} ).
\end{equation}

Therefore the quantile cross-spectral density measures dependence between two components of the multivariate process in different ranges of their joint distribution and across frequencies. Proceeding as in (\ref{gammaProcess}), the quantile cross-spectral density can be evaluated for every couple of components on a range of frequencies $\Omega$ and of quantile levels $\mathcal{T}$ in order to obtain a complete representation of the process, i.e., consider the set of matrices
\begin{equation}  \label{qcrossdens.X}
\bm{\mathfrak f}_{\bm{X}_t} \left(\Omega, \mathcal{T}\right) = 
\left\{ \bm{\mathfrak f} (\omega,\tau,\tau^{\prime}), \, \, \omega \in \Omega, \, \, \tau,\tau^{\prime} \in \mathcal{T} 
\right\}, 
\end{equation}
where $\bm{\mathfrak f} (\omega,\tau,\tau^{\prime})$ denotes the $d\times d$ matrix in $\mathbb{C}$
\begin{equation}   \label{qcrossdens}
\bm{\mathfrak f} (\omega,\tau,\tau^{\prime}) = \left( {\mathfrak f}_{j_1,j_2} (\omega, \tau,\tau^{\prime}) \right)_{1 \leq j_1,j_2 \leq d}.
\end{equation}

Representing $\bm{X}_t$ through $\bm{\mathfrak f}_{\bm{X}_t}$, a complete information on the general dependence structure of the process is available. Comprehensive discussions about the nice properties of the quantile cross-spectral density are given in \cite{Lee_SubbaRao:2012}, \cite{Dette_Hallin_Kley_Volgushev:2015} and \cite{Barunik_Kley:2019}, including invariance to monotone transformations, robustness and  capability to detect nonlinear dependence. It is also worth enhancing that the quantile cross-spectral density provides a full description of all copulas of pairs of components in $\bm{X}_t$, since the difference between the copula of an arbitrary couple $(X_{t,j_1},X_{t+l,j_2})$ evaluated in $(\tau,\tau^{\prime})$ and the independence copula at $(\tau,\tau^{\prime})$ can be written as 
\begin{equation}\label{copulaqcd}
\mathbb{P} \left(  X_{t,j_1} \leq q_{j_1}(\tau), X_{t+l,j_2} \leq q_{j_2}(\tau^{\prime}) \right) - \tau \tau^{\prime}
= \int_{-\pi}^{\pi} {\mathfrak f}_{j_1,j_2} (\omega, \tau,\tau^{\prime}) e^{il\omega} \, d \omega.
\end{equation}

According with the prior arguments, a dissimilarity measure between realizations of two multivariate processes, $\bm{X}_t$ and $\bm{Y}_t$, could be established by comparing their representations in terms of the quantile cross-spectral density matrices, $\bm{\mathfrak f}_{\bm{X}_t}$ and $\bm{\mathfrak f}_{\bm{Y}_t}$, respectively. For it, estimates of the quantile cross-spectral densities must be obtained. 

Let $\left\{ \bm{X}_{1},\ldots ,\bm{X}_{T} \right\}$ be a realization from the process $(\bm{X}_t)_{t\in\mathbb{Z}}$ so that $\bm{X}_{t}=(X_{t,1},\ldots,X_{t,d})$, $t=1,\ldots,T$. For arbitrary $j_1, j_2 \in \{1,\ldots,d\}$ and $(\tau,\tau^{\prime}) \in [0,1]^2$, \cite{Barunik_Kley:2019} propose to estimate ${\mathfrak f}_{j_1,j_2} (\omega, \tau,\tau^{\prime})$ considering a smoother of the cross-periodograms based on the indicator functions $I\{ \hat{F}_{T,j}(X_{t,j})\}$, where $\hat{F}_{T,j} (x) = T^{-1} \sum_{t=1}^{T} I\{ X_{t,j} \leq  x\}$ denotes the empirical distribution function of $X_{t,j}$. This approach extends to the multivariate case the estimator proposed by \cite{Kley_Volgushev_Dette_Hallin:2016} in the univariate setting. More specifically, the called \textit{rank-based copula cross periodogram} (CCR-periodogram) is defined by
\begin{equation}   \label{CCR}
I^{j_1,j_2}_{T,R} (\omega,\tau,\tau^{\prime}) = \frac{1}{2 \pi T} d^{j_1}_{T,R} (\omega,\tau) 
d^{j_2}_{T,R} (-\omega,\tau^{\prime}),
\end{equation}
where $d_{T,R}^{j} (\omega,\tau) = \sum_{t=1}^{T} I\{ \hat{F}_{T,j}(X_{t,j}) \leq \tau \} e^{-i\omega t}$.

The asymptotic properties of the CCR-periodogram are established in Proposition S4.1 of \cite{Barunik_Kley:2019}. Likewise the standard cross-periodogram, the CCR-periodogram is not a consistent estimator of ${\mathfrak f}_{j_1,j_2} (\omega, \tau,\tau^{\prime})$ \citep{Kley_Volgushev_Dette_Hallin:2016}. To achieve consistency, the CCR-periodogram ordinates (evaluated on the Fourier frequencies) are convolved with weighting functions $W_T(\cdot)$. The \textit{smoothed CCR-periodogram} takes the form
\begin{equation}   \label{smoothedCCR}
\hat{G}^{j_1,j_2}_{T,R} (\omega,\tau,\tau^{\prime}) = \frac{2 \pi}{T} \sum_{s=1}^{T-1} W_T \left( \omega- \frac{2 \pi s}{T} \right)  
I^{j_1,j_2}_{T,R} \left( \frac{2 \pi s}{T} ,\tau,\tau^{\prime} \right),
\end{equation}
where $W_T \left( u \right) = \sum_{v=-\infty}^{\infty} \frac{1}{h_T} W\left( 
\frac{u + 2 \pi v}{h_T} \right)$, with $h_T>0$ a sequence of bandwidths such that $h_T \rightarrow 0$ and $Th_T \rightarrow \infty$ as $T\rightarrow \infty$, and $W$ is a real-valued, even weight function with support $[-\pi, \pi]$. 
Consistency and asymptotic performance of the smoothed CCR-periodogram $\hat{G}^{j_1,j_2}_{T,R} (\omega,\tau,\tau^{\prime})$ are established in Theorem S4.1 of \cite{Barunik_Kley:2019}. 

The set of complex-valued matrices $\bm{\mathfrak f}_{\bm{X}_t} \left(\Omega, \mathcal{T}\right)$ in (\ref{qcrossdens.X}) characterizing the underlying process can be estimated by 
\begin{equation}  \label{est.qcrossdens.X}
\hat{\bm{\mathfrak f}}_{\bm{X}_t} \left(\Omega, \mathcal{T}\right) = 
\left\{ \hat{\bm{\mathfrak f}} (\omega,\tau,\tau^{\prime}), \, \, \omega \in \Omega, \, \, \tau,\tau^{\prime} \in \mathcal{T} 
\right\}, 
\end{equation}
where $\hat{\bm{\mathfrak f}} (\omega,\tau,\tau^{\prime})$ is the matrix
\begin{equation}   \label{est.qcrossdens}
\hat{\bm{\mathfrak f}} (\omega,\tau,\tau^{\prime}) = \left( 
\hat{G}^{j_1,j_2}_{T,R} (\omega,\tau,\tau^{\prime}) 
\right)_{1 \leq j_1,j_2 \leq d}.
\end{equation}

\subsection{A spectral dissimilarity measure between MTS}\label{subsectiondqcd}

A simple dissimilarity criterion between two $d$-variate time series $\bm X_t^{(1)}$ and $\bm X_t^{(2)}$ can be obtained by comparing their estimated sets of complex-valued matrices $\hat{\bm{\mathfrak f}}_{\bm{X}_t^{(1)}} \left(\Omega, \mathcal{T}\right)$ and $\hat{\bm{\mathfrak f}}_{\bm{X}_t^{(2)}} \left(\Omega, \mathcal{T}\right)$ evaluated on a common range of frequencies and quantile levels. Specifically, each time series $\bm X_t^{(u)}$, $u=1,2$, is characterized by means of a set of $d^2$ vectors \{$\bm \Psi_{j_1,j_2}^{(u)},1\leq j_1, j_2 \leq d \}$ constructed as follows. For a given set of $K$ frequencies $\Omega = \{\omega_1, \ldots, \omega_K\}$, and $r$ quantile levels $\mathcal{T}=\{\tau_1,\ldots,\tau_r\}$, each vector $\bm \Psi_{j_1,j_2}^{(u)}$ is given by
\begin{equation}\label{uj1j2}
\bm \Psi_{j_1,j_2}^{(u)}=(\bm \Psi_{1,j_1,j_2}^{(u)} , \ldots, \bm \Psi_{K,j_1,j_2}^{(u)}),
\end{equation}

\noindent where each $\bm \Psi_{k,j_1,j_2}^{(u)}$, $k=1,\ldots, K$ consists of a vector of length $r^2$ formed by rearranging by rows the matrix
\begin{equation}
(\hat{\bm{\mathfrak f}} (\omega_k,\tau_i,\tau_{i^{\prime}}); i,i^\prime=1,\ldots, r),
\end{equation}

\noindent with $\hat{\bm{\mathfrak f}} (\omega_k,\tau_i,\tau_{i^{\prime}}) \in \hat{\bm{\mathfrak f}}_{\bm{X}_t^{(u)}} \left(\Omega, \mathcal{T}\right)$.

Once the set of $d^2$ vectors $\bm \Psi_{j_1,j_2}^{(u)}$ is obtained, they are all concatenated in a vector $\bm \Psi^{(u)}$ in the same way as vectors $\bm \Psi_{k,j_1,j_2}^{(u)}$ constitute $\bm \Psi_{j_1,j_2}^{(u)}$ in \eqref{uj1j2}. In this manner, the dissimilarity between $\bm X_t^{(1)}$ and $\bm X_t^{(2)}$ is obtained by means of the Euclidean distance between $\bm \Psi^{(1)}$ and $\bm \Psi^{(2)}$
\begin{equation}\label{d_qcd}
\begin{split}
d_{QCD}(\bm X_t^{(1)}, \bm X_t^{(2)})= \Big[||\Re_v(\bm \Psi^{(1)})-\Re_v(\bm \Psi^{(2)})||^2+||\Im_v(\bm \Psi^{(1)})-\Im_v(\bm \Psi^{(2)})||^2\Big]^{1/2},
\end{split}
\end{equation}

\noindent where $\Re_v$ and $\Im_v$ denote the element-wise real and imaginary part operations, respectively, and $\hat G^{j_1,j_2}_{T,R}(\omega_{k}, \tau_{i}, \tau_{i^ {\prime}})^{(u)}$ is the corresponding element of the matrix given by \eqref{est.qcrossdens} for the series $\bm X_t^{(u)}$.





Computation of vectors $\bm \Psi^{(1)}, \ldots, \bm \Psi^{(n)}$ for every MTS in $S$ could be used to perform fuzzy clustering by means of an algorithm as fuzzy $C$-means or fuzzy $C$-medoids considering the distance $d_{QCD}$. This distance has been successfully applied to perform clustering on MTS in a crisp framework \citep{lopez2021quantile}.

\section{Robust fuzzy clustering based on the QCD-based distance}\label{sectionrobustclustering}

In this section, we introduce three novel robust model for fuzzy clustering of MTS, namely QCD-FCMn-E, QCD-FCMn-NC and QCD-FCMn-T. All of them are based on the non-robust fuzzy QCD-FCMn model devised in \cite{oriona2021c} and presented firstly in Section~\ref{subsectionqcdfcmn}. Then, the proposed techniques are  developed in Sections~\ref{subsectionexp}, \ref{subsectionnc} and \ref{subsectiontrimmed}.

\subsection{QCD-based fuzzy $C$-means clustering model (QCD-FCMn)} \label{subsectionqcdfcmn}

As in previous sections, consider a set $S$ of $n$ realizations of multivariate time series $\{\bm X_t^{(1)}, \ldots, \bm X_t^{(n)}\}$ and denote by $\bm \Psi=\{\bm \Psi^{(1)}, \ldots, \bm \Psi^{(n)}\}$ the corresponding vector of estimated quantile cross-spectral densities obtained as indicated in Section~\ref{subsectiondqcd}. Assume that all vectors $\bm \Psi^{(i)}$ have the same length, $2d^2r^2(\lfloor T/2 \rfloor+1)$, being $d$ the number of dimensions, $r$ the number of probability levels, $T$ the series length and $\lfloor \cdot \rfloor$ the floor function. This way, a pairwise dissimilarity matrix can be computed according to \eqref{d_qcd}. In this framework, we propose to perform partitional fuzzy clustering on $S$ by using the QCD-based fuzzy $C$-means clustering model (QCD-FCMn), whose aim is to find a set of centroids $\overline{\bm \Psi}=\{\overline{\bm \Psi}^{(1)}, \ldots, \overline{\bm \Psi}^{(C)}\}$, and the $n \times C$ matrix of fuzzy coefficients, $\bm U=(u_{ic})$, $i=1,\ldots,n$,  $c=1,\ldots,C$, which define the solution of the minimization problem
\begin{equation}\label{qcd_means}
\begin{dcases}
\min_{\overline{\bm \Psi}, \bm U}\sum_{i=1}^{n}\sum_{c=1}^{C}u_{ic}^m 
\norm{\bm \Psi^{(i)}-\overline{\bm \Psi}^{(c)}}^2, \quad \text{subject to:}  \\
\sum_{c=1}^{C}u_{ic}=1, \, u_{ic} \geq 0  \quad \text{and} \quad \overline{\bm \Psi}_L \le \overline{\bm \Psi}^{(c)}_k \le \overline{\bm \Psi}_U
\end{dcases}
\end{equation}
where $u_{ic} \in[0,1]$ represents the membership degree of the $i$-th series in the $c$-th cluster, $\overline{\bm \Psi}^{(c)}$ is the vector of estimated quantile cross-spectral densities with regards to the centroid series for the cluster $c$, $m>1$ is a parameter controlling the fuzziness of the partition, usually referred to as fuzziness parameter, and $\overline{\bm \Psi}^{(c)}_k$ is the $k$-th component of centroid $\overline{\bm \Psi}^{(c)}$. Constraints on $u_{ic}$ are standard requirements in fuzzy clustering. Specifically, that the sum of the membership degrees for each series equals one implies that all of them contribute with the same weight to the clustering process. The fuzziness parameter controls the level of fuzziness considered in the grouping procedure. In the naive case, when $m=1$, we have $u_{ic}=1$ if $c=\underset{{c' \in \{1,\ldots,c\}}}{\operatorname{\argmin}}d^2_{QCD}(\bm \Psi^{(i)}, \overline{\bm \Psi}^{(c')})$ and 0 otherwise so that the crisp version of the procedure is obtained. As the value of $m$ increases, the boundaries between clusters become softer and therefore the grouping is fuzzier. Note that the centroid of a cluster is the mean of all points in the cluster (here, the quantile cross-spectral features describing the MTS), weighted by the degree of belonging to the cluster. Hence, we can think of the centroids as the prototypes of each cluster, i.e, a feature vector artificially representing the characteristics of the time series belonging to that cluster with a high membership degree. In \eqref{qcd_means}, $\overline{\bm \Psi}_L$ an $\overline{\bm \Psi}_U$ stand for the possible lower and upper bound of $\overline{\bm \Psi}^{(c)}_k$, respectively. 

The goal of QCD-FCMn is to find a fuzzy partition into $C$ clusters such that the squared QCD-distance between the clusters and their prototypes is minimized. The quality of the clustering solution strongly depends on the capability of $d_{QCD}$ to identify different dependence structures. Unlike crisp clustering, here the non-stochastic uncertainty inherent to the assignment of series to clusters is incorporated to the procedure by means of the membership degrees. 

Regarding only the membership degree constraints, the optimization problem in \eqref{qcd_means} can be solved by the Lagragian multipliers method, given rise to a two-step iterative process. First, the minimization of the objective function with respect to $u_{ic}$ is carried out for $\bm {\overline \Psi}$ fixed.
\begin{equation}\label{updatememmeans}
u_{ic}=\Bigg[ \sum_{c'=1}^{C} \Bigg ( \frac{\norm{\bm \Psi^{(i)}-\overline{\bm \Psi}^{(c)}}^2}{ \norm{\bm \Psi^{(i)}-\overline{\bm \Psi}^{(c')}}^2}\Bigg )^{\frac{1}{m-1}} \Bigg]^{-1}, 
\end{equation} 
for $i=1,\ldots,n$ and $c=1,\ldots,C$.

The second step is based on the minimization of the objective function regarding $\bm {\overline \Psi}$, being $u_{ic}$ fixed
\begin{equation}\label{updatecent}
\overline{\bm \Psi}^{(c)}_k=\frac{\sum_{i=1}^{n}u_{ic}^m{\bm \Psi}^{(i)}_k}{\sum_{i=1}^{n}u_{ic}^m}, 
\end{equation}
for $k=1,\ldots, d^2r^2(\lfloor T/2 \rfloor+1)$ and $c=1,\ldots,C$.

Note that, in the prior iterative solutions, $\overline{\bm \Psi}^{(c)}_k$ already verifies that $\overline{\bm \Psi}_L \le \overline{\bm \Psi}^{(c)}_k \le \overline{\bm \Psi}_U$, since it inherits the possible constraints of the estimated quantile cross-spectral density features from the observed series, $\bm \Psi^{(i)}$ ($i=1,\ldots,n$), i.e., $\overline{\bm \Psi}_L \le {\bm \Psi}^{(i)}_k \le \overline{\bm \Psi}_U$. Indeed, from this inequality follows $\sum_{i=1}^{n}u_{ic}^m\overline{\bm \Psi}_L \le \sum_{i=1}^{n}u_{ic}^m{\bm \Psi}^{(i)}_k \le \sum_{i=1}^{n}u_{ic}^m\overline{\bm \Psi}_U$, and hence 
$\overline{\bm \Psi}_L \le \overline{\bm \Psi}^{(c)}_k \le \overline{\bm \Psi}_U$. An outline of the QCD-FCMn clustering algorithm is shown in Algorithm \ref{algorithm1}. 

\begin{algorithm}[h]
	\small
	\caption{The QCD-based fuzzy $C$-means clustering algorithm (QCD-FCMn)  \label{algorithm1}}
	\begin{algorithmic}[1]
		\State Fix $C$, $m$, \textit{max.iter}, \textit{tol}, \text{a matrix norm} $\norm{\cdot}_M$ 
		\State Set $iter \, =0$ 
		\State Initialize the membership matrix $\bm U = \bm U^{(0)}$
		\Repeat
		\State Set $\bm U_{\text{OLD}}= \bm U$   
		\Comment{Store the current membership matrix}
		\State Compute the centroids $\overline{\bm \Psi}^{(c)}$, $c=1,\ldots,C$, using \eqref{updatecent}
		\State Compute $u_{ic}$, $i=1,\ldots,n$, $c=1,\ldots,C$, using (\ref{updatememmeans})
		\hspace*{0.1cm} \Comment{Update the membership matrix}
		\State $iter \, \gets iter \, + 1$
		\Until{ \mbox{ $\norm{\bm U -\bm U_{\text{OLD}}}_M<tol \mbox{ or } iter \, = \, max.iter$ } }
	\end{algorithmic}
\end{algorithm}

The original QCD-FCMn model can be improved by performing the PCA transformation over the set of feature vectors $\bm \Psi=\{\bm \Psi^{(1)}, \ldots, \bm \Psi^{(n)}\}$ and applying the fuzzy $C$-means algorithm to the corresponding set of score vectors. In \cite{oriona2021c}, we showed that, by proceeding this way, the clustering algorithm generally increases its effectiveness because a lot of noise gets removed. Indeed, when there are well defined clusters, the use of PCA provides a partition with less overlap between groups, thus giving more informative solutions. Note that this is not surprising, since the QCD-based features are highly correlated due to the definition of the smoothed CCR-periodogram in \eqref{smoothedCCR}. Therefore, by considering the raw features, some variables could get a higher weight than others in the distance computation, thus creating a bias in the clustering algorithm. The PCA transformation avoids this problem by removing the underlying correlation between features, thus making the grouping process easier. For these reasons, from now on, the distance $d_{QCD}$ and the clustering procedure QCD-FCMn and those in subsequent sections are going to refer to the PCA-transformed features rather than the original features, although we maintain the notation for the sake of simplicity and readability. 

\subsection{QCD-based exponential fuzzy $C$-means clustering model (QCD-FCMn-E)}\label{subsectionexp}

The QCD-based exponential fuzzy $C$-means clustering model (QCD-FCMn-E) considers a different objective function given by:
\begin{equation}\label{minexp}
\begin{dcases}
\min_{\overline{\bm \Psi}, \bm U} \sum_{i=1}^{n} \sum_{c=1}^{C} u_{ic}^m\left[1-\text{exp}\left\{-\beta \norm{ \bm{\Psi}^{(i)}-\bm{\overline{\Psi}}^{(c)}}_2^2 \right\} \right]\\
\mbox{subject to:  } \sum_{c=1}^{C} u_{ic}=1 \mbox{  and  } u_{ic}\geq 0,
\end{dcases} 
\end{equation}
where $\beta>0$ is a constant.

The local optimal solution for \eqref{minexp} is given by (see \cite{wu2002alternative})
\begin{equation}\label{memexponential}
u_{ic}=\left(
\sum_{c'=1}^{C}
\left[
\frac{1-\text{exp}\left\{-\beta \norm{ \bm{\Psi}^{(i)}-\bm{\overline{\Psi}}^{(c)}}_2^2 \right\}}
{1-\text{exp}\left\{-\beta \norm{ \bm{\Psi}^{(i)}-\bm{\overline{\Psi}}^{(c')}}_2^2 \right\}}
\right]^{\frac{1}{m-1}}\right)^{-1}.
\end{equation}

The fuzzy clustering based on the Exponential distance is more robust than the one based on the Euclidean distance \citep{wu2002alternative}. Indeed, Exponential distance gives different weights to each element in accordance to whether an element is anomalous or not, namely small weights to outliers and larger weights to compact points in the dataset \citep{wu2002alternative, d2015time}.

An appropriate choice of the hyperparameter $\beta$ is totally crucial for a good performance of QCD-FCMn-E (see Section 4 in \cite{wu2002alternative} for further details). When $\beta$ tends to zero, the QCD-FCMn-E algorithm tends to the QCD-FCMn algorithm, which gives equal weight to all elements in the dataset regardless their outlying nature. The value of $\beta$ is usually determined as the inverse of the variability in the data (the more variability in the data, the less the value of $\beta$). This quantity has an impact on the membership degrees \eqref{memexponential} in terms of robustness to outliers. The following choice for $\beta$ has proven to be suitable for different types of datasets:
\begin{equation}  \label{selectbeta}
\beta=\left(
\frac{1}{n} \, \sum_{i=1}^{n}
\norm{\bm{\Psi}^{(i)}-\overline{\bm{\Psi}}^{(k)}}^2_2\right)^{-1},
\end{equation}
where $\overline{\bm{\Psi}}^{(k)}$ corresponds to the index $k$ such that
$\displaystyle  k=\argmin_{1 \leq i' \leq n}\sum_{i''=1}^{n}\norm{\bm{\Psi}^{(i'')}-\bm{\Psi}^{(i')}}^2_2$ (see \cite{d2015time} for more details).

In essence, QCD-FCMn-E adjusts the impact of anomalous series by smoothing their effect through suitable weights. This way, the membership degrees of the outliers are similarly distributed across the clusters but the true clustering structure is not seriously perverted because of their presence.

The QCD-FCMn-E robust clustering algorithm is implemented as outlined in Algorithm~\ref{Algorithm2}.

\begin{algorithm}[h]
	\small
	\caption{QCD-based exponential fuzzy $C$-means clustering algorithm (QCD-FCMn-E) \label{Algorithm2} }
	\begin{algorithmic}[1]
		\State Fix $C$, $m$ and \textit{max.iter}  
		\State Compute $\beta$ using \eqref{selectbeta}
		\State Set $iter \, =0$ 
		\State Initialize the membership matrix $\bm U = \bm U^{(0)}$
		\Repeat
		\State Set $\bm U_{\text{OLD}}= \bm U$   
		\Comment{Store the current membership matrix}
		\State Compute the centroids $\overline{\bm \Psi}^{(c)}$, $c=1,\ldots,C$, using \eqref{updatecent}
		\State Compute $u_{ic}$, $i=1,\ldots,n$, $c=1,\ldots,C$, using (\ref{memexponential})
		\hspace*{0.1cm} \Comment{Update the membership matrix}
		\State $iter \, \gets iter \, + 1$
		\Until{ \mbox{ $\norm{\bm U -\bm U_{\text{OLD}}}<tol \mbox{ or } iter \, = \, max.iter$ } }
	\end{algorithmic}
\end{algorithm}

\subsection{QCD-based fuzzy $C$-means clustering with noise cluster (QCD-FCMn-NC)}\label{subsectionnc}

The QCD-based fuzzy $C$-means clustering model with noise cluster (QCD-FCMn-NC) is specified as the following minimization problem:
\begin{equation}\label{minnc} 
\begin{dcases}
\min_{\bm{\overline{\Psi}}, \bm U}  \sum_{i=1}^{n}\sum_{c=1}^{C-1} u_{ic}^m \norm{ \bm{\Psi}^{(i)}-\bm{\overline{\Psi}}^{(c)}}_2^2 + \sum_{i=1}^{n} \delta^2 \left( 1-\sum_{c=1}^{C-1} u_{ic}\right)^m \\
\mbox{subject to:  } \displaystyle  \sum_{c=1}^{C} u_{ic}=1 \mbox{  and  } u_{ic}\geq 0,
\end{dcases} 
\end{equation}
where $\delta>0$ is the so-called noise distance. 

Note that the QCD-FCMn-NC model involves $C$ clusters, but only $(C-1)$ are ``valid'' clusters. The additional cluster, the noise cluster, is artificially created for outlier identification purposes. The aim is to locate the outliers and place them in the noise cluster, which is represented by a fictitious prototype time series that has a constant distance from every MTS (the noise distance, $\delta$). If the distance from a given MTS to a centroid series is lower than $\delta$, then the MTS is assigned to the real cluster; otherwise, it is located into the noise cluster. 

The difference $u_{i*}=1-\sum_{c=1}^{C-1}u_{ic}$ in the second term of the objective function in \eqref{minnc} expresses the membership degree of the $i$-th time series to the noise cluster, $i=1,\ldots,n$. This quantity is expected to take high values for series showing an outlying nature. Hence, by definition, the usual constraint on the membership degrees for the real clusters is here relaxed to $\sum_{c=1}^{C-1}u_{ic}<1$, which allows outlying time series to have small membership values in real clusters. 

Minimization of \eqref{minnc} with regards to the membership degrees can be carried out in a similar fashion to the noise clustering method suggested by \cite{dave1991characterization}, yielding:
\begin{equation}\label{memnc}
\begin{split}
u_{ic}=  \left( \sum_{c'=1}^{C}\left[
\frac{
	\norm{\bm{\Psi}^{(i)}-\bm{\overline{\Psi}}^{(c)}}_2^2}
{\norm{\bm{\Psi}^{(i)}-\bm{\overline{\Psi}}^{(c')}}_2^2}\right]^{\frac{1}{m-1}}+ 
\left[\frac{\norm{ \bm{\Psi}^{(i)}-\bm{\overline{\Psi}}^{(c)}}_2^2}{\delta^2}\right]^{\frac{1}{m-1}} \right)^{-1}.
\end{split}
\end{equation}

The hyperparameter $\delta$ regulates the degree of requirement to belong to the noise cluster so that a proper choice of $\delta$ is crucial to make the most of the QCD-FCMn-NC approach. When $\delta$ is too large, QCD-FCMn-NC degenerates to the non-robust version of the model and outliers are forced to belong to real clusters. On the other hand, if $\delta$ is too small, many objects can be considered as noise and misclassified into the noise cluster \citep{cimino2004noise}. Although there are some heuristic manners to estimate the proper value of $\delta$, its optimal choice under general conditions is still an open issue. \cite{dave2002robust} suggest to consider the dataset statistics and equate the computation of $\delta$ to the concept of ``scale'' in robust statistics \citep{dave1997robust}. However, as stated by \cite{cimino2004noise}, ``unfortunately, the proper estimation of [the] scale is not a trivial task \citep{dave1997noise} and requires some knowledge of the data, which cannot always be supposed in real clustering applications''. In the original noise clustering algorithm \citep{dave1991characterization}, the value of $\delta$ was set to:
\begin{equation}\label{eqn:delta}
\delta^2=\lambda \, \frac{1}{n(C-1)} \sum_{i=1}^{n}\sum_{c=1}^{C-1}\norm{ \bm{\Psi}^{(i)}-\bm{\overline{\Psi}}^{(c)}}_2^2,
\end{equation}
\noindent where $\lambda$ is a scale multiplier to be chosen in accordance with the nature of the data.

To select the most proper value of $\lambda$ (and consequently of $\delta$), \cite{cimino2004noise} proposes to execute the fuzzy clustering model with noise cluster with decreasing values of $\lambda$ and study the distribution of the percentage of objects assigned to the noise cluster. The distribution has a sudden change of slope (elbow) when the value of the noise distance is so small that elements naturally belonging to real clusters are grouped into the noise cluster. According to the elbow it is possible to figure out the optimal noise distance. The authors approximate the distribution of percentages with a Pareto distribution. 

The steps of the algorithm of QCD-FCMn-NC method are provided in Algorithm~\ref{Algorithm3}.

\begin{algorithm}[h]
	\small
	\caption{QCD-based fuzzy $C$-means clustering algorithm with noise cluster (QCD-FCMn-NC) \label{Algorithm3} }
	\begin{algorithmic}[1]
		\State Fix $C-1$, $m$ and \textit{max.iter}  
		\State Set $iter=0$ 
		\State Initialize the membership matrix $\bm U = \bm U^{(0)}$
		\Repeat
		\State Set $\bm U_{\text{OLD}}= \bm U$   \Comment{Store the current membership matrix}
		\State Compute $\delta$ using (\ref{eqn:delta}).
		\State Compute the centroids $\overline{\bm \Psi}^{(c)}$, $c=1,\ldots,C$, using \eqref{updatecent}
		\State Compute $u_{ic}$, $i=1,\ldots,n$, $c=1,\ldots,C$, using (\ref{memnc})
		\hspace*{0.1cm} \Comment{Update the membership matrix}
		\State $iter \, \gets iter \, + 1$
		\Until{ \mbox{ $\norm{\bm U -\bm U_{\text{OLD}}}<tol \mbox{ or } iter \, = \, max.iter$ } }
	\end{algorithmic}
\end{algorithm}

\subsection{QCD-based trimmed fuzzy $C$-means clustering (QCD-FCMn-T)}\label{subsectiontrimmed}

The QCD-based trimmed fuzzy $C$-means clustering model (QCD-FCMn-T) attains its robustness by removing a certain proportion of the series and requires the specification of the fraction $\alpha$ of the data to be trimmed. Then, all non-trimmed series are classified using the QCD-FCMn model.

For a fixed trimming ratio $\alpha$, the QCD-FCMn-T method can be formalized as the following minimization problem:
\begin{align} \label{mintrimmed}
\left\{ \begin{array}{l}
\textrm{min}_{\bm{Y},\bm U}  \displaystyle \sum_{i=1}^{H(\alpha)}\sum_{c=1}^{C} u_{ic}^m \norm{ \bm{\Psi}^{(i)}-\overline{\bm{\Psi}}^{(c)}}^2\\
\mbox{subject to:  } \displaystyle  \sum_{c=1}^{C} u_{ic}=1 \mbox{  and  } u_{ic}\geq 0.
\end{array} \right.
\end{align}
where $\bm{Y}$ ranges on all the subsets of $\bm{\Psi}$ of size $H(\alpha) = \lfloor n(1 - \alpha) \rfloor$. If $\alpha$ is set to 0, then none of the series is removed and the non-robust QCD-FCMn model is obtained. In fact, the local optimal solution for the estimation of the membership degrees $u_{ic}$ is provided by (\ref{updatecent}), with $i$ ranging in the subset of the non-trimmed series and $c=1,\dots,C$. By replacing the expression of the $u_{ic}$ in the objective function \eqref{mintrimmed}, a bit of algebra yields
\begin{align} 
\displaystyle \sum_{i=1}^{H(\alpha)}\left[\sum_{c=1}^{C} \left(\norm{ \bm{\Psi}^{(i)}-\bm{\overline{\Psi}}^{(c)}}^2\right)^{1/(1-m)} \right]^{1-m}=\sum_{i=1}^{H(\alpha)} h_i,
\end{align}
where 
\begin{align}
h_i = \displaystyle  \left[\sum_{c=1}^{C} \left(\norm{ \bm{\Psi}^{(i)}-\bm{\overline{\Psi}}^{(c)}}^2\right)^{1/(1-m)} \right]^{1-m}.
\end{align}

The objective function for QCD-FCMn-T takes the form  
\begin{equation}\label{eqn:hjp} 
\displaystyle  \sum_{i=1}^{H(\alpha)} h_{i:n},
\end{equation}
where $h_{i:n}$ represents the $i$-th item when $h_i$, $i=1,\dots,n$, are arranged in ascending order. Specification \eqref{eqn:hjp} is very advantageous to implement the procedure. 

Notice that, following \cite{krishnapuram2001low}, QCD-FCMn-T is defined by considering the Least Trimmed Squares approach. The value of $H(\alpha)<n$ is chosen depending on how many series we would like to eliminate in the clustering process. For instance, when $H(\alpha)=\lfloor n/2 \rfloor$, 50\% of the time series are not involved in the clustering process, and the objective function is minimized when the centroids are computed in such a way that the sum of harmonic mean (squared) Euclidean distance of 50 \% of the MTS to the corresponding centroids is as small as possible (further details in \cite{d2014robust} and \cite{d2017autoregressive}). 

Algorithm~\ref{Algorithm4} contains the steps concerning the QCD-FCMn-T model.

\begin{algorithm}[h]
	\small
	\caption{QCD-based trimmed fuzzy $C$-means clustering algorithm (QCD-FCMn-T)  \label{Algorithm4} }
	\begin{algorithmic}[1]
		\State Fix $C$, $m$, $\alpha$ and \textit{max.iter}  
		\State Set $iter \, =0$ 
		\State Initialize the membership matrix $\bm U = \bm U^{(0)}$
		\State Compute the initial centroids $\overline{\bm \Psi}=\{\overline{\bm \Psi}^{(1)}, \ldots, \overline{\bm \Psi}^{(C)}\}$, by means of \eqref{updatecent}
		\Repeat
		\State Set $\bm U_{\text{OLD}}= \bm U$  \Comment{Store the current membership matrix}
		\State Identify the subset $\bm{Y}$ formed by the $H(\alpha)=\left[ n(1-\alpha) \right]$ series closest to the centroids, i.e. minimizing $\sum_{i=1}^{H(\alpha)} h_{i:n}$
		\State Compute $u_{ic}$, $i=1,\ldots,n$, $c=1,\ldots,C$, using (\ref{memnc})\Comment{Update the membership matrix}
		\State Compute the centroids by means of
		\Statex \hspace*{0.45cm} $
		\overline{\bm \Psi}^{(c)}_k=\frac{\sum_{i=1}^{H(\alpha)}u_{ic}^m{\bm \Psi}^{(i)}_k}{\sum_{i=1}^{H(\alpha)}u_{ic}^m}, \, \, \text{for} \,
		k=1,\ldots, d^2r^2(\lfloor T/2 \rfloor+1) \, \, \text{and} \, \, c=1,\ldots,C
		$
		\State $iter \, \gets iter \, + 1$
		\Until{ \mbox{ $\overline{\bm{\Psi}}_{\text{OLD}} = \overline{\bm{\Psi}} \mbox{ or } iter \, = \, max.iter$ } }
	\end{algorithmic}
\end{algorithm}

\section{Simulation study}\label{sectionsimulationstudy}

This section is devoted to analyze the results from a wide simulation study designed to assess the effectiveness of the proposed robust methods. First, some alternative metrics considered for comparison purposes are described. Then, the simulation and assessment mechanisms are detailed and the main results discussed.

\subsection{Alternative metrics}
To shed light on the clustering accuracy of QCD-FCMn-E, QCD-FCMn-NC and QCD-FCMn-T, they were compared with some other robust clustering models based on alternative dissimilarities. Here, it is worthy to remark the lack in the literature of robust procedures to perform MTS clustering. However, the QCD-based fuzzy approach developed in Section~\ref{subsectionqcdfcmn} can be directly adjusted to involve other kind of extracted features. Specifically, an alternative fuzzy $C$-means model can be formalized as the minimization problem in \eqref{qcd_means} by replacing $\bm \Psi^{(i)}$, $\overline{\bm \Psi}$, $\overline{\bm \Psi}^{(c)}$, $\overline{\bm \Psi}_L$,  $\overline{\bm \Psi}_U$ by $\bm \varphi^{(i)}$, $\overline{\bm \varphi}$, $\overline{\bm \varphi}^{(c)}$, $\overline{\bm \varphi}_L$,  $\overline{\bm \varphi}_U$, respectively, with $\bm \varphi^{(i)}$ representing a vector of estimates of selected features for the $i$-th series, $i = 1\ldots, n$, and the remaining terms defined analogously. The iterative solutions are obtained through \eqref{updatememmeans} and \eqref{updatecent} by considering the corresponding features. In the same way, exponential, noise, and trimmed versions are directly constructed starting from \eqref{minexp}, \eqref{minnc} and \eqref{mintrimmed}, respectively. 

Thus, for comparison purposes, we introduced alternative robust fuzzy models based on other MTS features, which are described below.

\medskip

\noindent $\bullet$ \textit{Wavelet-based features}. \cite{d2012wavelets} introduced a squared Euclidean distance between wavelet features of two MTS considering estimates of wavelet variances and correlations. The estimates are obtained through the maximum overlap discrete wavelet transform, which requires choosing a wavelet filter of a given length and a number of scales. Thus, in this case the vector $\bm \varphi^{(i)}$ contains estimates of wavelet variances and wavelet correlations of a given MTS. The corresponding robust techniques are referred to as Wavelet-based Exponential Fuzzy $C$-Means Clustering model (W-FCMn-E), Wavelet-based Fuzzy $C$-Means Clustering model with Noise Cluster (W-FCMn-NC) and Wavelet-based Trimmed Fuzzy $C$-Means Clustering (W-FCMn-T). After performing some preliminary analyses, we concluded that the wavelet filter of length 4 of the Daubechies family, DB4, along with the maximum allowable number of scales, were the choices producing the best average results in the considered simulation scenarios (see Sections \ref{subsubsectionfirstassessment} and \ref{subsubsectionsecondassessment}). 

\medskip

\noindent $\bullet$ \textit{Correlation-based features}. In the univariate framework, \cite{d2009autocorrelation} proposed a fuzzy procedure based on the estimated autocorrelations of a UTS for lags between 1 and fixed $l$, which was extended to the multivariate context in \cite{oriona2021c}. Given an MTS and a fixed lag $l$, estimates of both the autocorrelations up to lag $l$ for each component (UTS) and the cross-correlations up to lag $l$ between each pair of components are calculated. This set of features defines the vector $\bm {\varphi}^{(i)}$ characterising the $i$-th MTS and used to perform clustering. We call the corresponding robust approaches Correlation-based Exponential Fuzzy $C$-Means Clustering model (C-FCMn-E), Correlation-based Fuzzy $C$-Means Clustering model with Noise Cluster (C-FCMn-NC) and Correlation-based Trimmed Fuzzy $C$-Means Clustering (C-FCMn-T). Since most of the considered generating processes contain only one significant lag (see Sections \ref{subsubsectionfirstassessment} and \ref{subsubsectionsecondassessment}), the hyperparameter $l$ was set to $l=1$  throughout the simulation study.

\subsection{Experimental design and results}
The simulation study was designed to reach general conclusions on the performance of the proposed robust methods, QCD-FCMn-E, QCD-FCMn-NC and QCD-FCMn-T. To this end, two different assessment schemes were planned. Both of them contain scenarios formed by two regular groups of MTS plus one or two outlying series. In the first scheme, the outlying series are anomalous because they are generated from processes different from the ones defining the regular clusters. In the second scheme, the atypical MTS share certain patterns with the MTS in regular clusters but are distorted in some specific way, which makes for a more challenging setting. This way, the different models are assessed under different degrees of difficulty in relation to outlier identification. 

\subsubsection{First assessment scheme}\label{subsubsectionfirstassessment}

With the aim of getting insights into the robustness of the proposed approaches, a wide range of generating processes were considered to build our simulation scenarios, including linear, nonlinear and conditionally heteroskedastic models. Each of the considered setups consisted of two well-established clusters (base scenario), with five realizations each from the same generating process, successively contaminated by adding one and two outlier series. The specific scenarios and the generation schemes are given below.

\medskip

\noindent \textbf{Robust fuzzy clustering of linear models} 

\noindent \textsc{Base Scenario 1}: A VAR(1) process given by
\begin{equation*}
\begin{pmatrix}
X_{t,1}  \\
X_{t,2} 
\end{pmatrix} = 
\begin{pmatrix}
0.2 & 0.2 \\
0.2 & 0.2 
\end{pmatrix} 
\begin{pmatrix}
X_{t-1,1}   \\
X_{t-1,2}
\end{pmatrix} +
\begin{pmatrix}
\epsilon_{t,1}   \\
\epsilon_{t,2}
\end{pmatrix},
\end{equation*}
\noindent and a VMA(1) process given by
\begin{equation*}
\begin{pmatrix}
X_{t,1}  \\
X_{t,2} 
\end{pmatrix} = 
\begin{pmatrix}
\epsilon_{t,1}   \\
\epsilon_{t,2}
\end{pmatrix} +
\begin{pmatrix}
-0.4 & -0.4 \\
-0.2 & -0.2 
\end{pmatrix} 
\begin{pmatrix}
\epsilon_{t-1,1}   \\
\epsilon_{t-1,2}
\end{pmatrix}.
\end{equation*}

\noindent \textsc{Scenario 1.1}: The ten series simulated from the Base Scenario 1 plus one outlier generated from a VARMA process with matrices of coefficients given in the Base Scenario 1.

\noindent \textsc{Scenario 1.2}: The eleven series from Scenario 1.1 plus one additional outlier time series simulated from the NAR process introduced in Scenario 2.1 below. 

\medskip

\noindent \textbf{Robust fuzzy clustering of nonlinear models}

\noindent \textsc{Base Scenario 2}: A EXPAR (exponential autoregressive) process given by 
\begin{equation*}
\begin{pmatrix}
X_{t,1}  \\
X_{t,2} 
\end{pmatrix} = 
\begin{pmatrix}
0.3-10\exp({-X_{t-1,1}^2-X_{t-1,2}^2})X_{t-1,2} \\
0.3-10\exp({-X_{t-1,1}^2-X_{t-1,2}^2})X_{t-1,1}
\end{pmatrix} 
+
\begin{pmatrix}
\epsilon_{t,1}   \\
\epsilon_{t,2}
\end{pmatrix},
\end{equation*}
\noindent and a BL (bilinear) process given by 
\begin{equation*}
\begin{pmatrix}
X_{t,1}  \\
X_{t,2} 
\end{pmatrix} = 
\begin{pmatrix}
0.6X_{t-1,1}+0.7X_{t-1,1}\epsilon_{t-1,2}+\epsilon_{t,1} \\
0.6X_{t-1,2}+0.7X_{t-1,2}\epsilon_{t-1,1}+\epsilon_{t,2}
\end{pmatrix} 
+
\begin{pmatrix}
\epsilon_{t,1}   \\
\epsilon_{t,2}
\end{pmatrix}.
\end{equation*}
\noindent \textsc{Scenario 2.1}: The ten series simulated from the Base Scenario 2 plus one outlier time series simulated from the NAR (nonlinear autoregressive) process
\begin{equation*}
\begin{pmatrix}
X_{t,1}  \\
X_{t,2} 
\end{pmatrix} = 
\begin{pmatrix}
0.7|X_{t-1,1}|/(|X_{t-1,2}|+1) \\ 
0.7|X_{t-1,2}|/(|X_{t-1,1}|+1)
\end{pmatrix} 
+
\begin{pmatrix}
\epsilon_{t,1}   \\
\epsilon_{t,2}
\end{pmatrix}.
\end{equation*}
\noindent \textsc{Scenario 2.2}: The series from Scenario 2.1 plus an additional outlier time series simulated from the VAR process
\begin{equation*}
\begin{pmatrix}
X_{t,1}  \\
X_{t,2} 
\end{pmatrix} = 
\begin{pmatrix}
0.1 & 0.1 \\
0.1 & 0.1 
\end{pmatrix} 
\begin{pmatrix}
X_{t-1,1}   \\
X_{t-1,2}
\end{pmatrix} +
\begin{pmatrix}
\epsilon_{t,1}   \\
\epsilon_{t,2}
\end{pmatrix},
\end{equation*}

\medskip

\noindent \textbf{\textbf{Robust fuzzy clustering of conditional heteroskedastic models}} 

\noindent \textsc{Base Scenario 3}: Consider
\begin{equation}
\begin{pmatrix}
X_{t,1}  \\
X_{t,2} 
\end{pmatrix} = \Sigma^{1/2}_t\begin{pmatrix}
\epsilon_{t,1}  \\
\epsilon_{t,2}
\end{pmatrix}.
\end{equation}
The data generating processes consist of two bivariate BEKK (Baba-Engle-Kraft-Kroner) models given by
\begin{equation}
\Sigma_t=C^\intercal C+A^\intercal\begin{pmatrix}
X_{t,1}  \\
X_{t,2} 
\end{pmatrix}\begin{pmatrix}
X_{t,1}  \\
X_{t,2} 
\end{pmatrix}^\intercal A+G^\intercal\Sigma_{t-1}G,
\end{equation}
where $C$ is a lower triangular matrix and $A$ and $G$ are $2\times 2$ matrices. In the first process, $C=\begin{pmatrix}
0.1 & 0  \\
0.1 & 0.1 
\end{pmatrix}$, $A=\begin{pmatrix}
0.2 & 1.2  \\
0.4 & 0.5 
\end{pmatrix}$ and  $G=\begin{pmatrix}
0.2 & -0.1  \\
-0.1 & -0.1
\end{pmatrix}$, whereas $C=\begin{pmatrix}
0.1 & 0  \\
0.1 & 0.1 
\end{pmatrix}$, $A=\begin{pmatrix}
0.5 & 0.4  \\
0.7 & -0.2 
\end{pmatrix}$ and  $G=\begin{pmatrix}
-0.5 & -0.4  \\
-0.1 & -0.4
\end{pmatrix}$, for the second generating process.

\noindent \textsc{Scenario 3.1}: The ten series simulated from the Base Scenario 3 plus one outlier time series simulated from a bivariate white noise process (WN).

\noindent \textsc{Scenario 3.2}: The eleven series from Scenario 3.1 plus one outlier time series simulated from the BL process of Scenarios 2.1 and 2.2.

In all cases, the error process $(\epsilon_{t,1},\epsilon_{t,2})^\intercal$ consisted of iid realizations from a bivariate standard Gaussian distribution. 

VARMA models are broadly used in many fields but determining the model order is a complex task. Fixing orders too small leads to inconsistent estimators whereas too large orders produce less accurate predictions. Note that our approach does not require prior modeling. Scenarios 2.1 and 2.2 consist of a multivariate extension of the univariate NAR process proposed in \cite{Zhang2001} and EXPAR and NL processes proposed in \cite{lafuente2018robust}. Nonlinear UTS arise in several application fields \citep{granger1993modelling, granger1978introduction, tong2009threshold}. Although nonlinear MTS have received much less attention than linear ones, there exist some fields as neurophysiology \citep{pereda2005nonlinear} and economy \citep{Koop1996} in which nonlinear analysis of MTS has proven to be critical. Thus, a good fuzzy clustering method should be able to specify proper membership degrees between different nonlinear generating processes. Scenarios 3.1 and 3.2 are motivated by the BEKK models introduced in \cite{engle1995multivariate}. BEKK models are a formulation of multivariate GARCH models frequently used to model MTS of financial variables as stock returns, commodity prices, inflation and exchange rates, among others \citep{chevallier2012time, rahman2012oil, heidari2013inflation}. Financial time series are known to exhibit empirical statistical regularities, so-called ``stylized-facts''. The most common stylized facts include: heavy tails and a peak center compared to the normal distribution, volatility clustering (periods of low volatility mingle with periods of high volatility), leverage effects (the quantities are negatively correlated with volatility), and autocorrelation at much longer horizons than expected. In our previous work \cite{lopez2021quantile}, the distance $d_{QCD}$ was tested either for clustering, classification and outlier detection purposes by considering a different extension of multivariate GARCH models, the so-called dynamic conditional correlation models (DCC) \citep{engle2002dynamic}, achieving great results. BEKK and DCC models are the most utilised formulations of multivariate GARCH models. It is worth remarking that each formulation has been shown to be is useful in its own right, and the choice of the proper formulation often depends on the empirical application \citep{caporin2012we}. Thus, in this work we have decided to evaluate the distance $d_{QCD}$ with BEKK models. Note that in Scenarios 1.2, 2.2 and 3.2, the second anomalous series is generated from a kind of process different from the ones concerning the remaining series. For instance, Scenario 2.2 three different types of nonlinear processes along with a linear one. The heterogeneity in terms of generating processes was introduced in these scenarios with the aim of making more demanding the clustering task.

\begin{figure}[h]
	\centering
	\includegraphics[width=0.8\textwidth]{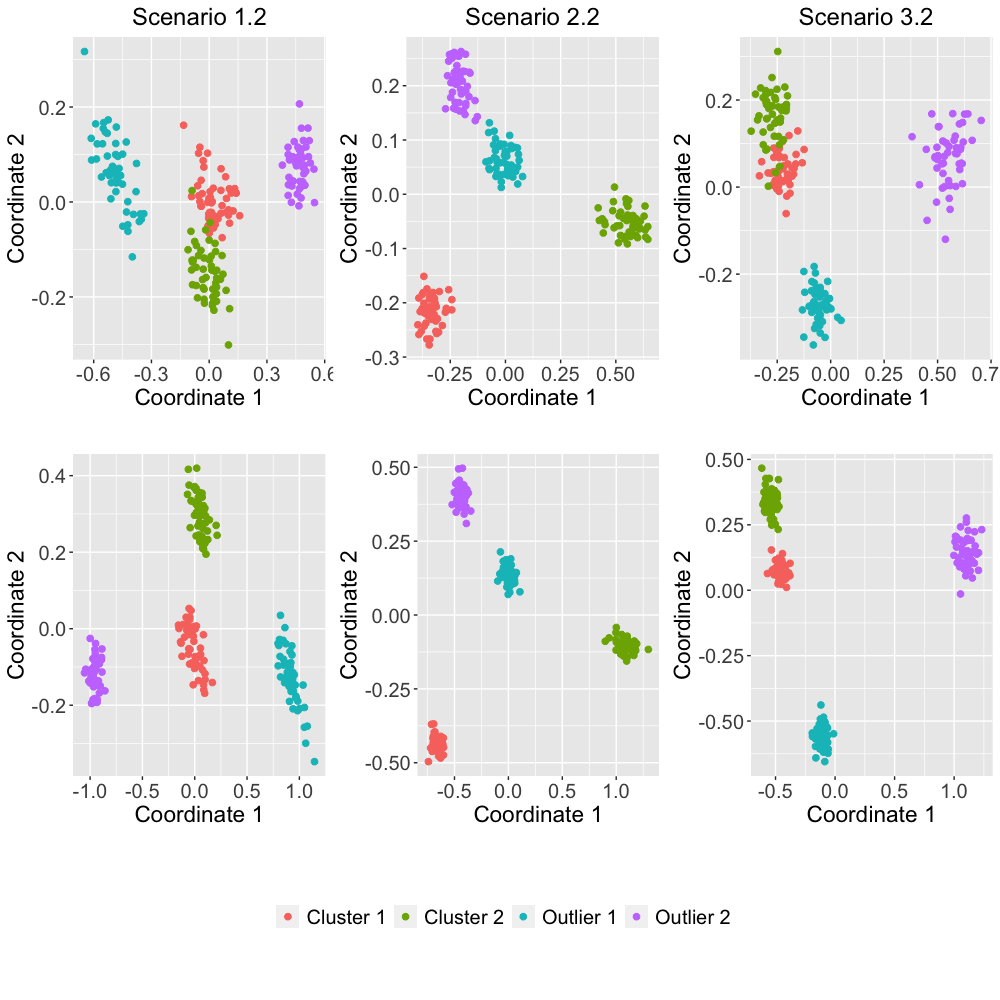}
	\vspace*{-1cm}
	\caption{Two-dimensional scaling based on QCD distances between simulated series in Scenarios 1.2, 2.2 and 3.2. The series length is $T=500$ (top panel) and $T=2000$ (bottom panel).}
	\label{mds}
\end{figure}

As a preliminary step, a metric two-dimensional scaling (2DS) based on the QCD distances was carried out to examine if $d_{QCD}$ is expected to perform robust fuzzy clustering correctly in the considered scenarios. 2DS represents the pairwise QCD distances in terms of Euclidean distances into a 2-dimensional space preserving the original distances as well as possible (by minimising a loss function). For Scenarios 1.2, 2.2 and 3.2, we simulated 50 MTS of equal length $T$ from each involved process, i.e. using the models proposed to generate the MTS of the two regular clusters and the two outlying MTS. The 2DS was performed for each set of 200 MTS and two values of $T$, $T=500$ and $T=2000$. The resulting 2-D scatter plots are depicted in Figure~\ref{mds}, where points coming from the same process exhibit the same colour. 





The quality of the embedding can be measured by the $R^2$ value, reporting the proportion of variance of the scaled data accounted for the 2DS procedure. We obtained the values 0.767 (Scenario 1.2), 0.751 (Scenario 2.2) and 0.746 (Scenario 3.2) when $T=500$ and 0.8905 (Scenario 1.2), 0.900 (Scenario 2.2) and 0.898 (Scenario 3.2) when $T=2000$. Notice that values above 0.6 are considered to provide an acceptable scaling procedure, whereas values above 0.8 mean a very good fit \citep{hair2009multivariate}. The values are very similar for the three considered scenarios, and all of them are above 0.6 for $T=500$, and above 0.8 for $T=2000$. Thus, we conclude that the plots in Figure \ref{mds} provide an accurate picture about the distance $d_{QCD}$ between the different underlying processes. 

Figure \ref{mds} shows different situations according to the scenario and the series length. For Scenarios 1.2 and 3.2 and $T=500$, the 100 series generated from the base scenario form a slightly overlapping cluster, whereas the groups of outlying series are well-separated. In the opposite way, for Scenario 2.2 and $T=500$, the two regular clusters are far away from one another and from the groups of anomalous series (which are very close to each other). When increasing the series length ($T=2000$), there are no longer overlapping clusters in any scenario, but the location of one given cluster in relation to the rest remains to same. This is expected since the QCD-based features are more accurately estimated by increasing $T$, and hence groups become more distant. In summary, Figure \ref{mds} uncovers that the robust approaches based on the QCD-based distance should be capable of discriminating between the regular clusters and the atypical series in Scenarios 1.2, 2.2 and 3.2 provided that the series length is large enough. Among all of them, Scenario 3.2 seems the more challenging scenario for a fixed $T$. 

Every proposed clustering procedure, QCD-FCMn-E, QCD-FCMn-NC and QCD-FCMn-T, and the robust versions of the alternative metrics were applied to cluster the series from Scenarios 1.1, 1.2, 2.1, 2.2, 3.1 and 3.2. In addition, the nonrobust QCD-FCMn model was also run to get a clear idea about to what extent the robust techniques are successful. The accuracy of each algorithm is measured by the proportion of times in which the series generated from the same process belong to the same cluster with a high membership degree. Robustness to outliers is inspected by studying to what extent the outlying series affect the resulting fuzzy partition. Concerning QCD-FCMn-T and QCD-FCMn-NC, the proportion of times that the anomalous series are trimmed or located in the noise cluster, respectively, are taken into consideration.

The number of clusters was set to $C=2$ as the three scenarios contain two base generating processes. As we are handling fuzzy partitions, it is necessary to specify a cutoff value in order to assign an MTS to a given cluster. For the nonanomalous series, we decided to assign the $i$-th element to the $c$-th cluster if $u_{ic}>0.7$. As for the outlying series, the assignment criterion was different among the three considered robust algorithms. By using the exponential cluster models, we assumed that the atypical series was correctly handled when $\underset{c \in \{1, 2\}}{\max}u_{ic}<0.7$, that is, when the outlying series exhibits similar membership degrees in both clusters. Regarding the noise cluster methods, we considered that an anomalous series was correctly identified if $u_{ic_{NC}}>0.5$, with ${c_{NC}}$ indicating the index of the noise cluster. Note that, in this case, the minimization problem in \eqref{minnc} involves three underlying clusters, so the value 0.5 is a stringent threshold. Finally, concerning the trimmed approaches, we examined if the anomalous series were correctly trimmed. It is worthy mentioning that these criteria and the chosen thresholds are in line with the suggestions recommended in the literature \citep{d2013noise, d2015time, d2009autocorrelation, maharaj2010wavelet, maharaj2011fuzzy, dembele2003fuzzy} and have already been considered in some works \citep{lafuente2018robust}. A given method is assumed to provide a correct clustering solution if it accurately handles both sets of series anomalous and nonanomalous. 

For obtaining the QCD-based features, 3 quantile levels were used, namely $\mathcal{T}=\{0.1, 0.5, 0.9\}$. This choice has proven to be enough to reach suitable results in a hard and soft clustering context \citep{lafuente2016clustering, vilar2018quantile, lopez2021quantile, oriona2021c}.

The fuzziness parameter $m$ plays a crucial role. When $m=1$, the crisp version of fuzzy $C$-means is obtained, and excessively large values of $m$ result in a partition with all memberships close to $1/C$, thus leading to a large degree of overlap between clusters. Thus, none of these values for $m$ is recommended \citep{arabie1981overlapping}. There is a broad range of literature on determining the appropriate fuzziness parameter. For instance, \cite{bezdek2013pattern} showed that values of $m$ between $1.5$ and $2.5$ are typically a good choice for the fuzzy-C means algorithm, which is also confirmed by \cite{cannon1986efficient} and \cite{hall1992comparison}. However, there seems to be no consensus about the optimal value for $m$ (see discussion in Section 3.1.6 of \cite{maharaj2011fuzzy}). In the context of time series clustering, the majority of works consider values of $m$ between 1.3 and 2.6 when performing simulation studies \citep{vilar2018quantile, d2012wavelets,d2009autocorrelation, maharaj2011fuzzy, lafuente2018robust}. Based on the previous considerations, we decided to use the values $m=1.8$, $2$, $2.2$ and $2.5$.

The three robust procedures involve hyperparameters, namely $\beta$, $\delta$ and $\alpha$, for exponential, noise and trimmed approaches, respectively. Our numerical experiments have revealed: (i) strong influence of these parameters on the clustering results, and (ii) their optimal values heavily rely on the selected value for $m$. Thus, we decided to proceed as follows. As far as the exponential and noise techniques, for a given method and value of $m$, the correct classification rate was recorded for a grid of equispaced values for the hyperparameter ranging from $0$ to $L$, $L$ being large enough to achieve near-zero rates. By doing so, we intend to assess not only the maximum rate of correct classification a method is able to achieve but also the sensitivity of the technique against the choice of the hyperparameter. Concerning the trimmed procedures, the trimming ratio $\alpha$ was chosen as to detect the real number of outliers, one in Scenarios 1.1, 2.1 and 3.1 and two in Scenarios 1.2, 2.2 and 3.2.

For each one of the nine scenarios, 100 trials were carried out by performing robust fuzzy clustering with all the described techniques. Then, the average percentages of correct classification over the 100 trials was computed as a measure of clustering effectiveness. In each scenario, two values for the series length were considered: $T\in\{750, 1500\}$ for Scenarios 1.1 and 1.2, $T\in\{600, 900\}$ for 2.1 and 2.2, and $T\in\{1500, 3000\}$ for 3.1 and 3.2. Since each scenario contains very distinct types of processes, it is reasonable that very different values of $T$ are needed to make an appropriate evaluation. In particular, the large values of $T$ in Scenarios 3.1 and 3.2 are due to the high variability inherent to the estimation process associated with heteroskedastic models, which requires to employ large realizations to generate accurate results, i.e. to allow $d_{QCD}$ to be capable of differentiate among the clusters. The 2DS plots in Figure \ref{mds} clearly support this argument by showing that, for a fixed value of $T$, Scenario 3.2 is expected to be the most demanding scenario. It is worth enhancing that this requirement is not necessarily a drawback since these sample sizes are often encountered in real MTS  fitted  by BEKK models \citep{wu2013dynamic, efimova2014energy}. Indeed multivariate series of stock returns and other  related  financial  quantities,  which  consist  of  measures  of  daily  or  even intra-daily data, are one common example of series fitted through this class of models. 

Curves of average rates of correct classification for the exponential-based approaches concerning Scenarios 1.1 and 1.2 are given in Figure~\ref{curvese11} for $T=750$ and in Figure~\ref{curvese12} in Appendix for $T=1500$.

\begin{figure}[h]
	\centering
	\includegraphics[width=0.85\textwidth]{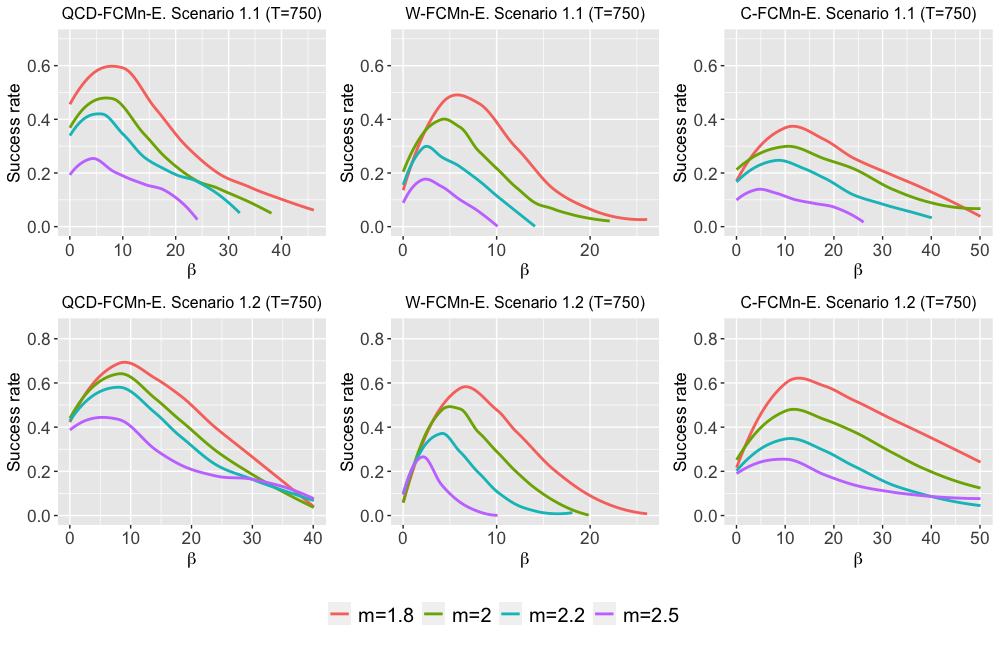}
	\vspace*{-0.75cm}
	\caption{Average rates of correct classification for QCD-FCMn-E, W-FCMn-E and C-FCMn-E as a function of $\beta$ in Scenarios 1.1 and 1.2 for series of length $T=750$ and four fuzziness levels $m$.}
	\label{curvese11}
\end{figure}

Figures \ref{curvese11} and \ref{curvese12} illustrate that the three procedures performed pretty well when dealing with linear processes. For $T=750$, QCD-FCMn-E achieved a maximum classification rate considerably above that of W-FCMn-E and C-FCMn-E in both scenarios. For $T=1500$, the three techniques showed a maximum average score close to one for $m=1.8$, $2$ and $2.2$ and a bit worse for $m=2.5$. The only exception to this fact was C-FCMn-E in Scenario 1.1, displaying a maximum average accuracy quite below that of its competitors for all values of $m$. This method showed the most stability to the choice of $\beta$, usually attaining acceptable classification rates for large values of $\beta$ such that the remaining two techniques obtained poor results. This fact is not surprising, as the sample autocorrelations are known to be highly discriminatory features in the matter of linear models. 

The results of Figures \ref{curvese11} and \ref{curvese12} are summarized in Table~\ref{tablee1}, where the maximum accuracy values and the areas under the curves (AUC) are given. To make the results easier to interpret in terms of comparison purposes, the AUC quantities were normalized with respect to the maximum value of AUC attained by a method for fixed $T$ and $m$. This way, the displayed AUC scores are bounded between 0 and 1. Note that the results in Table~\ref{tablee1} corroborate our previous remarks. In short, the three methods seem fairly capable of performing robust clustering of linear models, QCD-FCMn-E displaying better results for shorter lengths and C-FCMn-E exhibiting the least instability to the selection of the parameter $\beta$. 
\begin{table}[ht]
	\centering
	\small 
	\resizebox{8cm}{!}{
		\begin{tabular}{cc rr r c r rr}
			\hline
			\multicolumn{2}{l}{Exponential-based}	&  \multicolumn{3}{c}{Maximum} & & \multicolumn{3}{c}{AUC} \\ \cline{3-5} \cline{7-9}
			\multicolumn{2}{l}{approach}  &  \multicolumn{1}{c}{QCD} & \multicolumn{1}{c}{W} & \multicolumn{1}{c}{C}  & & \multicolumn{1}{c}{QCD} & \multicolumn{1}{c}{W} & \multicolumn{1}{c}{C} \\ \hline
			\multicolumn{9}{l}{\textsc{Scenario 1.1}}\\ \hline 
			$T=750$ & $m=1.8$ & \textbf{0.60} & 0.50 & 0.38 &  & 0.94 & 0.45 & \textbf{1} \\ 
			& $m=2.0$ & \textbf{0.49} & 0.41 & 0.29 & & 0.97 & 0.48 & \textbf{1}\\ 
			& $m=2.2$ & \textbf{0.43} & 0.27 & 0.23 & & \textbf{1} & 0.41 & 0.89 \\ & $m=2.5$ & \textbf{0.25} & 0.14 & 0.12 & & \textbf{1} & 0.32 & 0.41 \\   \hline 
			$T=1500$ & $m=1.8$ & 0.97 & \textbf{0.98} & 0.84 & & 0.54 & 0.54 & \textbf{1} \\ 
			& $m=2.0$ & 0.95 & \textbf{0.96} & 0.73 & & 0.51 & 0.52 & \textbf{1} \\ & $m=2.2$ & 0.90 & \textbf{0.92} & 0.65 & & 0.53 & 0.53 & \textbf{1} \\ & $m=2.5$ & \textbf{0.83} & 0.79 & 0.50 & & 0.66 & 0.58 & \textbf{1} \\  \hline 
			\multicolumn{9}{l}{\textsc{Scenario 1.2}}\\ \hline 
			$T=750$ & $m=1.8$ & \textbf{0.71} & 0.59 & 0.63 & & 0.77 & 0.30 & \textbf{1}  \\ 
			& $m=2.0$ & \textbf{0.63} & 0.49 & 0.48 & & 0.81 & 0.29 & \textbf{1} \\ & $m=2.2$ & \textbf{0.57} & 0.39 & 0.37 & & 0.87 & 0.26 & \textbf{1} \\ & $m=2.5$ & \textbf{0.44} & 0.26 & 0.25 & & \textbf{1} & 0.23 & 0.94 \\ \hline 
			$T=1500$ & $m=1.8$ & \textbf{0.99} & 0.96 & 0.96 & & 0.64 & 0.53 & \textbf{1} \\ 
			& $m=2.0$ & \textbf{0.98} & 0.95 & 0.89 & & 0.55 & 0.45 & \textbf{1} \\ 
			& $m=2.2$ & \textbf{0.95} & 0.90 & 0.83 & & 0.48 & 0.40 & \textbf{1} \\ 
			& $m=2.5$ & \textbf{0.88} & 0.78 & 0.67 & & 0.54 & 0.41 & \textbf{1} \\ \hline
		\end{tabular}
	}
	\caption{Maximum correct classification rate and normalized AUC for QCD-FCMn-E, W-FCMn-E and C-FCMn-E. Scenarios 1.1 and 1.2.} 
	\label{tablee1}
	\small
\end{table}


Results for the noise approaches regarding Scenarios 1.1 and 1.2 are shown in Table~\ref{tablenc1}, Figure \ref{curvesnc11} for $T=500$ and Figure~\ref{curvesnc12} in Appendix for $T=1500$. Again, the three strategies performed well, although the results are slightly distinct from the ones obtained with the exponential approach. For $T=1500$ and $m=1.8$, we obtained excellent rates of correct classification near one. When the value of $m$ got larger, the overall performance substantially decreased, specially for $m=2.5$ with very poor scores. When $m=2.5$, all the procedures produced very blurry partitions in which the outlying series were generally allocated in the noise cluster with a membership degree less than 0.5, thus causing failed trials. Table~\ref{tablenc1} shows that the scores from QCD-FCMn-NC are clearly above the ones concerning W-FCMn-NC and C-FCMn-NC, thus concluding that QCD-FCMn-NC was the best performing model in terms of both maximum accuracy and robustness to the choice of $\delta$.
\begin{table}[h]
	\centering
	\small 
	\resizebox{8cm}{!}{
		\begin{tabular}{cc rr r c r rr}
			\hline
			\multicolumn{2}{l}{Noise cluster}	&  \multicolumn{3}{c}{Maximum} & & \multicolumn{3}{c}{AUC} \\ \cline{3-5} \cline{7-9}
			\multicolumn{2}{l}{approach}  &  \multicolumn{1}{c}{QCD} & \multicolumn{1}{c}{W} & \multicolumn{1}{c}{C}  & & \multicolumn{1}{c}{QCD} & \multicolumn{1}{c}{W} & \multicolumn{1}{c}{C} \\ \hline
			\multicolumn{9}{l}{\textsc{Scenario 1.1}}\\ \hline 
			$T=750$ & $m=1.8$ & \textbf{0.63} & 0.55 & 0.58 & & 0.98 & \textbf{1} & 0.70 \\ 
			& $m=2.0$ & \textbf{0.43} & 0.29 & 0.36 &  & \textbf{1} & 0.86 & 0.72 \\ 
			& $m=2.2$ & \textbf{0.26} & 0.12 & 0.17 &  & \textbf{1} & 0.44 & 0.52 \\ 
			& $m=2.5$ & \textbf{0.07} & 0.01 & 0.01 & & \textbf{1} & 0.11 & 0 \\ 
			\hline 
			$T=1500$ & $m=1.8$ & \textbf{0.95} & 0.93 & 0.94 & & \textbf{1} & 0.83 & 0.59 \\ 
			& $m=2.0$ & 0.88 & 0.840 & \textbf{0.91}  &  & \textbf{1} & 0.75 & 0.59 \\ 
			& $m=2.2$ & \textbf{0.73} & 0.60 & 0.67   &  & \textbf{1} & 0.63 & 0.52 \\ 
			& $m=2.5$ & \textbf{0.31} & 0.15 & 0.20   &  & \textbf{1} & 0.31 & 0.35 \\ 
			\hline 
			\multicolumn{9}{l}{\textsc{Scenario 1.2}}\\ \hline 
			$T=750$ & $m=1.8$ & \textbf{0.61} & 0.34 & 0.58 & & \textbf{0.142} & 0.063 & 0.122 \\ 
			& $m=2.0$ & \textbf{0.43} & 0.15 & 0.34 &  & \textbf{1} & 0.25 & 0.75 \\ 
			& $m=2.2$ & \textbf{0.20} & 0.01 & 0.18 &  & \textbf{1} & 0 & 0.70 \\ 
			& $m=2.5$ & \textbf{0.05} & 0 & 0.02 & & \textbf{1} & 0 & 0.50 \\ 
			\hline 
			$T=1500$ & $m=1.8$ & \textbf{0.95} & 0.89 & 0.93 &  & \textbf{1} & 0.48 & 0.62 \\ 
			& $m=2.0$ & \textbf{0.85} & 0.68 & 0.81 &  & \textbf{1} & 0.39 & 0.57 \\ 
			& $m=2.2$ & \textbf{0.74} & 0.40 & 0.54 & & \textbf{1} & 0.28 & 0.48 \\ 
			& $m=2.5$ & \textbf{0.32} & 0.05 & 0.16 & & \textbf{1} & 0.08 & 0.32 \\ 
			\hline
		\end{tabular}
	}
	\caption{Maximum correct classification rate and normalized AUC for QCD-FCMn-NC, W-FCMn-NC and C-FCMn-NC, Scenarios 1.1 and 1.2.}
	\label{tablenc1} 
\end{table}
\begin{figure}[H]
	\centering
	\includegraphics[width=0.85\textwidth]{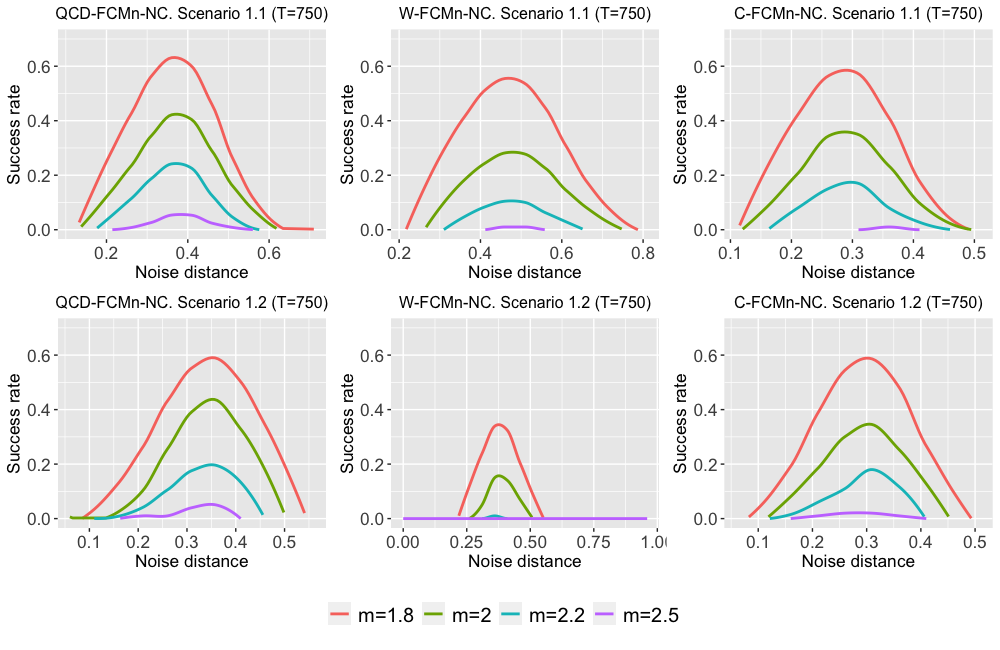}
	\vspace*{-0.75cm}
	\caption{Average rates of correct classification for QCD-FCMn-NC, W-FCMn-NC and C-FCMn-NC as a function of $\delta$ in Scenarios 1.1 and 1.2 for series of length $T=750$ and four fuzziness levels $m$.}
	\label{curvesnc11}
\end{figure}


The three trimming-based procedures accomplished very similar proportions of successful trials in the linear scenarios (Table \ref{tabletrimmed1}). QCD-FCMn-T overall outperformed W-FCMn-T and C-FCMn-T, specially in Scenario 1.2 with $T=750$. By comparing the results in Tables~\ref{tablee1},  \ref{tablenc1} and \ref{tabletrimmed1}, we conclude that the trimmed approaches are lightly more effective than their counterparts approaches based on the exponential and noise techniques in Scenarios 1.1 and 1.2.
\begin{table}[h]
	\centering
	\small 
	\resizebox{8.5cm}{!}{
		\begin{tabular}{cc rr r c r rr}
			\hline
			\multicolumn{2}{l}{Trimmed approach}	&  \multicolumn{3}{c}{\textsc{Scenario 1.1}} & & \multicolumn{3}{c}{\textsc{Scenario 1.2}} \\ \cline{3-5} \cline{7-9}
			\multicolumn{2}{l}{}  &  \multicolumn{1}{c}{QCD} & \multicolumn{1}{c}{W} & \multicolumn{1}{c}{C}  & & \multicolumn{1}{c}{QCD} & \multicolumn{1}{c}{W} & \multicolumn{1}{c}{C} \\ \hline
			$T=750$ & $m=1.8$ & 0.67 & 0.65 & \textbf{0.71} & &  \textbf{0.73} & 0.67 & 0.69  \\ 
			& $m=2.0$ & \textbf{0.59} & 0.55 & 0.56 & & \textbf{0.66} & 0.59 & 0.61 \\ 
			& $m=2.2$ & \textbf{0.52} & 0.44 & 0.44 & & \textbf{0.60} & 0.48 & 0.49 \\ 
			& $m=2.5$ & \textbf{0.33} & 0.30 & 0.23 & & \textbf{0.47} & 0.31 & 0.30  \\  \hline
			$T=1500$  & $m=1.8$ & \textbf{1} & 0.98 & 0.95 & & 0.98 & \textbf{0.99} & 0.97\\ 
			& $m=2.0$ & 0.97 & \textbf{0.98} & 0.93 & & \textbf{0.95} & 0.93 & 0.91 \\ 
			& $m=2.2$ & 0.95 & \textbf{0.98} & 0.87& & \textbf{0.93} & 0.91 & 0.87 \\ 
			& $m=2.5$ & 0.95 & \textbf{0.98} & 0.87 & &  \textbf{0.93} & 0.91 & 0.87  \\ \hline 
		\end{tabular}
	}
	\caption{Average rates of correct classification for QCD-FCMn-T, W-FCMn-T and C-FCMn-T. Scenarios 1.1 and 1.2.}
	\label{tabletrimmed1}
\end{table}


Results of QCD-FCMn-E, W-FCMn-E and C-FCMn-E in Scenarios 2.1 and 2.2, with nonlinear models, are given in Table~\ref{tablee2}, Figure~\ref{curvese21} for $T=600$ and Figure~\ref{curvese22} in Appendix for $T=900$. In all cases, QCD-FCMn substantially outmatched the alternative approaches. The correlation-based strategy C-FCMn-E was able to make some correct classifications for a broader range of values for $\beta$, but attaining an overall accuracy much less than QCD-FCMn-E. W-FCMn-E accomplished very poor results for all values of $m$, $\beta$ and $T$. Indeed, the average rates of successful trials attained by our proposed strategy were always close to one. 
\begin{table}[ht]
	\centering
	\small 
	\resizebox{8cm}{!}{
		\begin{tabular}{cc rr r c r rr}
			\hline
			\multicolumn{2}{l}{Exponential-based}	&  \multicolumn{3}{c}{Maximum} & & \multicolumn{3}{c}{AUC} \\ \cline{3-5} \cline{7-9}
			\multicolumn{2}{l}{approach}  &  \multicolumn{1}{c}{QCD} & \multicolumn{1}{c}{W} & \multicolumn{1}{c}{C}  & & \multicolumn{1}{c}{QCD} & \multicolumn{1}{c}{W} & \multicolumn{1}{c}{C} \\ \hline
			\multicolumn{9}{l}{\textsc{Scenario 2.1}}\\ \hline 
			$T=600$ & $m=1.8$ & \textbf{0.93} & 0.30 & 0.35 &  & \textbf{1} & 0.02 & 0.32 \\ 
			& $m=2.0$ &  \textbf{0.94} & 0.27 & 0.40 & & \textbf{1} & 0.01 & 0.43 \\ 
			&  $m=2.2$ & \textbf{0.96} & 0.16 & 0.44 &  & \textbf{1} & 0.01 & 0.54 \\ 
			&   $m=2.5$ & \textbf{0.98} & 0.09 & 0.46 & & \textbf{1} & 0.01 & 0.74 \\ 
			\hline 
			$T=900$ &  $m=1.8$ & \textbf{0.98} & 0.35 & 0.32 & & \textbf{1} & 0.03 & 0.38 \\ 
			&  $m=2.0$ & \textbf{0.99} & 0.36 & 0.43 &  & \textbf{1} & 0.02& 0.55 \\ 
			&  $m=2.2$ & \textbf{0.99} & 0.31 & 0.52 & & \textbf{1} & 0.02 & 0.74 \\ 
			&  $m=2.5$ &  \textbf{1} & 0.17 & 0.58 &  & 0.98 & 0.01 & \textbf{1} \\ 
			\hline 
			\multicolumn{9}{l}{\textsc{Scenario 2.2}}\\ \hline 
			$T=600$ &   $m=1.8$ & \textbf{0.99}& 0.27 & 0.27 &  & \textbf{1} & 0.02 & 0.32 \\ 
			&  $m=2.0$ & \textbf{0.99} & 0.28 & 0.35 &  & \textbf{1} & 0.01 & 0.42 \\ 
			&   $m=2.2$ & \textbf{0.99} & 0.21 & 0.37 & & \textbf{1} & 0.01 & 0.47 \\ 
			&   $m=2.5$ & \textbf{0.99} & 0.07 & 0.45  &  & \textbf{1} & 0.01 & 0.53 \\ 
			\hline 
			$T=900$ &  $m=1.8$ & \textbf{1} & 0.28 & 0.22 & & \textbf{1} & 0.03 & 0.31 \\ 
			&  $m=2.0$ & \textbf{1} & 0.26 & 0.29 & & \textbf{1}& 0.02 & 0.46 \\ 
			&  $m=2.2$ & \textbf{1} & 0.24 & 0.35 & & \textbf{1} & 0.01 & 0.58 \\ 
			&   $m=2.2$ & \textbf{1} & 0.12 & 0.46 & & \textbf{1} & 0.01 & 0.70 \\ 
			\hline
		\end{tabular}
	}
	\caption{Maximum correct classification rate and normalized AUC for QCD-FCMn-E, W-FCMn-E and C-FCMn-E. Scenarios 2.1 and 2.2.} 
	\label{tablee2}
\end{table}
\begin{figure}[H]
	\centering
	\includegraphics[width=0.85\textwidth]{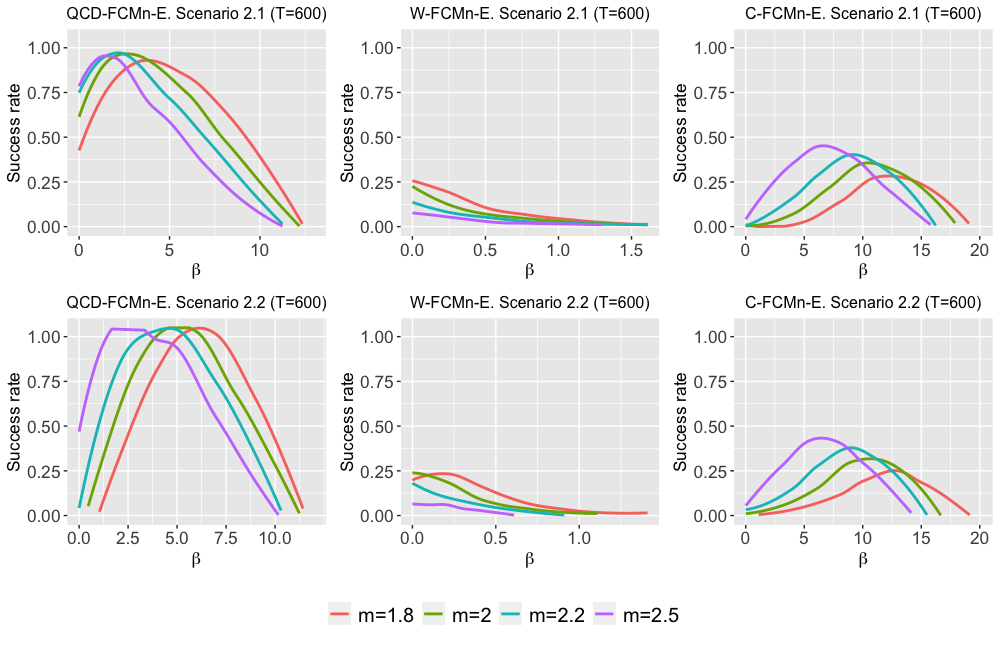}
	\vspace*{-0.75cm}
	\caption{Average rates of correct classification for QCD-FCMn-E, W-FCMn-E and C-FCMn-E as a function of $\beta$ in Scenarios 2.1 and 2.2 for series of length $T=600$ and four fuzziness levels $m$.}
	\label{curvese21}
\end{figure}


In Scenarios 2.1 and 2.2, the noise-based approaches lead to very similar results (Table~\ref{aucnc2}, see also Figures~\ref{curvesnc21} and \ref{curvesnc22} in Appendix). Again, QCD-FCMn-NC clearly attained the best average scores by far. The wavelet-based method W-FCMn-NC was incapable of making correct classifications, and C-FCMn-NC achieved very low success rates in Scenario 2.1, specially for larger values of $m$, and was totally ineffective in Scenario 2.2. 
\begin{table}[ht]
	\centering
	\small 
	\resizebox{8cm}{!}{
		\begin{tabular}{cc rr r c r rr}
			\hline
			\multicolumn{2}{l}{Noise cluster}	&  \multicolumn{3}{c}{Maximum} & & \multicolumn{3}{c}{AUC} \\ \cline{3-5} \cline{7-9}
			\multicolumn{2}{l}{approach}  &  \multicolumn{1}{c}{QCD} & \multicolumn{1}{c}{W} & \multicolumn{1}{c}{C}  & & \multicolumn{1}{c}{QCD} & \multicolumn{1}{c}{W} & \multicolumn{1}{c}{C} \\ \hline
			\multicolumn{9}{l}{\textsc{Scenario 2.1}}\\ \hline 
			$T=600$ &  $m=1.8$ & \textbf{0.53} & 0 & 0.14 & & \textbf{1} & 0 & 0.14 \\ 
			& $m=2.0$ & \textbf{0.34} & 0 & 0.09 & & \textbf{1} & 0 & 0.10 \\ 
			&  $m=2.2$ & \textbf{0.29} & 0 & 0.03 & & \textbf{1} & 0 & 0.07 \\ 
			&  $m=2.5$ & \textbf{0.17} & 0 & 0 & & \textbf{1} & 0 & 0 \\  
			\hline 
			$T=900$ & $m=1.8$ & \textbf{0.88} & 0 & 0.23 & & \textbf{1} & 0 & 0.09 \\ 
			& $m=2.0$ & \textbf{0.81} & 0 & 0.14 & &\textbf{1} & 0 & 0.07 \\ 
			& $m=2.2$ &  \textbf{0.70} & 0 & 0.07 & & \textbf{1} & 0 & 0.03 \\ 
			& $m=2.5$ &  \textbf{0.58}& 0 & 0.01 & & \textbf{1} & 0 & 0 \\ 
			\hline
			\multicolumn{9}{l}{\textsc{Scenario 2.2}}\\ \hline 
			$T=600$ &  $m=1.8$ & \textbf{0.97} & 0 & 0 & & \textbf{1} & 0 & 0 \\ 
			& $m=2.0$ & \textbf{0.94} & 0 & 0 & & \textbf{1} & 0 & 0 \\ 
			& $m=2.2$ & \textbf{0.87} & 0 & 0 &  & \textbf{1} & 0 & 0 \\ 
			&  $m=2.5$ & \textbf{0.66} & 0 & 0 &  & \textbf{1} & 0 & 0 \\ 
			\hline 
			$T=900$ & $m=1.8$ & \textbf{1} & 0 & 0 & & \textbf{1} & 0 & 0 \\ 
			& $m=2.0$ & \textbf{1} & 0 & 0 & & \textbf{1} & 0 & 0 \\ 
			& $m=2.2$ & \textbf{0.99} & 0 & 0 & & \textbf{1} & 0 & 0  \\ 
			& $m=2.5$ & \textbf{0.91} & 0 & 0 & & \textbf{1} & 0 & 0 \\ 
			\hline
		\end{tabular}
	}
	\caption{Maximum correct classification rate and normalized AUC for QCD-FCMn-NC, W-FCMn-NC and C-FCMn-NC, Scenarios 2.1 and 2.2.}
	\label{aucnc2} 
\end{table}

From Table \ref{aucnc2} clearly follows that W-FCMn-NC and C-FCMn-NC are not suitable techniques for handling the nonlinear processes defining Scenarios 2.1 and 2.2. To unravel the reasons behind of this poor behavior, we carried out a detailed examination of the clustering solutions. It was observed that W-FCMn-NC often allocated the single or both outlying series in the noise cluster with large membership degrees. Moreover, some of the series generated from the BL process were also placed in the noise cluster, thus provoking unsuccessful trials. When increasing the value of noise distance $\delta$, the anomalous MTS get assigned to regular clusters, but at least one of the BL series remains located in the noise cluster until $\delta$ is large enough. As a result, W-FCMn-NC is not able to perform successful classifications. A similar situation occurs in Scenario 2.2 for C-FCMn-NC. The partitions returned by C-FCMn-NC in this scenario are characterised by the allocation of at least one of the BL series in the noise cluster, hence producing failure trials. In Scenario 2.1, although some BL series are also frequently placed in the noise cluster, there exist some trials in which all these series are correctly classified whereas the single outlying series is clearly located into the noise cluster. This accounts for the low success rate achieved by C-FCMn-NC in Scenario 2.1. In short, the procedures W-FCMn-NC and C-FCMn-NC find it hard to classify the series generated from the BL process in Scenarios 2.1 and 2.2, which in turn leads to these methods being incapable of counterbalancing the effects of the abnormal MTS. 

\begin{figure}[h]
	\centering
	\includegraphics[width=0.7\textwidth]{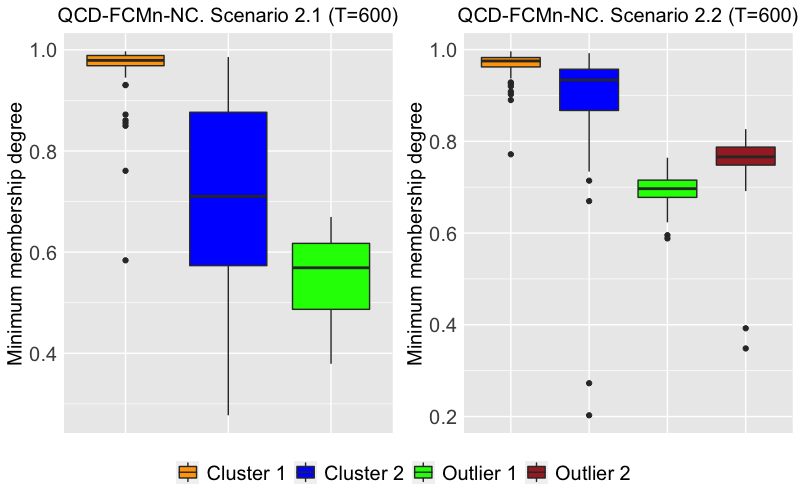}
	\vspace*{-0.75cm}
	\caption{Minimum membership degrees of the series in cluster 1, the series in cluster 2 and the outlying series in the noise cluster.}
	\label{boxplotsnc2}
\end{figure}

Interestingly enough, the QCD-based approach substantially ameliorated its efficacy when dealing with two outliers. To gain insights into this fact, we fixed $T=600$, $m=1.8$, and took the membership matrices produced by QCD-FCMn-NC in each one of the trials with the optimal values of $\delta$ (see Figure~\ref{curvesnc21} in Appendix), namely $\delta=0.38$ for Scenario 2.1 and $\delta=0.3$  for Scenario 2.2. In each trial, we first recorded the maximum membership value for each MTS, and then group these values separately by kind of series (EXPAR and BL) and selected the minimum at each group. These minimum values are crucial to know if the method made a correct classification. In addition, the membership values of the outlying series in relation to the noise cluster were also recorded. The boxplots of these minimum membership degrees over all the trials are shown in Figure~\ref{boxplotsnc2}. 

The left panel of Figure~\ref{boxplotsnc2} illustrates that the failure trials of QCD-FCMn-NC in Scenario 2.1 are mainly due to the series generated from the BL process (Cluster 2). Almost half of the times at least one BL series was assigned to Cluster 1 with a membership value less than 0.7. Although some outlier series were located in the noise cluster with a membership degree less than 0.5, the number of misclassifications due to this fact was negligible. The success rate of QCD-FCMn-NC in this setting was 0.53. When the second atypical series from the VAR process was included (right panel of Figure~\ref{boxplotsnc2}), the minimum membership degrees of the BL series considerably increased and both anomalous series were generally allocated in the noise cluster with membership values far beyond 0.5. Thus, including the VAR series made QCD-FCMn-NC better capable of both neutralizing the effects of the outliers and detecting the true clustering structure, thus attaining a greater success rate (0.97). 
The corresponding graph for Scenario 2.2 in the 2DS scatter plots of Figure~\ref{mds} help us to understand  the improvement exhibited by QCD-FCMn-NC when dealing with two outliers. We observe that the first anomalous series is closer to the regular clusters than the second one. Thus, when the second outlying series, which is close to the first one, is introduced, a small ``cluster'' of two outliers is formed, making it easier for the technique to distinguish between the true clusters and the abnormal MTS.


The scores generated by the trimmed techniques in Scenarios 2.1 and 2.2 are given in Table \ref{tabletrimmed2}. Not surprisingly, QCD-FCMn-T obtained the best scores by a large degree. The method C-FCMn-T attained markedly greater results than C-FCMn-E and C-FCMn-NC, but they are not comparable with the ones of QCD-FCMn-T. It is worth noting that, in the context of nonlinear processes, the exponential approach QCD-FCMn-E displayed a better performance than QCD-FCMn-NC and QCD-FCMn-T. 
\begin{table}[h]
	\centering
	\small 
	\resizebox{8cm}{!}{
		\begin{tabular}{cc rr r c r rr}
			\hline
			\multicolumn{2}{l}{Trimmed approach}	&  \multicolumn{3}{c}{\textsc{Scenario 2.1}} & & \multicolumn{3}{c}{\textsc{Scenario 2.2}} \\ \cline{3-5} \cline{7-9}
			\multicolumn{2}{l}{}  &  \multicolumn{1}{c}{QCD} & \multicolumn{1}{c}{W} & \multicolumn{1}{c}{C}  & & \multicolumn{1}{c}{QCD} & \multicolumn{1}{c}{W} & \multicolumn{1}{c}{C} \\ \hline
			$T=600$ & $m=1.8$ & \textbf{0.78} & 0.13 & 0.54 & & \textbf{0.98} & 0.21 & 0.53  \\ 
			& $m=2.0$ & \textbf{0.76} & 0.15 & 0.54 & &  \textbf{0.98} & 0.17 & 0.54 \\ 
			& $m=2.2$ & \textbf{0.74} & 0.10 & 0.53 & & \textbf{0.98} & 0.14 & 0.52 \\ 
			& $m=2.5$ & \textbf{0.69} & 0.08 & 0.53 & &  \textbf{0.98} & 0.12 & 0.51  \\  \hline
			$T=900$  & $m=1.8$ & \textbf{0.96} & 0.22 & 0.68 & &\textbf{1} & 0.17 & 0.63\\ 
			& $m=2.0$ & \textbf{0.96} & 0.18 & 0.68 & &  \textbf{1} & 0.17 & 0.62 \\ 
			& $m=2.2$ & \textbf{0.95} & 0.18 & 0.68 & & \textbf{1} & 0.14 & 0.63 \\ 
			& $m=2.5$ & \textbf{0.95} & 0.18 & 0.68 & & \textbf{1} & 0.14 & 0.63  \\ \hline 
		\end{tabular}
	}
	\caption{Average rates of correct classification for QCD-FCMn-T, W-FCMn-T and C-FCMn-T. Scenarios 2.1 and 2.2.}
	\label{tabletrimmed2}
\end{table}


With regards to the results for Scenarios 3.1 and 3.2, remark that both the wavelet-based and the correlation-based methods achieved very poor scores. In fact, the exponential and the noise approaches led to success rates close to zero. The four strategies W-FCMn-E, C-FCMn-E, W-FCMn-NC and C-FCMn-NC are prone to the same types of error. Depending on the value of $\beta$ or $\delta$, they usually (i) are not capable of clearly differentiating between both BEKK processes since the ten MTS in the base scenario obtained membership degrees close to 0.5 in both clusters (the same occurred for the atypical series), or (ii) locate all the BEKK series  in one cluster with high memberships and the one or two outlying series in another cluster, also with high memberships. In addition, the noise cluster techniques W-FCMn-NC and C-FCMn-NC were sometimes capable of assigning correctly the outlying series into the noise cluster, but in this case the BEKK series are mixed. In conclusion, it is clear that these approaches are not able to properly distinguish between BEKK models nor neutralize the effect of the abnormal series. Therefore, in Scenarios 3.1 and 3.2, we decided only to show the results based on the QCD approaches. In particular, the results for QCD-FCMn-E and QCD-FCMn-NC are jointly shown in Table~\ref{tablee3nc3} (see also Figures~\ref{curvese31}, \ref{curvese32}, \ref{curvesnc31} and \ref{curvesnc32} in Appendix). Note that in this case the AUC values were not normalized because only the results based on QCD are considered.
\begin{table}[h]
	\centering
	\small
	\resizebox{8cm}{!}{
		\begin{tabular}{cc rr c rr}
			\hline 
			&  &  \multicolumn{2}{c}{QCD-FCMn-E} & & \multicolumn{2}{c}{QCD-FCMn-NC} \\ \cline{3-4} \cline{6-7}
			&  & Max & AUC & & Max & AUC \\ \hline
			\multicolumn{7}{l}{\textsc{Scenario 3.1}}\\ \hline 
			$T=1500$ & $m=1.8$  & 0.99 & 30.83  & & 0.990 & 0.419  \\ 
			& $m=2.0$           & 0.99 & 27.18  & & 0.970 & 0.349  \\ 
			& $m=2.2$           & 0.99 & 22.88  & & 0.900 & 0.309 \\ 
			& $m=2.5$           & 0.99 & 18.03  & & 0.710 & 0.157 \\ 
			\hline 
			$T=3000$  & $m=1.8$ & 1    & 21.45  & & 1 & 0.659 \\ 
			& $m=2.0$           & 1    & 19.15  & & 1 & 0.582 \\ 
			& $m=2.2$           & 0.99 & 17.08  & & 1 & 0.513 \\ 
			& $m=2.5$           & 1    & 13.66  & & 0.950 &  0.396 \\ 
			\hline 
			\multicolumn{7}{l}{\textsc{Scenario 3.2}}\\ \hline 
			$T=1500$ & $m=1.8$  & 0.56 & 4.87   & & 0.920  & 0.294 \\ 
			& $m=2.0$           & 0.53 & 4.26   & & 0.850  & 0.242 \\ 
			& $m=2.2$           & 0.47 & 3.18   & & 0.750  & 0.191 \\
			& $m=2.5$           & 0.42 & 1.99   & & 0.480  & 0.103 \\ 
			\hline 
			$T=3000$ & $m=1.8$  & 0.59 &  3.10 & &  1 & 0.494 \\ 
			& $m=2.0$           & 0.62 &  2.61 & &  1 & 0.440 \\ 
			& $m=2.2$           & 0.57 &  2.01  & &  0.970 & 0.374 \\ 
			& $m=2.5$           & 0.47 &  1.16  & &  0.880 & 0.282 \\ 
			\hline
		\end{tabular}
	}
	\caption{Maximum correct classification rate and AUC for QCD-FCMn-E and QCD-FCMn-NC. Scenarios 3.1 and 3.2.}
	\label{tablee3nc3}
\end{table}

The results in Table~\ref{tablee3nc3} indicate that both models performed very well when only one outlier was present (Scenario 3.1), reaching perfect scores regardless of the value of $m$. The scores worsened when the additional outlier given by the BL process was added (Scenario 3.2). It seems that this second outlying series made much more challenging the clustering task, particularly for QCD-FCMn-E. In fact, in Scenario 3.2, the noise cluster approach performed clearly better than the exponential one for both values of $T$ and all values of $m$. For $T=3000$, QCD-FCMn-NC attained perfect scores when $m=1.8$, 2, 2.2 for some values of $\delta$. 

In these scenarios involving conditional heteroskedastic processes, the trimmed method QCD-FCMn-T exhibited the same pattern as QCD-FCMn-E and QCD-FCMn-NC (see Table \ref{tabletrimmed3} in Appendix): high scores in Scenario 3.1 and worse behaviour in Scenario 3.2. Notice that, in the case of BEKK processes, the exponential-based strategy defeated both QCD-FCMn-NC and QCD-FCMn-T.


It can be concluded that each one of the approaches, exponential, noise and trimmed, has shown to be the most appropriate choice for coping with a different type of generating processes. This insight is illuminating, as it indicates that no method dominates the remaining ones under all circumstances. Thus, practitioners should always take into account the three proposed techniques when performing robust clustering of MTS. 

To better understand to what extent the QCD-based robust approaches are useful, Table~\ref{tableqcdraw2} simultaneously provides the rates of correct classification reached by the nonrobust QCD-FCMn method and the highest one obtained among the three robust approaches (in brackets). On balance, the robust techniques are clearly beneficial when there exist outlier series in the dataset. Although the differences are more pronounced for the smallest values of $m$, they hold for all $m$, tus concluding that the proposed robust strategies are far less dependent on the suitable choice of $m$. This is a very desirable property given the difficulty of properly selecting $m$ in practice. The results for Scenario 3.2 are particularly noteworthy since, while the method QCD-FCMn was incapable of performing correct classification, the robust versions still took advantage of their powerful neutralizing capacity and produced high scores. 
\begin{table}[ht]
	\centering
	\small
	\resizebox{9cm}{!}{
		\begin{tabular}{ccrrrr}
			\hline
			Scenario & $T$ & $m=1.8$ & $m=2$ & $m=2.2$ & $m=2.5$ \\ \hline 
			\textsc{1.1} & 750 & 0.37 (\textbf{0.67}) & 0.39 (\textbf{0.59}) & 0.38 (\textbf{0.52}) & 0.24 (\textbf{0.33}) \\ 
			& 1500  & 0.33 (\textbf{1}) & 0.47 (\textbf{0.97}) & 0.53 (\textbf{0.95}) & 0.63 (\textbf{0.95}) \\ \hline 
			\textsc{1.2} & 750 &  0.48 (\textbf{0.73}) & 0.56 (\textbf{0.66}) & 0.55 (\textbf{0.60}) & 0.43 (\textbf{0.47}) \\ 
			& 1500  & 0.87 (\textbf{0.99}) & 0.95 (\textbf{0.98}) & \textbf{0.95} (\textbf{0.95}) & 0.88 (\textbf{0.93}) \\ \hline 
			\textsc{2.1} & 600 &  0.55 (\textbf{0.93}) & 0.79 (\textbf{0.94}) & 0.89 (\textbf{0.96}) & \textbf{0.98} (\textbf{0.98}) \\ 
			& 900  & 0.67 (\textbf{0.98}) & 0.87 (\textbf{0.99}) & 0.96 (\textbf{0.99}) & \textbf{1} (\textbf{1}) \\ \hline 
			\textsc{2.2} & 600 &  0.29 (\textbf{0.99}) & 0.53 (\textbf{0.99}) & 0.78 (\textbf{0.99}) & 0.94 (\textbf{0.99}) \\ 
			& 900  & 0.32 (\textbf{1}) & 0.56 (\textbf{1}) & 0.81 (\textbf{1}) & 0.99 (\textbf{1}) \\ \hline 
			\textsc{3.1} & 1500  & 0.76 (\textbf{1})  & 0.97 (\textbf{0.99}) & \textbf{1} (0.99) & \textbf{1} (0.99) \\ 
			& 3000 & 0.70 (\textbf{1}) & \textbf{1} (\textbf{1}) & \textbf{1} (\textbf{1}) & \textbf{1} (\textbf{1}) \\ \hline 
			\textsc{3.2} & 1500 & 0 (\textbf{0.92}) & 0 (\textbf{0.85}) & 0 (\textbf{0.75}) & 0 (\textbf{0.48}) \\ 
			& 3000  & 0 (\textbf{1}) & 0 (\textbf{1}) & 0 (\textbf{0.97}) & 0 (\textbf{0.88}) \\ 
			\hline
		\end{tabular}
	}
	\caption{Rates of correct classification reached by QCD-FCMn and by the best performing  robust QCD approach (in brackets).}
	\label{tableqcdraw2}
\end{table}


All simulation were replicated by considering heavy tails in the error distribution since this feature often arises in real time series, specially in finance \citep{harvey2013dynamic, bernardi2017multiple, rachev2003handbook, mikosch2003modeling}. Specifically, the innovations in all scenarios were simulated from a multivariate $t$ distribution with 3 degrees of freedom. Results for exponential, noise, and trimmed techniques are provided in Appendix, in Tables~\ref{expt}, \ref{nct} and \ref{trimmedt}, respectively. It is clearly observed that the quantile-based approaches exhibited the most robustness against the effect of heavy tails. For instance, whereas the three robust approaches concerning W-FCMn and C-FCMn performed quite accurately when handling linear processes with normal innovations (Tables \ref{tablee1}-\ref{tabletrimmed1}), they totally failed when some amount of fat-tailedness was introduced. Only the correlation-based technique was capable of attaining some successful trials, particularly the trimmed approach (C-FCMn-T). Similar behaviour is observed for the rest of scenarios. Interestingly the trimmed-based approaches W-FCMn-T and C-FCMn-T, although outperformed by QCD-FCMn-T, obtained high scores in Scenario 3.1 involving BEKK processes, even improving their results with Gaussian innovations (Table \ref{tabletrimmed3}). However, for Scenario 3.2, QCD-FCMn-T was the only approach attaining high success rates. 

Overall, the change in the error distribution did not affect the performance of the proposed approaches in Scenarios 1.1 and 1.2, and slightly decreased their scores in Scenarios 2.1, 2.2, 3.1 and 3.2, specially for the shortest value of the series length. In any case, the robust QCD-based versions have proven to be the best ones in terms of robustness not only against outlying series, but also against the distributional form of the error terms.

\subsubsection{Second assessment scheme}\label{subsubsectionsecondassessment}

In this section, we analyze robustness against a new type of atypical series, namely outliers constructed by distorting series pertaining to a regular cluster. More precisely, we consider two class of outliers which are pervasive in the MTS literature: the multivariate innovational outliers (MIO) and the multivariate temporary (or transitory) changes (MTC) (see e.g. \cite{tsay2000outliers, galeano2006outlier}). Note that although these types of outliers are usually defined in the context of VARMA processes, their extension to general types of processes is straightforward. Generally speaking, MIO appear when the noise distribution of the original series is perverted and a MTC occurs when the former series receives a sudden impact that disappears gradually with time. Both outliers are assumed to emerge at some particular time point, say $t_0$. Specifically, if $\bm X_t$ denotes an arbitrary $d$-dimensional realization generated from a given cluster with innovations following a distribution $F$, then a MIO series $\bm X_t'$ is generated identically as  $\bm X_t$ for $t=1,\ldots,t_0-1$, but with innovations following a distribution $F'$ for $t=t_0,\ldots,T$. On the other hand, a MTC series  $\bm X_t''$ is constructed as $\bm X_t''=\bm X_t$ if $t<t_0$ and $\bm X_{t_0+k}''=\bm X_{t_0}+\eta^k\bm w$ for $k=0, 1, \ldots T-t_0$, where $\bm w=(w_1, \ldots, w_d)^{\intercal}$ is the size of the outlier and $0<\eta<1$ is a constant which regulates the propagation of the anomalous effect in subsequent observations. 

Six new simulation scenarios, referred to as MTC 1, MTC 2, MTC 3, MIO 1, MIO 2 and MIO 3, were designed to investigate how the robust techniques deal with this class of outliers. For $i \in \{1, 2, 3\}$, Scenario MTC (MIO) $i$ consisted of $5$ realizations of each generating process in Base Scenario $i$ plus an MTC (MIO) outlier constructed from the first generating process in Base Scenario $i$. The outlying effect was introduced at $t_0=T/2$ in all cases. The outlier sizes used in the definition of the MTC outliers were $\bm w=(5, -5)^{\intercal}$ in Scenarios MTC 1 and MTC 2, and $\bm w=(1, -1)^{\intercal}$ in Scenario MTC 3. The parameter $\eta$ was always set to $0.99$. As for the MIO outliers, the distribution $F'$ was chosen to be the $\chi^2$ distribution with $3$ degrees of freedom in Scenarios MIO 1 and MIO 2, and the $\chi^2$ with $3/10$ degrees of freedom in Scenario MIO 3. Note that the selected parameter concerning Scenarios MTC 3 and MIO 3 produce less pronounced changes in the corresponding outlying series. The reason is that the marginal variance of both components in the first generating process of Base Scenario 3 is far less than that of their counterparts in Base Scenarios 1 and 2. Hence, a modification in both the size of the outliers and the mean and variance of the innovations needed to be made to maintain the difficulty in the outlier detection task in Scenarios MTC 3 and MIO 3. The remaining simulation features (cut-off values for fuzzy membership degrees, sample sizes, input parameters,...) and the alternative procedures are exactly the same as in Section~\ref{subsubsectionfirstassessment}.

Note that, unlike the outlying series handled in Section~\ref{subsubsectionfirstassessment}, MIO and MTC outliers can be detected by visual inspection of the realizations (see Figure~\ref{geometricoutlier} in Appendix to elucidate this assertion). However, in practice, by dealing with a dataset with hundreds or thousands of MTS, it is often unfeasible to perform outlier detection by visual examination. Thus, it is desirable to have available a proper robust clustering algorithm to be able to detect and neutralize the effect of these types of atypical series.

Table~\ref{tablesecondoutliers} contains the average correct classification rates concerning the six new scenarios. For the sake of simplicity, we have included the results only for the largest value of the series length $T$ and $m=1.8$ and $2.2$. It is evident from Table~\ref{tablesecondoutliers} that, overall, the QCD-based robust approaches outperform the other two methods in terms of outlier neutralization. Only in one occasion a QCD-based method did not attain the best rate of correct classification. This was QCD-FCMn-E in Scenario MTC 1 with $m=2.2$, being outmatched by the wavelet-based procedure. Unlike the alternative techniques, the proposed methods achieved perfect classification rates in most of the settings. The correlation-based methods acquired very similar successful rates on some scenarios, but totally failed in several others. Specifically, Scenario MTC 1 along with the scenarios concerning conditional heteroskedastic processes were particularly challenging for this metric. The wavelet-based strategies reached very poor results.
\begin{table}[ht]
	\centering
	\resizebox{9cm}{!}{
		\setlength{\tabcolsep}{1.75pt}
		\begin{tabular}{lcc rrr c rrr c rrr}\hline 
			& & & \multicolumn{3}{c}{Exponential (E)} & & \multicolumn{3}{c}{Noise (NC)} & & \multicolumn{3}{c}{Trimmed (T)} \\ \cline{4-6} \cline{8-10} \cline{12-14}
			Scenario & $T$ & $m$ & \multicolumn{1}{r}{QCD} & \multicolumn{1}{r}{W} & \multicolumn{1}{r}{C} & &  
			\multicolumn{1}{r}{QCD} & \multicolumn{1}{r}{W} & 
			\multicolumn{1}{r}{C} & &  \multicolumn{1}{r}{QCD} & 
			\multicolumn{1}{r}{W} & \multicolumn{1}{r}{C}  \\ \hline
			MTC 1  & 1500 & $1.8$ & \textbf{0.61} & 0.55 & 0.13 & & \textbf{1} & 0 & 0.91 & & \textbf{0.99}& 0.5 & 0.97 \\ 
			&      & $2.2$ & 0.48 & \textbf{0.58} & 0.14 & & \textbf{0.97} & 0 & 0.67 &  & \textbf{0.99} & 0.33 & 0.87 \\ 
			\hline 
			MTC 2  & 900  & $1.8$ & \textbf{1} & 0 & \textbf{1} & &\textbf{1} & 0 & 0.89 &  & \textbf{1} & 0 & 0.99 \\ 
			&      & $2.2$ & \textbf{1} & 0 & 0.99 & & \textbf{0.98} & 0 & 0.42 & & \textbf{1} & 0 & 0.99 \\ 
			\hline 
			MTC 3 & 3000  & $1.8$ & \textbf{1} & 0.3 & 0 &  & \textbf{1} & 0 & 0 &  & \textbf{1} & 0.53 & 0.02 \\
			&       & $2.2$ & \textbf{1} & 0.25 & 0 & & \textbf{1} & 0 & 0 & & \textbf{1} & 0.34 & 0.01 \\ 
			\hline 
			MIO 1 & 1500  & $1.8$ & \textbf{0.83} & 0.01 & 0.72 & & \textbf{0.99} & 0.95 & 0.91 & & \textbf{0.99} & 0.97 & 0.96\\ 
			&        & $2.2$ & \textbf{0.73}& 0 & 0.65 & & \textbf{0.91} & 0.7 & 0.45 & & \textbf{0.98} & 0.93 & 0.93 \\ 
			\hline 
			MIO 2 & 900 & $1.8$ & \textbf{1} & 0.14 & 0.99 & & \textbf{1} & 0 & 0.99 &  & \textbf{1} & 0.23 & 0.99 \\ 
			&     & $2.2$ & \textbf{1} & 0.09 & 0.98 & & \textbf{0.99} & 0 & 0.97 &  & \textbf{1} & 0.19 & 0.99 \\ 
			\hline 
			MIO 3 & 3000 & $1.8$ & \textbf{1} & 0.01 & 0 & & \textbf{1} & 0.41 & 0.02 & & \textbf{1} & 0.48 & 0.04 \\ 
			&      & $2.2$ & \textbf{1} & 0.01 & 0 & & \textbf{1} & 0.16 & 0 &  & \textbf{1} & 0.35 & 0 \\ 
			\hline 
		\end{tabular}
	}
	\caption{Averages rates of correct classification for robust approaches in Scenarios MTC 1, MTC 2, MTC 3, MIO 1, MIO 2 and MIO 3.}
	\label{tablesecondoutliers}
\end{table}

From scores in Table~\ref{tablesecondoutliers} follows that the three proposed robust procedures are highly capable of dealing with the types of anomalous series introduced in this new simulation study. Overall, the trimmed-based method yielded the best classification rates, closely followed by the noise approach. The results reached by the exponential procedure, although with quite high success rates, were worse by working with linear processes (Scenarios MTC 1 and MIO 1). Of course, the scores could be increased by using a less stringent cut-off value (e.g. 0.6 instead of 0.7). It is worth revealing that, although not shown in the manuscript, we have repeated the simulation study for different parameters concerning the definition of MTC and MIO type outliers. Particularly, we considered $\bm w=(2.5, -2.5)^{\intercal}$ ($\bm w=(0.5, -0.5)^{\intercal}$ in Scenario MTC 3) and $\eta=0.9$ in the case of MTC type outliers, and 1.5 degrees of freedom for the $\chi^2$ distribution ($1.5/10$ in Scenario MIO 3) in the case of MIO type outliers. As expected, the corresponding rates of correct classification were for all methods worse than those provided in Table \ref{tablesecondoutliers}, but the QCD-based procedures were again the best performing ones. Additionally, simulations were carried out with the same parameters than in the original simulation study but introducing the outlying effect in $t_0=3T/4$. As the anomalous period is shorter, the results were again worse than those in Table \ref{tablesecondoutliers} but one more time the quantile-based metric generally defeated its competitors. In short, the exponential, noise and trimmed approaches based on $d_{QCD}$ have proven to be successful in counterbalancing the effects of distorted individual MTS. 

\section{Applications}\label{sectionapplications}

In this section, the three robust approaches are employed to perform clustering in two real MTS datasets involving financial and environmental time series data, respectively. We applied the QCD-FCMn clustering algorithm to both databases in \cite{oriona2021c}, obtaining very meaningful solutions. The goal here is to take a step further, by considering the robust algorithms to make a proper treatment of potentially outlying series which may have affected the solutions given in \cite{oriona2021c}. Notice that, in both cases, the target is to show the usefulness of the proposed algorithms without intending to give financial or environmental advice. 

\subsection{Robust fuzzy clustering of the top 20 companies of S\&P 500 index}  \label{App1}

The first dataset was taken from the finance section of the Yahoo website \texttt{https://es.finance.yahoo.com}. It contains daily stock returns and trading volume of the current top 20 companies of the S\&P 500 index according to market capitalization. The sample period spans from 6th July 2015 to 7th February 2018, thus resulting serial realizations of length $T=655$. The S\&P 500 is a stock market index that tracks the stocks of 500 large-cap U.S. companies. The top 20 involves some of the most important companies in the world, as Apple, Google, Facebook or Berkshire Hathaway. 

It is important to highlight that the relationship between price and volume has been extensively analyzed in the literature \citep{karpoff1987relation, campbell1993trading, gebka2013causality} and constitutes itself a topic of great financial interest. Prices and trading volume are known to exhibit some empirical linkages over the fluctuations of stock markets. Thus, it is interesting to describe each of the considered companies by means of these two quantities, and our goal is to consider the joint behaviour of prices and volume to perform fuzzy clustering. We assume that two companies show a similar behaviour if the corresponding bivariate time series exhibit similar dependence structures. 

\begin{figure}[ht]
	\centering
	\includegraphics[width=0.8\textwidth]{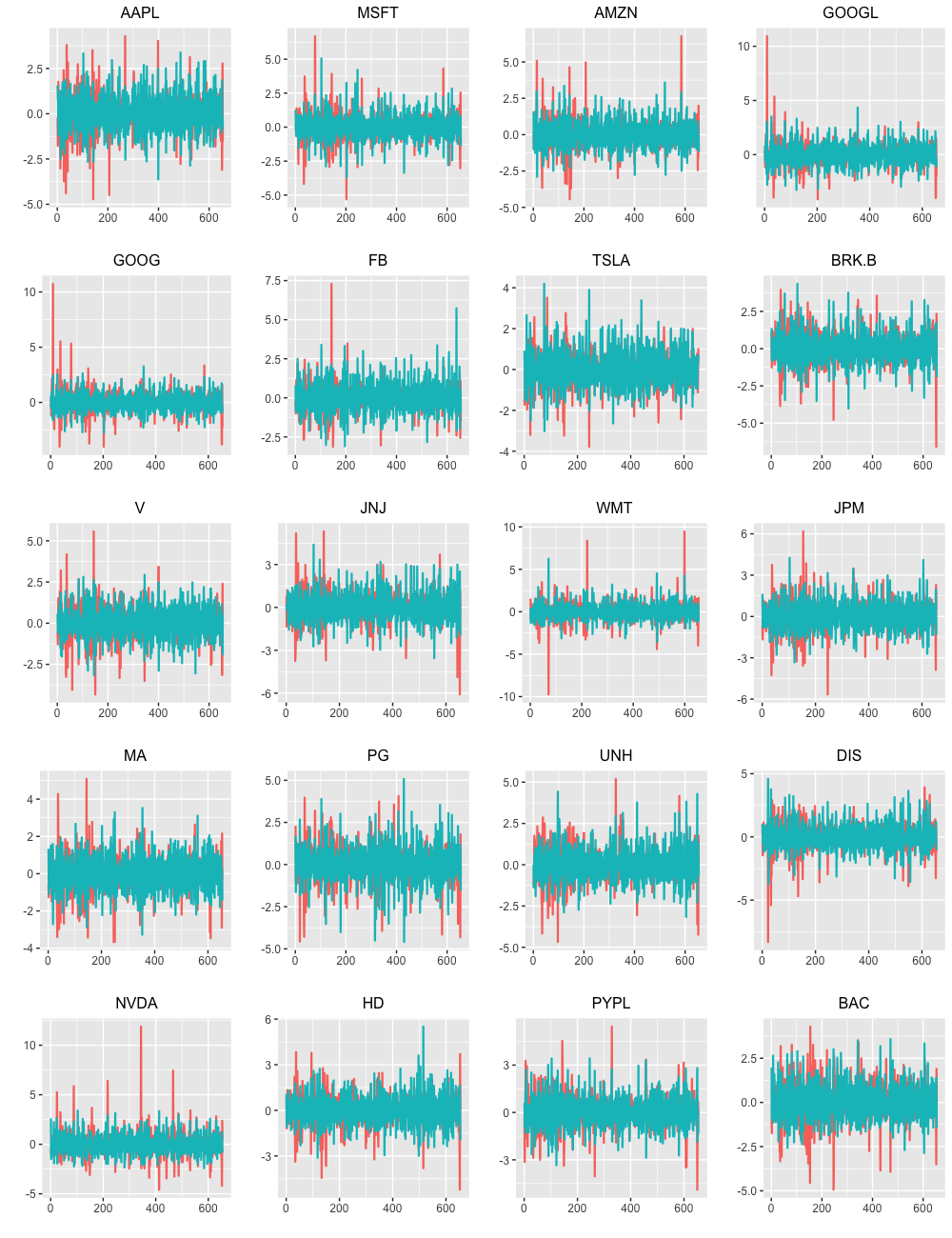}
	\vspace*{-0.25cm}
	\caption{Daily returns (red colour) and change in volume (blue colour) of the top 20 companies in the S\&P 500 index.}
	\label{top20}
\end{figure}

Since the UTS of prices and trading volume are non-stationary in mean, all of them were transformed by taking the first differences of the natural logarithm of the original values. This way, prices give rise to stock returns, and volume to what we call change in volume. Next, all UTS were normalized to have zero mean and unit variance. The resulting MTS are depicted in Figure~\ref{top20}. Overall, plots in Figure~\ref{top20} exhibit common traits of financial time series. There is a substantial degree of heteroskedasticity and both components exhibit the so-called phenomenon of volatility clustering: large values (positive or negative) tend to group together, resulting in a marked persistence. These particular properties of financial time series, usually referred to as stylized facts, are generally accounted for by modelling the series by way of multivariate GARCH-type models, for instance the BEKK models considered in Scenarios 3.1 and 3.2. It is worth remembering that the robust fuzzy clustering algorithms demonstrated their efficacy when coping with this type of models, specially when the error distribution possesses some amount of fat-tailedness. This property also relates to the stylized facts \citep{harvey2013dynamic, schmitt2013non, bradley2003financial}. Therefore, it is expected that QCD-FCMn-E, QCD-FCMn-NC and QCD-FCMn-T can provide a meaningful fuzzy partition determining groups of companies following a similar behavioural pattern and isolated MTS exhibiting atypical dependence structures. 

\begin{figure}[h]
	\centering
	\includegraphics[width=0.7\textwidth]{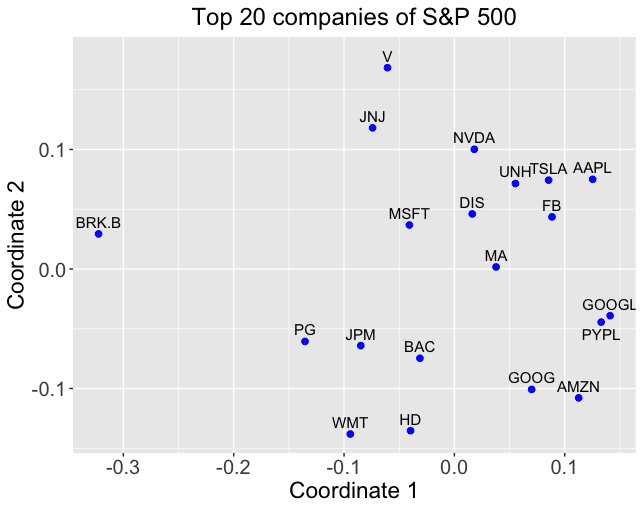}
	\vspace*{-0.25cm}
	\caption{Two-dimensional scaling plane based on the QCD-based distance for the daily returns and change in volume of the top 20 companies in the S\&P 500 index.}
	\label{mdstop20}
\end{figure}

In \cite{oriona2021c}, the number of clusters $C$ and the fuzziness parameter $m$ were simultaneously selected according to the minimization of four internal clustering validity indexes. The optimal values were $C=6$ and $m=1.9$. In fact, the 6-cluster solution provided by QCD-FCMn (see Table~\ref{tablefuzzyqcd} in Appendix) gave rise to meaningful groups, putting together companies according to underlying characteristic as the market capitalization or the company nature, among others. For instance, there was a cluster containing the two giants of information technology, Apple (AAPL) and Microsoft (MSFT), and another cluster grouping the two branches of Google (GOOGL and GOOG) and Amazon (AMZN). Figure~\ref{mdstop20} shows a metric 2-dimensional scaling plot of the companies according to the pairwise QCD-based distance matrix. Since the associated R-squared is 0.7251, the scatter plot can be considered an acceptable representation of the underlying distance configuration \citep{hair2009multivariate}. 

From Figure~\ref{mdstop20}, it could be concluded that the company Berkshire Hathaway (BRK.B) is by far the most outlying one. In fact, Table~\ref{tablefuzzyqcd} shows that this company constitutes an isolated cluster in the resulting partition, $C_3$, where the remaining companies display negligible membership values. Thus, it is expected that BRK.B was detected as an outlier by the robust approaches. Figure~\ref{mdstop20} also shows any other potentially anomalous elements, located partly far from the bulk of the data points, as Visa (V) or Walmart (WMT). These companies could be partially considered as atypical by the robust procedures. 

Based on the previous remarks, it is interesting to analyse the results achieved by the three QCD-based robust versions. Concerning the hyperparameters of each procedure, the coefficient $\beta$ in QCD-FCMn-E was chosen by means of \eqref{selectbeta} in Section \ref{subsectionexp}, resulting $\beta=174.85$. The trimming rate $\alpha$ associated with QCD-FCMn-T was selected by considering a grid of values for $\alpha$ and choosing the one giving rise to the least average value of the indexes considered in \cite{oriona2021c}, resulting $\alpha=0.15$. As for QCD-FCMn-NC, the noise distance $\lambda$ was set in the following manner: QCD-FCMn-NC was ran several times for decreasing values of $\delta$ and the proportion of series placed in the noise cluster was recorded. Then, we chose the value of $\delta$ associated with a sudden change in this proportion. The underlying rationale here is that, by gradually decreasing the value of $\delta$, an appropriate threshold will be figured out since substantially low values of $\delta$ lead to partitions where non-outlying elements are located in the noise cluster. 

Table~\ref{tablefuzzyqcdexp} contains the fuzzy partition obtained by the QCD-FCMn-E algorithm. Compared to the results using QCD-FCMn (Table~\ref{tablefuzzyqcd}), the main difference is that QCF-FCMn-E establishes that BRK.B is an outlier company since its membership degrees are split across the clusters almost uniformly. By dealing here with a 6-cluster solution, we decided to determine the $i$-th MTS as an anomalous one if $u_{ic}\in[0.05, 0.35]$, $c=1,\ldots, 6$. According to this criterion, no additional outliers were found by QCD-FCMn-E. PayPal (PYPL) was almost determined as an anomalous series but its membership degree is slightly above 0.35 in cluster $C_2$. Thus, PYPL could be seen as a potential outlier requiring individual analysis. This is not surprising since PayPal is also the MTS  displaying the most spread out membership values in the cluster solution reported by QCD-FCMn (Table~\ref{tablefuzzyqcd}).  
\begin{table}[t]
\centering
	\resizebox{8cm}{!}{
		\begin{tabular}{ccccccc}
			\hline
			Company & $C_1$ & $C_2$ & $C_3$ & $C_4$ & $C_5$ & $C_6$ \\ 
			\hline
			AAPL & 0.118 & 0.205 & 0.061 & \textbf{0.469} & 0.094 & 0.053 \\ 
			MSFT & 0.134 & 0.072 & 0.203 & \textbf{0.397} & 0.125 & 0.069 \\ 
			AMZN & \textbf{0.854} & 0.027 & 0.029 & 0.044 & 0.018 & 0.028 \\ 
			GOOGL & \textbf{0.666} & 0.053 & 0.055 & 0.137 & 0.044 & 0.044 \\ 
			GOOG & \textbf{0.905} & 0.014 & 0.023 & 0.030 & 0.011 & 0.016 \\ 
			FB & 0.005 & \textbf{0.966} & 0.005 & 0.009 & 0.009 & 0.007 \\ 
			TSLA & 0.064 & 0.041 & 0.029 & \textbf{0.808} & 0.038 & 0.020 \\ 
			\textbf{BRK.B} & 0.126 & 0.121 & 0.273 & 0.147 & 0.183 & 0.151 \\ 
			V & 0.010 & 0.024 & 0.017 & 0.029 & \textbf{0.907} & 0.012 \\ 
			JNJ & 0.006 & 0.016 & 0.012 & 0.015 & \textbf{0.944} & 0.008 \\ 
			WMT & 0.014 & 0.017 & 0.042 & 0.013 & 0.015 & \textbf{0.899} \\ 
			JPM & 0.013 & 0.010 & \textbf{0.918} & 0.015 & 0.012 & 0.031 \\ 
			MA & 0.137 & 0.164 & 0.110 & \textbf{0.415} & 0.092 & 0.081 \\ 
			PG & 0.008 & 0.007 & \textbf{0.944} & 0.010 & 0.010 & 0.021 \\ 
			UNH & 0.012 & \textbf{0.902} & 0.014 & 0.023 & 0.032 & 0.017 \\ 
			DIS & 0.067 & 0.133 & 0.083 & \textbf{0.533} & 0.133 & 0.052 \\ 
			NVDA & 0.044 & 0.039 & 0.040 & \textbf{0.786} & 0.070 & 0.022 \\ 
			HD & 0.009 & 0.012 & 0.020 & 0.008 & 0.008 & \textbf{0.942} \\ 
			PYPL & 0.185 & \textbf{0.359} & 0.089 & 0.162 & 0.082 & 0.123 \\ 
			BAC & 0.074 & 0.084 & 0.257 & 0.077 & 0.061 & \textbf{0.446} \\ 
			\hline
		\end{tabular}
	}
	\caption{Membership degrees for top 20 companies in the S\&P 500 index by considering the QCD-FCMn-E.}
	\label{tablefuzzyqcdexp}
\end{table}

The partition obtained with the QCD-FCMn-NC algorithm is given in Table~\ref{tablefuzzyqcdnc}. Four outlying companies were identified, namely BRK.B, V, Johnson \& Johnson (JNJ) and PYPL. As QCD-FCMn-NC is here handling a 7-cluster solution, we resolved to consider a series as anomalous when its maximum membership corresponded to the noise cluster and was above 0.25. BRK.B displayed the highest membership value in the noise cluster, followed by V and JNJ, and lastly by PYPL. It is worth remarking that three of the top five companies, AAPL, MSFT and GOOGL exhibited membership values above 0.20 in the noise cluster, thus suggesting that these companies, specially AAPL (whose highest membership value is 0.267 in $C_1$) could be seen as plausible outliers. Compared to the partition provided by QCD-FCMn-E, the main difference is that the cluster $C_4$, formed by AAPL, MSFT, Tesla (TSLA), Mastercard (MA), Walt Disney (DIS) and Nvidia (NVDA), is split by QCD-FCMn-NC into two clusters, $C_1$ (AAPL-MA-DIS) and $C_2$ (MSFT-TSLA-NVDA). In short, the fuzzy partition determined by QCD-FCMn-NC is also consistent with the 2D plot in Figure~\ref{mdstop20}, where BRK.B constitutes the most isolated point and V and JNJ are located at the top of the graph, moderately distant from the rest of the points.
\begin{table}[t]
	\centering
	\resizebox{9cm}{!}{
		\begin{tabular}{cccccccc}
			\hline
			Company & $C_1$ & $C_2$ & $C_3$ & $C_4$ & $C_5$ & $C_6$ & $NC$ \\ 
			\hline
			AAPL & \textbf{0.267} & 0.247 & 0.076 & 0.132 & 0.036 & 0.028 & 0.214 \\ 
			MSFT & 0.171 & \textbf{0.335 }& 0.079 & 0.036 & 0.13 & 0.034 & 0.215 \\ 
			AMZN & 0.032 & 0.022 & \textbf{0.822} & 0.014 & 0.018 & 0.016 & 0.076 \\ 
			GOOGL & 0.072 & 0.093 & \textbf{0.516} & 0.03 & 0.035 & 0.023 & 0.231 \\ 
			GOOG & 0.021 & 0.019 & \textbf{0.872} & 0.008 & 0.016 & 0.01 & 0.054 \\ 
			FB & 0.015 & 0.006 & 0.004 & \textbf{0.941} & 0.004 & 0.005 & 0.025 \\ 
			TSLA & 0.127 & \textbf{0.657} & 0.05 & 0.03 & 0.022 & 0.012 & 0.102 \\ 
			\textbf{BRK.B} & 0.054 & 0.055 & 0.036 & 0.033 & 0.127 & 0.053 & \textbf{0.642} \\ 
			\textbf{V} & 0.176 & 0.167 & 0.038 & 0.127 & 0.08 & 0.046 & \textbf{0.366} \\ 
			\textbf{JNJ} & 0.203 & 0.129 & 0.037 & 0.147 & 0.106 & 0.061 & \textbf{0.317} \\ 
			WMT & 0.011 & 0.006 & 0.007 & 0.009 & 0.030 & \textbf{0.890} & 0.047 \\ 
			JPM & 0.013 & 0.007 & 0.007 & 0.004 & \textbf{0.932} & 0.015 & 0.022 \\ 
			MA & \textbf{0.812} & 0.05 & 0.027 & 0.029 & 0.024 & 0.015 & 0.043 \\ 
			PG & 0.015 & 0.009 & 0.008 & 0.006 & \textbf{0.900} & 0.022 & 0.04 \\ 
			UNH & 0.015 & 0.006 & 0.003 & \textbf{0.942} & 0.004 & 0.005 & 0.025 \\ 
			DIS & \textbf{0.936} & 0.026 & 0.005 & 0.01 & 0.007 & 0.003 & 0.013 \\ 
			NVDA & 0.03 & \textbf{0.924} & 0.007 & 0.006 & 0.007 & 0.003 & 0.023 \\ 
			HD & 0.008 & 0.004 & 0.005 & 0.007 & 0.014 & \textbf{0.935} & 0.027 \\ 
			\textbf{PYPL} & 0.184 & 0.078 & 0.124 & 0.21 & 0.053 & 0.073 & \textbf{0.278} \\  BAC & 0.136 & 0.045 & 0.055 & 0.058 & 0.253 & \textbf{0.298} & 0.155 \\ 
			\hline
		\end{tabular}
	}
	\caption{Membership degrees for top 20 companies in the S\&P 500 index by considering the QCD-FCMn-NC.}
	\label{tablefuzzyqcdnc}
\end{table}

Concerning QCD-FCMn-T, the optimal trimmed ratio was $\alpha=0.15$ so that three series, BRK.B, WMT and The Home Depot (HD), were trimmed away. Table \ref{tablefuzzyqcdt} includes the membership degrees for the series in Figure \ref{top20} achieved by the trimmed-based strategy QCD-FCMn-T. The companies whose membership values are blank correspond to series that were removed in the trimming procedure. Note that the WMT and HD were not determined as anomalous by QCD-FCMn-E and QCD-FCMn-NC. However, Figure~\ref{mdstop20} suggests that these companies could exhibit a distinct behaviour than that of the remaining firms, as the corresponding points are located in the lower part of the graph, far from the bulk.

\begin{table}
	\centering
	\resizebox{7.5cm}{!}{
	\begin{tabular}{ccccccc}
		\hline
		Company & $C_1$ & $C_2$ & $C_3$ & $C_4$ & $C_5$ & $C_6$ \\ 
		\hline
		AAPL & 0.084 & 0.146 & 0.298 & \textbf{0.365} & 0.066 & 0.041 \\ 
		MSFT & 0.108 & 0.049 & 0.212 & \textbf{0.356} & 0.099 & 0.176 \\ 
		AMZN & \textbf{0.865} & 0.018 & 0.050 & 0.032 & 0.010 & 0.025 \\ 
		GOOGL & \textbf{0.682} & 0.033 & 0.091 & 0.128 & 0.025 & 0.040 \\ 
		GOOG & \textbf{0.902} & 0.010 & 0.032 & 0.027 & 0.008 & 0.022 \\ 
		FB & 0.002 & \textbf{0.983} & 0.007 & 0.003 & 0.003 & 0.002 \\ 
		TSLA & 0.023 & 0.013 & 0.055 & \textbf{0.885} & 0.013 & 0.010 \\ 
		\textbf{BRK.B} & - & - & - & - & - & - \\ 
		V & 0.004 & 0.014 & 0.016 & 0.016 & \textbf{0.941} & 0.009 \\ 
		JNJ & 0.004 & 0.016 & 0.019 & 0.013 & \textbf{0.937} & 0.012 \\ 
		\textbf{WMT} & - & - & - & - & - & - \\ 
		JPM & 0.002 & 0.001 & 0.004 & 0.002 & 0.002 & \textbf{0.989} \\ 
		MA & 0.005 & 0.006 & \textbf{0.968} & 0.011 & 0.004 & 0.006 \\ 
		PG & 0.015 & 0.012 & 0.028 & 0.016 & 0.019 & \textbf{0.909} \\ 
		UNH & 0.006 & \textbf{0.924} & 0.027 & 0.012 & 0.022 & 0.008 \\ 
		DIS & 0.020 & 0.039 & \textbf{0.772} & 0.098 & 0.042 & 0.030 \\ 
		NVDA & 0.025 & 0.021 & 0.085 & \textbf{0.803} & 0.043 & 0.024 \\ 
		\textbf{HD} & - & - & - & - & - & - \\ 
		PYPL & 0.155 & \textbf{0.301} & 0.298 & 0.114 & 0.057 & 0.075 \\ 
		BAC & 0.076 & 0.086 & 0.226 & 0.066 & 0.060 & \textbf{0.485} \\ 
		\hline
	\end{tabular}
	}
	\caption{Membership degrees for top 20 companies in the S\&P 500 index by considering the QCD-FCMn-T.}
	\label{tablefuzzyqcdt}
\end{table}

In summary, each of the approaches QCD-FCMn-E, QCD-FCMn-NC and QCD-FCMn-T provides a different conclusion in terms of identification of outlying series, but all of them are consistent with the plot in Figure~\ref{mdstop20}. Moreover, there was consensus on establishing BRK.B as an outlier, which is quite reasonable from a financial point of view. BRK.B is the only company among the top 20 whose main task consists of investing in the remaining companies. Consequently, it is unsurprising that this firm exhibits a particular behaviour in terms of returns and trading volume. 

\subsection{Robust fuzzy clustering of air pollution data}  
\label{subsectionenvironmental}

The second study case concerns clustering of geographical zones in terms of their temporal records of air pollutants. We considered trivariate time series of hourly concentrations of nitrogen dioxide (NO$_2$), ozone (O$_3$) and nitrogen monoxide (NO) during the whole year 2018 in 20 different stations located in Galicia, an autonomous community of Spain. The choice of these pollutants was mainly based on the fact that several studies have uncovered serious health effects associated with the continuous exposure to high levels of NO$_2$ and O$_3$. The data were sourced from the website of Ministry for the Ecological Transition and the Demographic challenge \texttt{https://www.miteco.gob.es/}, which also contains information about the type of location, namely ``urban'', ``suburban'', ``rural'' and ``near power plant''. Thus, from an environmental point of view, it is reasonable to think that the joint behaviour of the concentration of these  pollutants is different depending on where the station is situated. 

\begin{figure}[!t]
	\centering
	\includegraphics[width=0.85\textwidth]{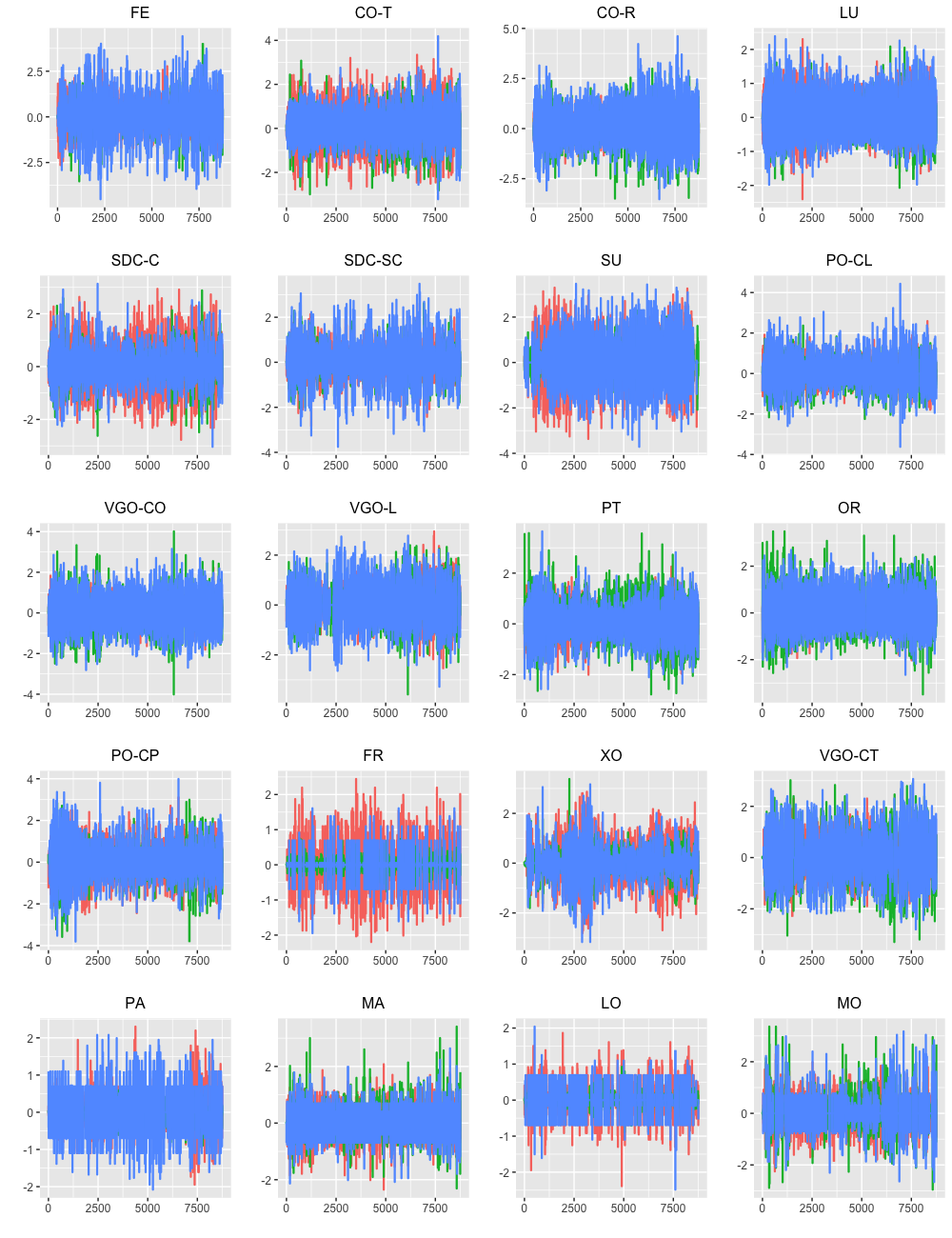}
	\caption{Transformed levels of NO$_2$ (red colour) O$_3$ (green colour) and NO (red colour) in the 20 monitoring stations of Galicia.}
	\label{top20stations}
\end{figure}

\begin{figure}[ht]
	\centering
	\includegraphics[width=0.85\textwidth]{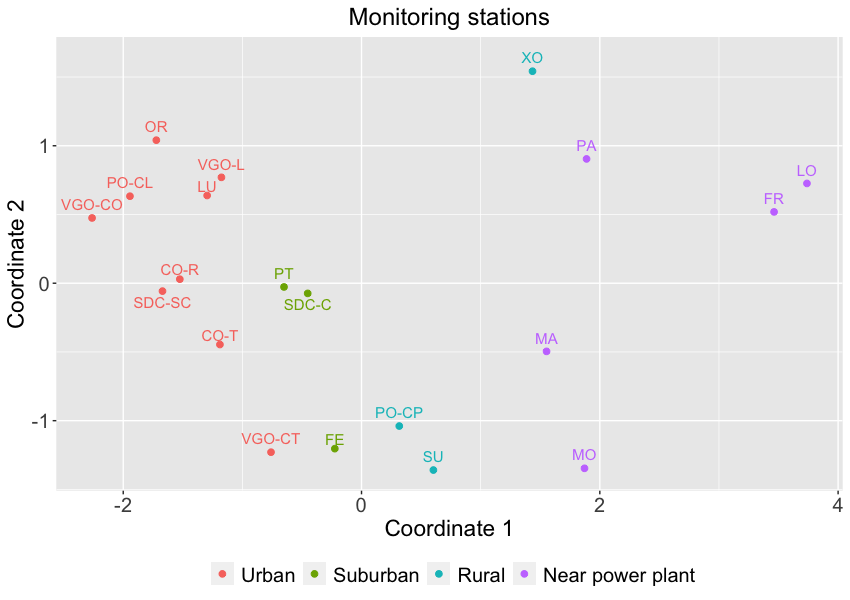}
	\vspace*{-0.5cm}
	\caption{2D scaling plane based on the QCD distances for the hourly levels of NO$_2$, O$_3$ and NO in the 20 monitoring stations of Galicia.}
	\label{mdstop20stations}
\end{figure}

The 20 MTS available are formed by $T=8760$ hourly records and, given their non-stationarity in mean, we proceeded to transform them by taking first differences. The new set of series, which was subject to clustering in \cite{oriona2021c} are depicted in Figure \ref{top20stations}. The resulting 2DS plane based on the QCD distance is given in Figure \ref{mdstop20stations}, where the points have been coloured according to the mentioned categories. 

Proceeding as in Section~\ref{App1}, the optimal values for $C$ and $m$ found in \cite{oriona2021c} were $C=3$ and $m=1.9$. The 3-cluster solution attained by QCD-FCMn by considering 0.6 as cutoff (Table~\ref{fuzzytablestations} in Appendix) identified groups of stations sharing the same location category, as it was expected from Figure~\ref{mdstop20stations}. Specifically, a cluster ($C_1$) grouped all the urban stations except for VGO-CT, other cluster ($C_2$) was formed by VGO-CT, one suburban (FE) and two rural (SU and PO-CP) stations, and the cluster $C_3$ involved three stations located near a power plant. The remaining 5 stations (SDC-C, PT, XO, MA and MO) displayed a relative high membership value in two clusters. These results reveal that the standard fuzzy model QCD-FCMn produced meaningful groups from a point of view of geographical location, but it was not able to identify anomalous series since no station displays membership degrees uniformly distributed among the three clusters. However, the 2D plot in Figure~\ref{mdstop20stations} clearly indicates the existence of some MTS whose behaviour deviates significantly from the majority, as LO, FR or XO. Thus, it should be desirable to use a robust method capable of simultaneously preserving the true cluster structure while identifying outlier series.

The membership matrices associated with the proposed robust models, QCD-FCMn-E, QCD-FCMn-NC and QCD-FCMn-T, are jointly given in Table \ref{fuzzytablestationsexpNCT}.

\begin{table}[ht]
	\centering
	\normalsize
	\resizebox{9cm}{!}{
		\setlength{\tabcolsep}{2.5pt}
		\begin{tabular}{c ccc c cccc c ccc}
			\hline
			& \multicolumn{3}{c}{QCD-FCMn-E} & & \multicolumn{4}{c}{QCD-FCMn-NC} & & \multicolumn{3}{c}{QCD-FCMn-T} \\ \cline{2-4} \cline{6-9} \cline{11-13}
			Station   & $C_1$ & $C_2$ & $C_3$ & & $C_1$ & $C_2$ & $C_3$ & $NC$ & & $C_1$ & $C_2$ & $C_3$  \\ 
			\hline
			FE & 0.114 & 0.046 & \textbf{0.840} & & 0.043 & 0.022 & \textbf{0.813} & 0.122 &  & 0.064 & 0.042 & \textbf{0.894}\\ 
			CO-T & \textbf{0.656} & 0.065 & 0.279 & & \textbf{0.357} & 0.040 & 0.274 & 0.329 & & \textbf{0.604} & 0.071 & 0.325\\ 
			CO-R & \textbf{0.745} & 0.071 & 0.184 & & \textbf{0.431} & 0.040 & 0.145 & 0.384 & & \textbf{0.748} & 0.068 & 0.184\\ 
			LU & \textbf{0.945} & 0.014 & 0.041 & & \textbf{0.893} & 0.009 & 0.030 & 0.068 & & \textbf{0.931} & 0.022 & 0.047\\ 
			SDC-C & 0.431 & 0.057 & \textbf{0.513} & & 0.250 & 0.048 & \textbf{0.493} & 0.209 & & 0.34 & 0.101 & \textbf{0.559} \\ 
			SDC-SC & \textbf{0.900} & 0.024 & 0.076 & & \textbf{0.684} & 0.020 & 0.098 & 0.198 & & \textbf{0.899} & 0.023 & 0.078 \\ 
			SU & 0.107 & 0.100 & \textbf{0.793} & & 0.059 & 0.077 & \textbf{0.554} & 0.310 & & 0.089 & 0.144 & \textbf{0.767} \\ 
			PO-CL & \textbf{0.947} & 0.016 & 0.037 & & \textbf{0.927} & 0.005 & 0.015 & 0.053 & & \textbf{0.961} & 0.012 & 0.027  \\ 
			VGO-CO & \textbf{0.906} & 0.029 & 0.065 & & \textbf{0.779} & 0.013 & 0.042 & 0.166 & & \textbf{0.932} & 0.021 & 0.047 \\ 
			VGO-L & \textbf{0.887} & 0.030 & 0.083 & & \textbf{0.783} & 0.020 & 0.059 & 0.138  & & \textbf{0.866} & 0.047 & 0.087\\ 
			PT & \textbf{0.454} & 0.094 & 0.452 & & 0.273 & 0.060 & 0.297 & \textbf{0.370} &  & 0.387 & 0.164 & \textbf{0.449} \\ 
			OR & \textbf{0.923} & 0.025 & 0.052 & & \textbf{0.871} & 0.010 & 0.025 & 0.094 &  & \textbf{0.937} & 0.023 & 0.040 \\ 
			PO-CP & 0.030 & 0.020 & \textbf{0.950} & & 0.039 & 0.036 & \textbf{0.778} & 0.147&  & 0.037 & 0.063 & \textbf{0.900}\\ 
			FR & 0.041 & \textbf{0.900} & 0.059 &  & 0.027 & 0.449 & 0.049 & \textbf{0.475} & & - & - & - \\ 
			XO & 0.213 & \textbf{0.514} & 0.273 & & 0.071 & 0.349 & 0.087 & \textbf{0.493} & & 0.091 & \textbf{0.792} & 0.11\\ 
			VGO-CT & 0.226 & 0.063 & \textbf{0.711} & & 0.108 & 0.031 & \textbf{0.610} & 0.251 &  & 0.148 & 0.065 & \textbf{0.787} \\ 
			PA & 0.056 & \textbf{0.847} & 0.097 & & 0.006 & \textbf{0.934} & 0.011 & 0.049 & & 0.029 & \textbf{0.918} & 0.053 \\ 
			MA & 0.109 & 0.383 & \textbf{0.508} & & 0.051 & \textbf{0.381}& 0.228 & 0.340 &  & 0.052 & \textbf{0.711} & 0.237 \\ 
			LO & 0.055 & \textbf{0.869} & 0.076 & & 0.027 & 0.387 & 0.047 & \textbf{0.539} & & - & - & -  \\ 
			MO & 0.155 & 0.321 & \textbf{0.524} & & 0.053 & 0.176 & 0.204 & \textbf{0.567} & & 0.100 & \textbf{0.500} & 0.400\\ 
			\hline
		\end{tabular}
	}
	\caption{Membership degrees for the 20 monitoring stations in Galicia by considering QCD-FCMn-E, QCD-FCMn-NC and QCD-FCMn-T models.}
	\label{fuzzytablestationsexpNCT}
\end{table}

QCD-FCMn-E produced a very similar partition than the one given by the standard algorithm, although all the series have membership degrees slightly more spread out. As we are dealing with 3 clusters, the $i$-th station was deemed anomalous if $u_{ic}>0.2$, $c=1,\ldots,3$. In accordance with this benchmark, only XO was identified as outlier, which corresponds to the highest point in Figure~\ref{mdstop20stations}. 

The models QCD-FCMn-NC and QCD-FCMn-T led to different results. The former identified 4 outlying stations, FR, XO, LO and MO, according to the assignment rule of locating a station in the noise cluster if the corresponding membership is above 0.4. These locations correspond to a rural station and three stations located near a power plant. Cluster $C_1$ showed the same composition as in the solutions given by QCD-FCMn and QCD-FCMn-E, and $C_3$ was also similar except for PT and MO (now outliers). There is also a new cluster $C_2$ formed by MA and PA, two stations situated near a power plant, having a lot of overlap with the noise cluster. Finally, the partition provided by QCD-FCMn-T shares traits with those of both QCD-FCMn and QCD-FCMn-E but is slightly different. The cluster $C_1$ is the standard urban cluster, but clusters $C_2$ and $C_3$ are constituted differently from clusters $C_2$ and $C_3$ in the mentioned models. The optimal $\alpha$ was determined to be $\alpha=0.2$ so that two MTS corresponding to the most peripheral points in Figure~\ref{mdstop20stations}, FR and LO, were trimmed away. 

\section{Concluding remarks}\label{sectionconcludingremarks}

In this work we have introduced three robust approaches for MTS clustering built on the QCD-based fuzzy $C$-means clustering model (QCD-FCMn) defined in \cite{oriona2021c}. All of them are aimed at counterbalancing the negative effects that outlying series (i.e, series generated by a different stochastic process than those of the regular clusters) provoke in the clustering solution. The proposed strategies are based on three robust clustering methodologies suggested in the literature, namely the metric, noise and trimmed approaches. The metric approach (QCD-FCMn-E) acquires its robustness with respect to outliers by considering a exponential distance measure. The noise approach (QCD-FCMn-NC) obtains its robustness against outlying series by introducing an artificial cluster represented by a noise prototype. The trimmed approach (QCD-FCMn-T) removes a certain proportion of anomalous time series data. Fundamentally, we intended to take advantage of three pivotal elements:
\begin{itemize}
	\item The versatility of the fuzzy logic by allowing overlapping clusters. This permits to appropriately handling situations with inherent uncertainty, as it is often the case in MTS clustering.
	\item The powerful properties of the QCD-based distance to discriminate between general dependence structures (see \cite{oriona2021c}). By construction, this metric inherits the nice properties of the quantiles, hence exhibiting robustness against outliers and heavy tails. 
	\item The high capability of the metric, noise and trimmed approaches to neutralize the disruptive impact of atypical series.  
\end{itemize}

A broad simulation study involving different types of generating processes, namely linear, nonlinear and BEKK models, was carried out to evaluate the performance of the proposed approaches. In order to make the assessment task fairly general, two types of anomalous series were considered: (i) series entirely generated from a different process and (ii) series suffering an abrupt change at a particular time point. For comparison purposes, two alternative dissimilarity measures were also taken into account. Overall, the suggested techniques outperformed the remaining metrics in terms of classification accuracy for the two classes of outlying series. Furthermore, they were the least sensitive to the selection of the corresponding hyperparameters. None of the three robust approaches proved to be better than the remaining ones in all the considered scenarios. Thus, each one of the techniques is useful in its own right. The three of them outperformed the the nonrobust version QCD-FCMn in terms of outlier neutralization by a large degree, which highlights the usefulness of the robust algorithms. For illustrative purposes, we applied the methods to two MTS datasets containing financial and environmental series. Our analyses showed that the proposed techniques were able to identify some series showing anomalous dynamic patterns, leading to interesting conclusions. 

Although we have proposed three accurate methods for robust fuzzy clustering of MTS based on generating processes, there is still room for further research in this topic. First, introducing suitable ways for automatically selecting the input parameters of these techniques would be highly desirable, as the performance of the methods is substantially dependent on an appropriate choice. Second, some extensions of the introduced strategies regarding a combination of the QCD-based metric and a shape-based dissimilarity could be appropriate to perform robust clustering in some specific scenarios. For instance, by dealing with nonstationary series or datasets in which the different patterns are characterised by both generating processes and geometric profiles. Third, note that in our approach each MTS is characterized by a set of curves of the form $\left\{ W \left( \hat{G}^{j_1,j_2}_{T,R} (\omega, \tau,\tau^{\prime}) \right), 1\le j_1, j_2 \le d, \tau, \tau^{\prime} \in \mathcal{T}\right\}$, where $W(\cdot)$ is used interchangeably to denote the real part and the imaginary part operator. Our numerical studies have revealed that some of these curves contain far more information than others. Thus, it would be reasonable to create a robust fuzzy clustering algorithm giving more importance to the functions with more discriminatory power. This could be naturally accomplished by introducing weights in the objective function \eqref{qcd_means}. Finally, spatial extensions and possibilistic versions of the robust models introduced here could be constructed. The former techniques could be useful when dealing with series containing geographical information, as the ones in Section~\ref{subsectionenvironmental}. The mentioned topics for further research will be properly addressed in the future.

\section*{Declarations}

\subsection*{Acknowledgements}

This research has been supported by the Ministerio de Economía y Competitividad (MINECO) grants MTM2017-82724-R and PID2020-113578RB-100, the Xunta de Galicia (Grupos de Referencia Competitiva ED431C-2020-14), and the Centro de Investigación del Sistema Universitario de Galicia ‘‘CITIC’’ grant ED431G 2019/01; all of them through the European Regional Development Fund (ERDF). This work has received funding for open access charge by Universidade da Coruña/CISUG.

\subsection*{Conflicts of interest}

The authors declare that they have no known competing financial interests or personal relationships that could have appeared to influence the work reported in this paper.

\subsection*{Availability of data and material}

Not applicable.

\subsection*{Code availability}

Not applicable.

\section*{Appendix}

\subsection*{Simulation study: additional figures}

\begin{figure}[!ht]
	\centering
	\includegraphics[width=0.85\textwidth]{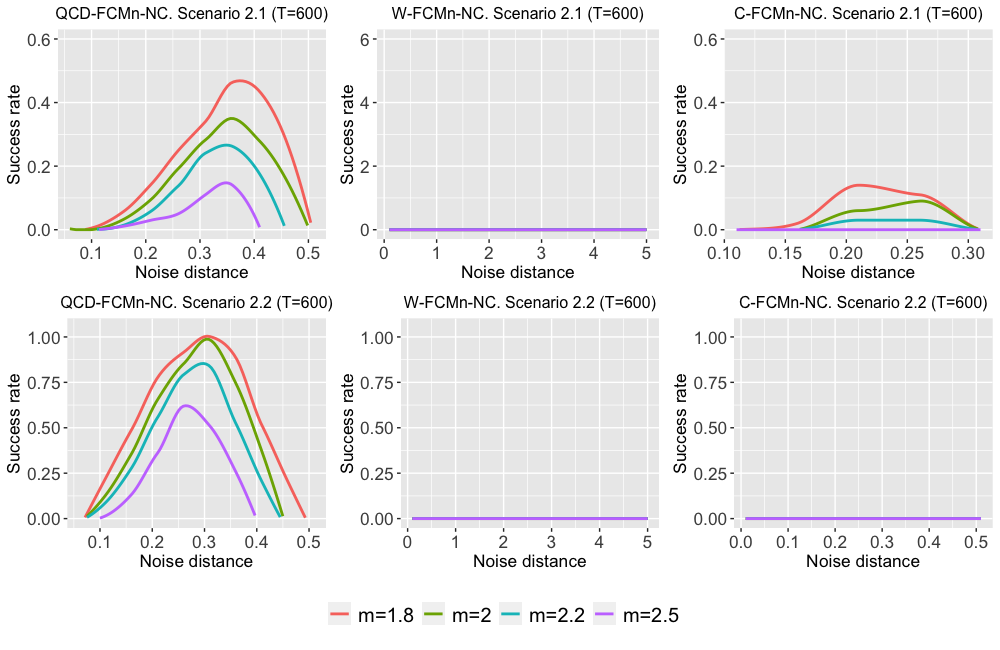}
	\vspace*{-0.25cm}
	\caption{Average rates of correct classification for QCD-FCMn-NC, W-FCMn-NC and C-FCMn-NC as a function of $\beta$ in Scenarios 2.1 and 2.2 for series of length $T=600$ and four fuzziness levels $m$.}
	\label{curvesnc21}
\end{figure}

\begin{figure}[!ht]
	\centering
	\includegraphics[width=0.85\textwidth]{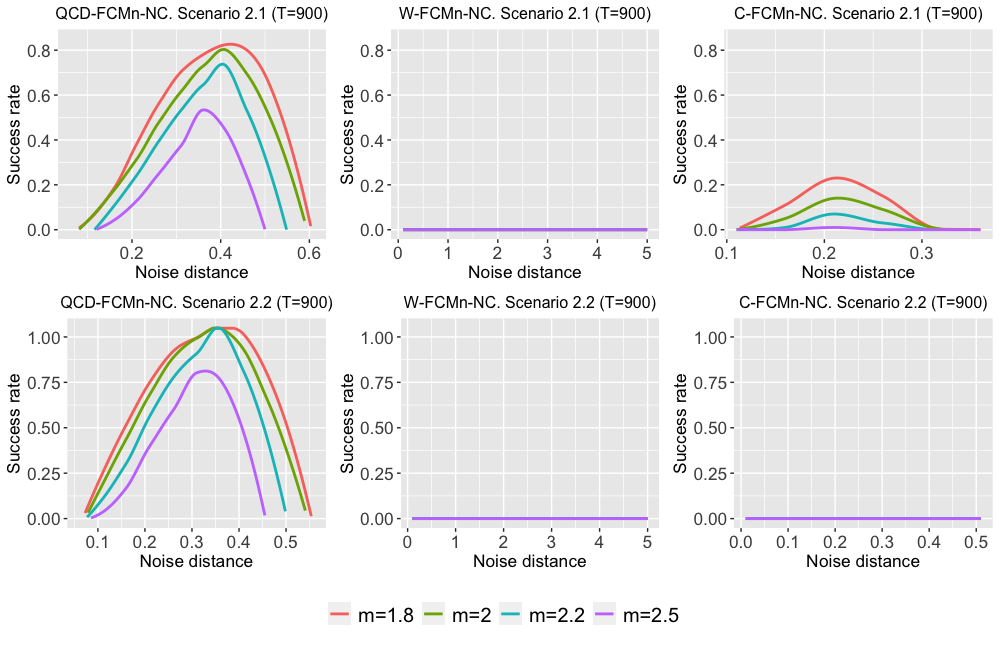}
	\vspace*{-0.25cm}
	\caption{Average rates of correct classification for QCD-FCMn-NC, W-FCMn-NC and C-FCMn-NC as a function of $\beta$ in Scenarios 2.1 and 2.2 for series of length $T=900$ and four fuzziness levels $m$.}
	\label{curvesnc22}
\end{figure}

\begin{figure}[!ht]
	\centering
	\includegraphics[width=0.85\textwidth]{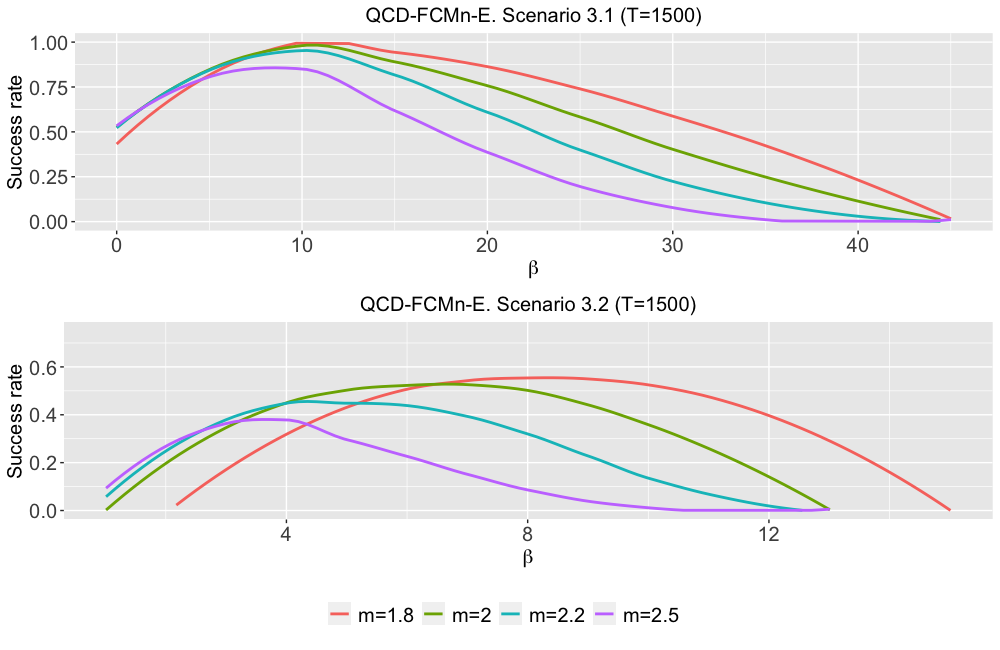}
	\vspace*{-0.25cm}
	\caption{Average rates of correct classification for QCD-FCMn-E as a function of $\beta$ in Scenarios 3.1 and 3.2 for series of length $T=1500$ and four fuzziness levels $m$.}
	\label{curvese31}
\end{figure}

\begin{figure}[!ht]
	\centering
	\includegraphics[width=0.85\textwidth]{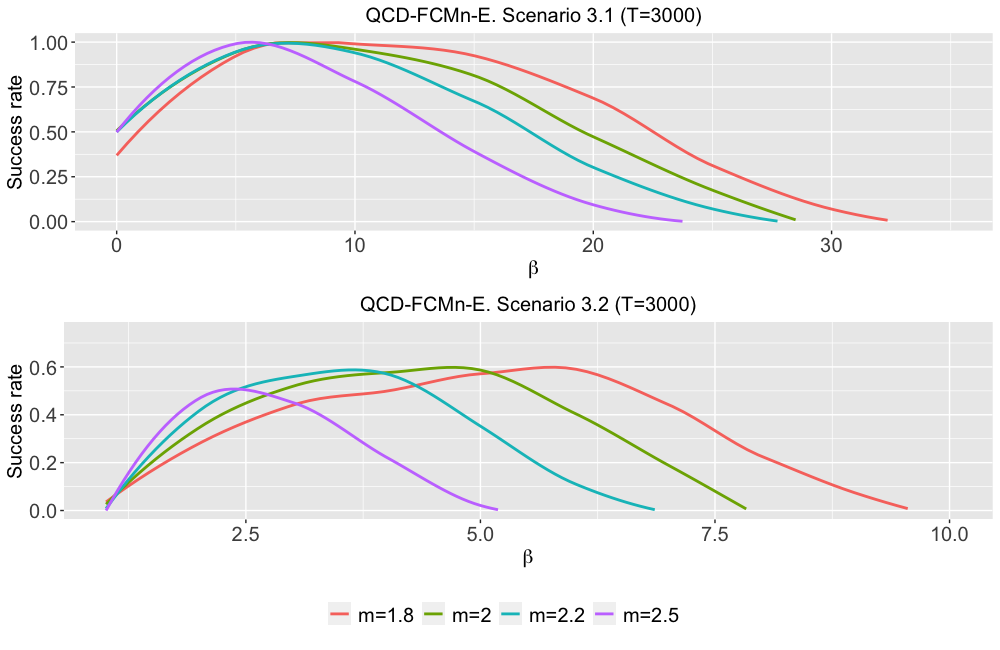}
	\vspace*{-0.25cm}
	\caption{Average rates of correct classification for QCD-FCMn-E as a function of $\beta$ in Scenarios 3.1 and 3.2 for series of length $T=3000$ and four fuzziness levels $m$.}
	\label{curvese32}
\end{figure}

\begin{figure}[!ht]
	\centering
	\includegraphics[width=0.85\textwidth]{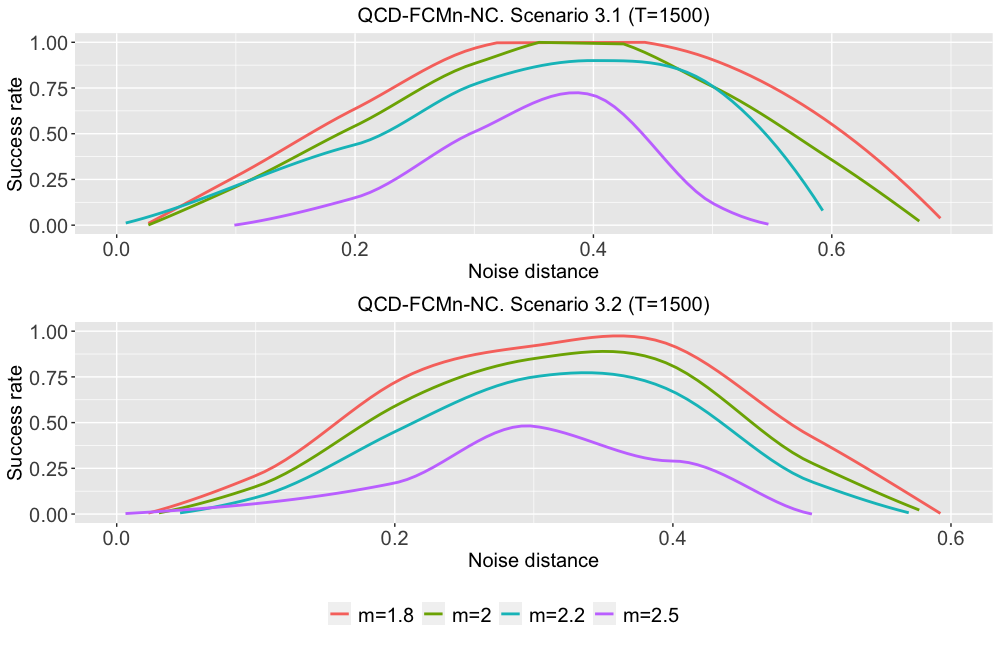}
	\vspace*{-0.25cm}
	\caption{Average rates of correct classification for QCD-FCMn-NC as a function of $\beta$ in Scenarios 3.1 and 3.2 for series of length $T=1500$ and four fuzziness levels $m$.}
	\label{curvesnc31}
\end{figure}

\begin{figure}[!ht]
	\centering
	\includegraphics[width=0.85\textwidth]{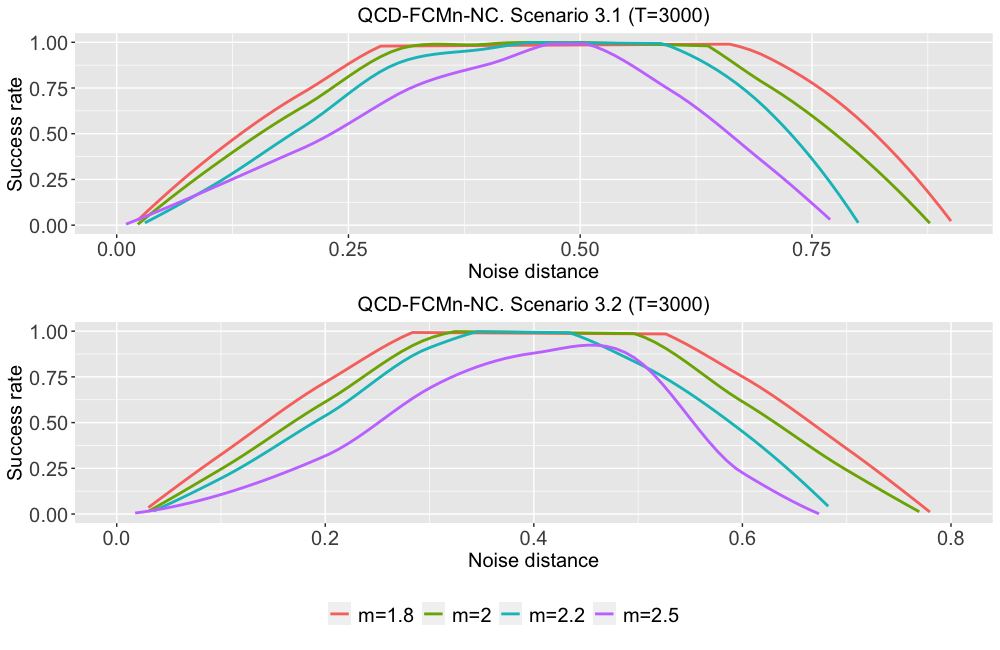}
	\vspace*{-0.25cm}
	\caption{Average rates of correct classification for QCD-FCMn-NC as a function of $\beta$ in Scenarios 3.1 and 3.2 for series of length $T=3000$ and four fuzziness levels $m$.}
	\label{curvesnc32}
\end{figure}

Additional figures of the simulation study carried out throughout Section \ref{sectionsimulationstudy} are given in this Section of the Appendix. Figures \ref{curvese12}, \ref{curvesnc12}, \ref{curvese22}, \ref{curvesnc21}, \ref{curvesnc22}, \ref{curvese31}, \ref{curvese32}, \ref{curvesnc31}, and \ref{curvesnc32} contain curves of average rates of correct classifications for exponential and noise approaches for some scenarios and values of the series length. 
\begin{figure}[!ht]
	\centering
	\includegraphics[width=0.85\textwidth]{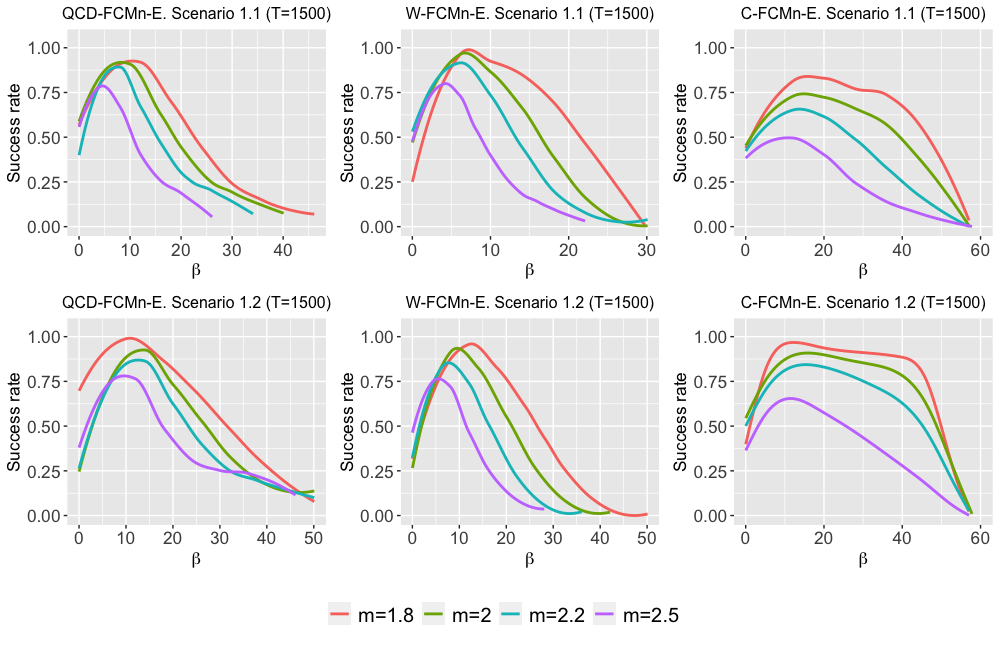}
	\vspace*{-0.25cm}
	\caption{Average rates of correct classification for QCD-FCMn-E, W-FCMn-E and C-FCMn-E as a function of $\beta$ in Scenarios 1.1 and 1.2 for series of length $T=1500$ and four fuzziness levels $m$.}
	\label{curvese12}
\end{figure}
\begin{figure}[!ht]
	\centering
	\includegraphics[width=0.85\textwidth]{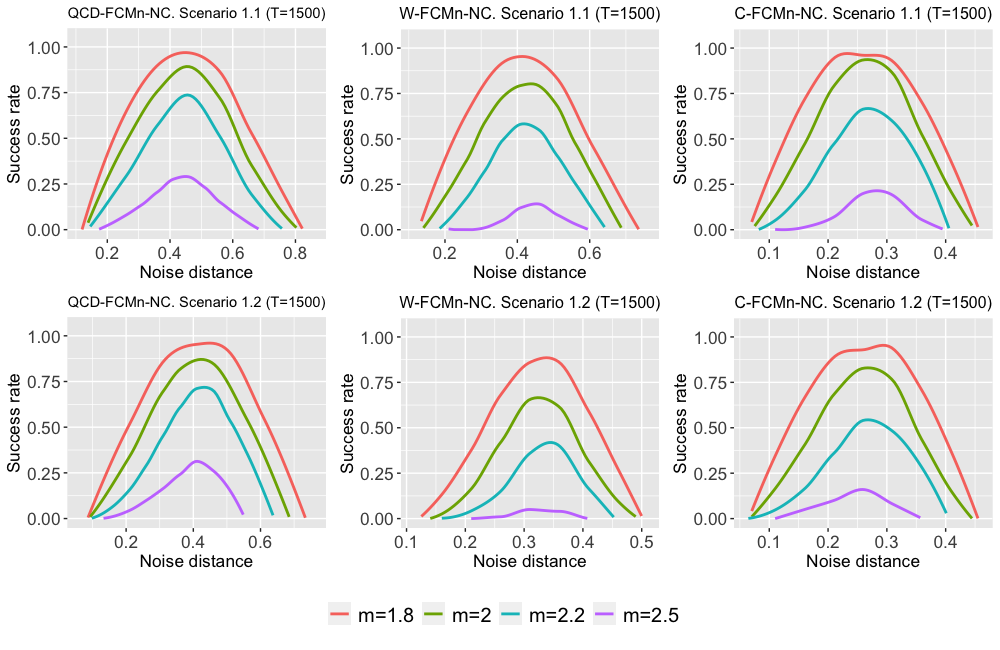}
	\vspace*{-0.25cm}
	\caption{Average rates of correct classification for QCD-FCMn-NC, W-FCMn-NC and C-FCMn-NC as a function of $\beta$ in Scenarios 1.1 and 1.2 for series of length $T=1500$ and four fuzziness levels $m$.}
	\label{curvesnc12}
\end{figure}
\begin{figure}[!ht]
	\centering
	\includegraphics[width=0.85\textwidth]{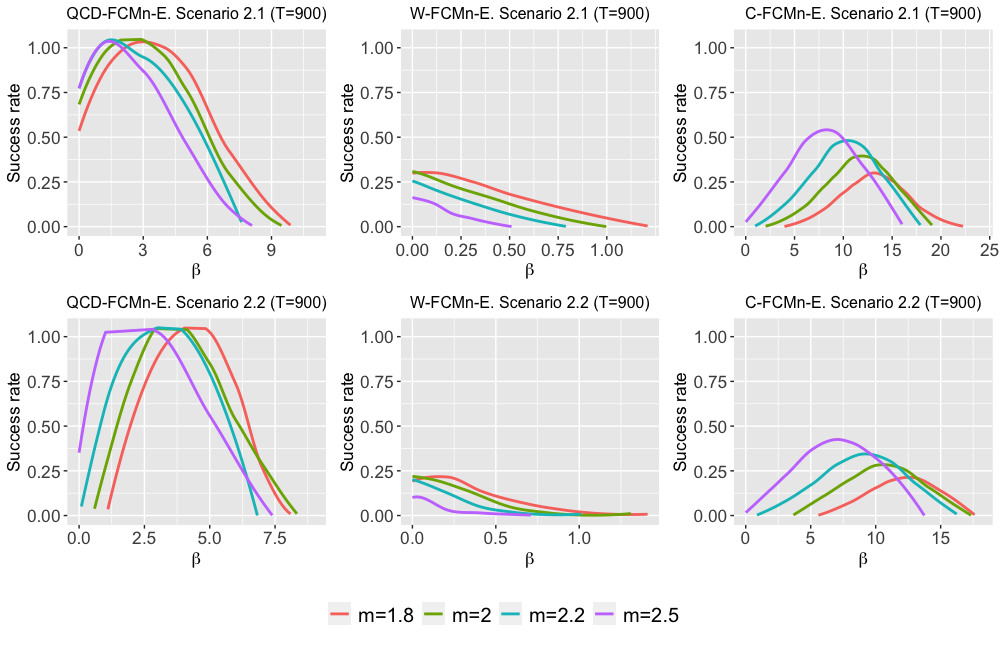}
	\vspace*{-0.25cm}
	\caption{Average rates of correct classification for QCD-FCMn-E, W-FCMn-E and C-FCMn-E as a function of $\beta$ in Scenarios 2.1 and 2.2 for series of length $T=900$ and four fuzziness levels $m$.}
	\label{curvese22}
\end{figure}

\clearpage

\begin{figure}[!ht]
	\centering
	\includegraphics[width=0.85\textwidth]{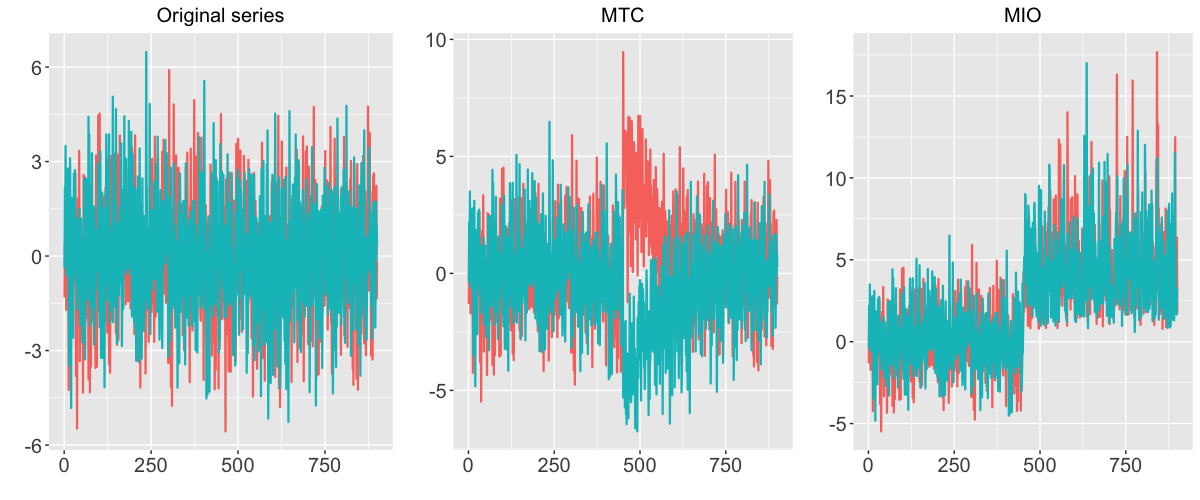}
	\caption{Realizations of the first process in Base Scenario 1 (left panel), the contaminated series in Scenario MTC 2 (middle panel) and the contaminated series in Scenario MIO 2 (right panel).}
	\vspace*{-0.25cm}
	\label{geometricoutlier}
\end{figure}

In order to show that MIO and MTC series can usually be detected by visual inspection of the serial realizations, we have depicted in Figure~\ref{geometricoutlier} the profiles of one realization of the first process in Base Scenario 2 of Section \ref{subsubsectionsecondassessment} along with one realization of the corresponding contaminated series in relation to Scenarios MTC 2 and MIO 2. Note that the difference between the original series (left panel) and its perverted versions through MTC and MIO (middle and left panel, respectively) is very clear. The MTC causes a change in the level of the series to both the upper and the lower sides of the plot, which is expected given the size of the outlier, $\bm w=(5, -5)^{\intercal}$. This change decreases gradually with time until the series gets back to its normal behaviour. On the other hand, the MIO provokes an abrupt change in the series when $t=450$, which is held until the end, $t=T=900$. The period from $t=450$ to  $t=900$ is characterised by the series taking only positive values. By comparing the middle and right panels of Figure \ref{geometricoutlier} with the left panel, one can easily deduce that the series in the former panels have an outlying nature.

\subsection*{Additional tables concerning Simulation study and Application sections}

This section of the Appendix contains tables with some meaningful results regarding the simulation study performed in Section \ref{sectionsimulationstudy} and the two applications with real data considered in Section~\ref{sectionapplications}. Specifically,

\begin{itemize}
    \item Table~\ref{tabletrimmed3} contains the results for the trimmed-based approaches in Scenarios 3.1 and 3.2.
    \item Tables \ref{expt}, \ref{nct} and \ref{trimmedt} include the rates of correct classification for the exponential, noise and trimmed approaches, respectively, for the simulation study carried out in Section \ref{subsubsectionfirstassessment} with innovations drawn from a multivariate $t$ distribution with 3 degrees of freedom.
    \item Concerning the robust fuzzy clustering of the top 20 companies of S\&P 500 index in Section~\ref{App1}, Table~\ref{tablefuzzyqcd} contains the membership degrees for the series in Figure~\ref{top20} according to the 6-cluster solution attained by the standard algorithm QCD-FCMn.
    \item Concerning the robust fuzzy clustering of air pollution data in Section~\ref{subsectionenvironmental}, Table~\ref{fuzzytablestations} shows the fuzzy 3-cluster partition for the 20 monitoring stations in Galicia when the standard QCD-FCMn model was applied using 0.6 as cutoff.  For a given station, the highest membership degree is shown in bold when it is larger than 0.6. When the three membership values are below this cutoff, the corresponding quantities are written in italic font.
\end{itemize}


\vspace*{0.75cm}

\begin{table}[!h]
	\centering
	\resizebox{9cm}{!}{
		\begin{tabular}{ccrrr}
			\hline
			& & QCD-FCMn-T & W-FCMn-T & C-FCMn-T \\ \hline 
			\multicolumn{5}{l}{\textsc{Scenario 3.1}}\\ \hline 
			$T=1500$ & $m=1.8$ & \textbf{1} & 0.21 & 0.02 \\ 
			& $m=2$ & \textbf{0.99} & 0.14 & 0.01 \\ 
			& $m=2.2$ & \textbf{0.99} & 0.05 & 0.01 \\ 
			& $m=2.5$ & \textbf{0.95} & 0.01 & 0 \\ \hline 
			$T=3000$ & $m=1.8$ & \textbf{1} & 0.45 & 0.05 \\ 
			& $m=2$ & \textbf{1} & 0.32 & 0.03 \\ 
			& $m=2.2$ & \textbf{1} & 0.25 & 0.02 \\ 
			& $m=2.5$ & \textbf{1} & 0.25 & 0.02 \\ \hline 
			\multicolumn{5}{l}{\textsc{Scenario 3.2}}\\ \hline 
			$T=1500$ & $m=1.8$ & \textbf{0.44} & 0 & 0.01 \\ 
			& $m=2$ & \textbf{0.41} & 0 & 0 \\ 
			& $m=2.2$ & \textbf{0.40} & 0 & 0 \\ 
			& $m=2.5$ & \textbf{0.38} & 0 & 0 \\ \hline 
			$T=3000$ & $m=1.8$ & \textbf{0.82} & 0 & 0 \\ 
			& $m=2$ & \textbf{0.77} & 0 & 0 \\ 
			& $m=2.2$ & \textbf{0.72} & 0 & 0 \\ 
			& $m=2.5$ & \textbf{0.72} & 0 & 0 \\ 
			\hline
		\end{tabular}
	}
	\caption{Maximum correct classification rate for QCD-FCMn-T, W-FCMn-T and C-FCMn-T. Scenarios 3.1 and 3.2.}
	\label{tabletrimmed3}
\end{table}

\begin{table}[!t]
	\centering
		\resizebox{9cm}{!}{
	\begin{tabular}{ccrrr}
		\hline
		&  & QCD-FCMn-E & W-FCMn-E & C-FCMn-E \\ \hline 
		\multicolumn{5}{l}{\textsc{Scenario 1.1}}\\ \hline 
		$T=750$ & $m=1.8$ & \textbf{0.75} & 0 & 0.04 \\ 
		& $m=2.2$ & \textbf{0.46} & 0 & 0.01 \\ \hline 
		$T=1500$ & $m=1.8$ & \textbf{0.94} & 0 & 0.19 \\ 
		& $m=2.2$ & \textbf{0.77} & 0 & 0.11 \\ \hline 
		\multicolumn{5}{l}{\textsc{Scenario 1.2}}\\ \hline 
		$T=750$ & $m=1.8$ & \textbf{0.71} & 0 & 0.07 \\ 
		& $m=2.2$ & \textbf{0.55} & 0 & 0.03 \\ \hline 
		$T=1500$ & $m=1.8$ & \textbf{0.96} & 0 & 0.13 \\ 
		& $m=2.2$ & \textbf{0.93} & 0 & 0.08 \\ \hline 
		\multicolumn{5}{l}{\textsc{Scenario 2.1}}\\ \hline 
		$T=600$ & $m=1.8$ & \textbf{0.30} & 0 & 0.06 \\ 
		& $m=2.2$ & \textbf{0.45} & 0 & 0.01 \\ \hline 
		$T=900$ & $m=1.8$ & \textbf{0.84} & 0 & 0.09 \\ 
		& $m=2.2$ & \textbf{0.86} & 0 & 0.05 \\ \hline 
		\multicolumn{5}{l}{\textsc{Scenario 2.2}}\\ \hline 
		$T=600$  & $m=1.8$ & \textbf{0.63} & 0 & 0.04 \\ 
		& $m=2.2$ & \textbf{0.84} & 0 & 0.01 \\ \hline 
		$T=900$  & $m=1.8$ & \textbf{0.95} & 0 & 0.02 \\ 
		& $m=2.2$ & \textbf{0.91} & 0 & 0 \\ \hline 
		\multicolumn{5}{l}{\textsc{Scenario 3.1}}\\ \hline 
		$T=1500$ & $m=1.8$ & \textbf{0.93} & 0 & 0 \\ 
		& $m=2.2$ & \textbf{0.82} & 0 & 0 \\ \hline 
		$T=3000$ & $m=1.8$ & \textbf{0.98} & 0 & 0 \\ 
		& $m=2.2$ & \textbf{0.95} & 0 & 0 \\ \hline 
		\multicolumn{5}{l}{\textsc{Scenario 3.2}}\\ \hline 
		$T=1500$  & $m=1.8$ & \textbf{0.82} & 0 & 0 \\ 
		& $m=2.2$ & \textbf{0.80} & 0 & 0 \\ \hline 
		$T=3000$  & $m=1.8$ & \textbf{0.94} & 0 & 0 \\ 
		& $m=2.2$ & \textbf{0.29} & 0 & 0 \\ 
		\hline
	\end{tabular}
	}
	\caption{Averages rates of correct classification with the optimal selection of hyperparameter $\beta$ for QCD-FCMn-E, W-FCMn-E and C-FCMn-E when innovations were drawn from a multivariate $t$ distribution with 3 degrees of freedom.}
	\label{expt}
\end{table}

\begin{table}[!t]
	\centering 
	\resizebox{9cm}{!}{
		\begin{tabular}{ccrrr} \hline 
			&  & QCD-FCMn-NC & W-FCMn-NC & C-FCMn-NC \\ \hline 
			\multicolumn{5}{l}{\textsc{Scenario 1.1}}\\ \hline 
			$T=750$ & $m=1.8$ & \textbf{0.56} & 0 & 0.05 \\ 
			& $m=2.2$ & \textbf{0.15} & 0 & 0 \\ \hline 
			$T=1500$ & $m=1.8$ & \textbf{0.96} & 0 & 0.13 \\ 
			& $m=2.2$ & \textbf{0.72} & 0 & 0.01 \\ \hline 
			\multicolumn{5}{l}{\textsc{Scenario 1.2}}\\ \hline 
			$T=750$ & $m=1.8$ & \textbf{0.61} & 0 & 0.05 \\ 
			& $m=2.2$ & \textbf{0.20} & 0 & 0 \\ \hline 
			$T=1500$ & $m=1.8$ & \textbf{0.95} & 0 & 0.07 \\ 
			& $m=2.2$ & \textbf{0.76} & 0 & 0 \\ \hline 
			\multicolumn{5}{l}{\textsc{Scenario 2.1}}\\ \hline 
			$T=600$ & $m=1.8$ & \textbf{0.16}& 0 & 0 \\ 
			& $m=2.2$ & \textbf{0.05} & 0 & 0 \\ \hline 
			$T=900$ & $m=1.8$ & \textbf{0.68} & 0 & 0 \\
			& $m=2.2$ & \textbf{0.50} & 0 & 0 \\ \hline 
			\multicolumn{5}{l}{\textsc{Scenario 2.2}}\\ \hline 
			$T=600$ & $m=1.8$ & \textbf{0.80} & 0 & 0 \\ 
			& $m=2.2$ & \textbf{0.52} & 0 & 0 \\ \hline 
			$T=900$ & $m=1.8$ & \textbf{0.99} & 0 & 0 \\ 
			& $m=2.2$ & \textbf{0.91} & 0 & 0 \\ \hline  
			\multicolumn{5}{l}{\textsc{Scenario 3.1}}\\ \hline 
			$T=1500$ & $m=1.8$ & \textbf{0.78} & 0 & 0 \\ 
			& $m=2.2$ & \textbf{0.41} & 0 & 0 \\ \hline 
			$T=3000$ & $m=1.8$ & \textbf{1} & 0 & 0 \\ 
			& $m=2.2$ & \textbf{0.87} & 0 & 0 \\ \hline 
			\multicolumn{5}{l}{\textsc{Scenario 3.2}}\\ \hline  
			$T=1500$  & $m=1.8$ & \textbf{0.31} & 0 & 0 \\ 
			& $m=2.2$ & \textbf{0.12} & 0 & 0 \\ \hline 
			$T=3000$  & $m=1.8$ & \textbf{0.99} & 0 & 0 \\ 
			& $m=2.2$ & \textbf{0.93} & 0 & 0 \\ 
			\hline
		\end{tabular}
	}
	\caption{Averages rates of correct classification with the optimal selection of hyperparameter $\delta$ for QCD-FCMn-NC, W-FCMn-NC and C-FCMn-NC when innovations were drawn from a multivariate $t$ distribution with 3 degrees of freedom.}
	\label{nct}
\end{table}

\begin{table}[!t]
	\centering
		\resizebox{9cm}{!}{
	\begin{tabular}{ccrrr} \hline 
		&  & QCD-FCMn-T & W-FCMn-T & C-FCMn-T \\ \hline 
		\multicolumn{5}{l}{\textsc{Scenario 1.1}}\\ \hline 
		$T=750$ & $m=1.8$ & \textbf{0.81} & 0.02 & 0.08 \\ 
		& $m=2.2$ & \textbf{0.62} & 0 & 0.03 \\ \hline 
		$T=1500$ & $m=1.8$ & \textbf{0.98} & 0 & 0.32 \\ 
		& $m=2.2$ & \textbf{0.94} & 0 & 0.14 \\ \hline 
		\multicolumn{5}{l}{\textsc{Scenario 1.2}}\\ \hline 
		$T=750$ & $m=1.8$ & \textbf{0.77} & 0.01 & 0.13 \\ 
		& $m=2.2$ & \textbf{0.64} & 0.01 & 0.03 \\ \hline 
		$T=1500$ & $m=1.8$ & \textbf{0.98} & 0 & 0.19 \\ 
		& $m=2.2$ & \textbf{0.95} & 0 & 0.10 \\ \hline 
		\multicolumn{5}{l}{\textsc{Scenario 2.1}}\\ \hline 
		$T=600$ & $m=1.8$ & \textbf{0.36} & 0 & 0.13 \\ 
		& $m=2.2$ & \textbf{0.35} & 0 & 0.09 \\ \hline 
		$T=900$ & $m=1.8$ & \textbf{0.79} & 0 & 0.16 \\ 
		& $m=2.2$ & \textbf{0.79} & 0 & 0.12 \\ \hline 
		\multicolumn{5}{l}{\textsc{Scenario 2.2}}\\ \hline 
		$T=600$ & $m=1.8$ & \textbf{0.80} & 0 & 0.05 \\ 
		& $m=2.2$ & \textbf{0.78} & 0 & 0.05 \\ \hline 
		$T=900$ & $m=1.8$ & \textbf{1} & 0 & 0.10 \\ 
		& $m=2.2$ & \textbf{1} & 0 & 0.11 \\ \hline 
		\multicolumn{5}{l}{\textsc{Scenario 3.1}}\\ \hline 
		$T=1500$ & $m=1.8$ & \textbf{0.96} & 0.73 & 0.59 \\ 
		& $m=2.2$ & \textbf{0.92} &  0.59 & 0.36  \\ \hline 
		$T=3000$ & $m=1.8$ & \textbf{1} & 0.88 & 0.72 \\ 
		& $m=2.2$ & \textbf{1} & 0.83 & 0.66 \\ \hline 
		\multicolumn{5}{l}{\textsc{Scenario 3.2}}\\ \hline 
		$T=1500$  & $m=1.8$ & \textbf{0.39} & 0  & 0 \\ 
		& $m=2.2$ & \textbf{0.32} & 0 & 0 \\ \hline 
		$T=3000$  & $m=1.8$ & \textbf{0.76} & 0 & 0.01  \\ 
		& $m=2.2$ & \textbf{0.64} & 0 & 0.01 \\ 
		\hline
	\end{tabular}
	}
	\caption{Averages rates of correct classification for QCD-FCMn-T, W-FCMn-T and C-FCMn-T when innovations were drawn from a multivariate $t$ distribution with 3 degrees of freedom.}
	\label{trimmedt}
\end{table}



\begin{table}[!ht]
	\centering
	\resizebox{8.5cm}{!}{
		\begin{tabular}{ccccccc}
			\hline
			Company & $C_1$ & $C_2$ & $C_3$ & $C_4$ & $C_5$ & $C_6$ \\ 
			\hline
			AAPL & 0.115 & 0.213 & 0.016 & \textbf{0.516} & 0.090 & 0.050 \\ 
			MSFT & 0.147 & 0.068 & 0.060 & \textbf{0.494} & 0.137 & 0.095 \\ 
			AMZN & \textbf{0.891} & 0.022 & 0.006 & 0.041 & 0.012 & 0.028 \\ 
			GOOGL & \textbf{0.757} & 0.039 & 0.013 & 0.127 & 0.028 & 0.036 \\ 
			GOOG & \textbf{0.923} & 0.012 & 0.005 & 0.032 & 0.009 & 0.019 \\ 
			FB & 0.003 & \textbf{0.983} & 0.001 & 0.005 & 0.005 & 0.004 \\ 
			TSLA & 0.068 & 0.040 & 0.009 & \textbf{0.824} & 0.038 & 0.021 \\ 
			BRK.B & 0.000 & 0.000 & \textbf{1} & 0.000 & 0.000 & 0.000 \\ 
			V & 0.006 & 0.019 & 0.006 & 0.025 & \textbf{0.933} & 0.010 \\ 
			JNJ & 0.004 & 0.013 & 0.004 & 0.013 & \textbf{0.959} & 0.008 \\ 
			WMT & 0.032 & 0.045 & 0.022 & 0.032 & 0.035 & \textbf{0.834} \\ 
			JPM & 0.096 & 0.067 & 0.055 & 0.122 & 0.089 & \textbf{0.572} \\ 
			MA & 0.129 & 0.160 & 0.017 & \textbf{0.499} & 0.087 & 0.109 \\ 
			PG & 0.087 & 0.071 & 0.120 & 0.117 & 0.118 & \textbf{0.486} \\ 
			UNH & 0.010 & \textbf{0.905} & 0.004 & 0.026 & 0.037 & 0.018 \\ 
			DIS & 0.057 & 0.120 & 0.014 & \textbf{0.619} & 0.128 & 0.061 \\ 
			NVDA & 0.044 & 0.038 & 0.013 & \textbf{0.802} & 0.077 & 0.025 \\ 
			HD & 0.035 & 0.052 & 0.015 & 0.032 & 0.030 & \textbf{0.836} \\ 
			PYPL & 0.189 & \textbf{0.420} & 0.019 & 0.172 & 0.071 & 0.129 \\ 
			BAC & 0.035 & 0.042 & 0.011 & 0.041 & 0.029 & \textbf{0.841} \\ 
			\hline
		\end{tabular}
	}
	\caption{Membership degrees for top 20 companies in the S\&P 500 index by considering the QCD-FCMn model and a 6-cluster partition.}
	\label{tablefuzzyqcd}
\end{table}



\begin{table}[!ht]
	\centering
	\resizebox{8.5cm}{!}{
		\begin{tabular}{ccccc}
			\hline
			Station  & Area & $C_1$ & $C_2$ & $C_3$ \\ 
			\hline
			FE & Suburban & 0.113 & 0.035 & \textbf{0.853} \\ 
			CO-T & Urban & \textbf{0.701} & 0.041 & 0.258 \\ 
			CO-R & Urban & \textbf{0.799} & 0.042 & 0.159 \\ 
			LU & Urban & \textbf{0.952} & 0.010 & 0.039 \\ 
			SDC-C & Suburban & \textit{0.452} & \textit{0.044} & \textit{0.504} \\ 
			SDC-SC & Urban & \textbf{0.928} & 0.012 & 0.060 \\ 
			SU & Rural & 0.089 & 0.076 & \textbf{0.835} \\ 
			PO-CL & Urban & \textbf{0.963} & 0.008 & 0.029 \\ 
			VGO-CO & Urban & \textbf{0.938} & 0.014 & 0.048 \\ 
			VGO-L & Urban & \textbf{0.903} & 0.020 & 0.076 \\ 
			PT & Suburban & \textit{0.469} & \textit{0.069} & \textit{0.462} \\ 
			OR & Urban & \textbf{0.944} & 0.014 & 0.042 \\ 
			PO-CP & Rural & 0.022 & 0.012 & \textbf{0.966}\\ 
			FR & Near power plant & 0.018 & \textbf{0.947} & 0.036 \\ 
			XO & Rural & \textit{0.191 }& \textit{0.538} & \textit{0.271} \\ 
			VGO-CT & Urban & 0.227 & 0.043 & \textbf{0.730} \\ 
			PA & Near power plant & 0.049 & \textbf{0.850} & 0.101 \\ 
			MA & Near power plant & \textit{0.088} & \textit{0.345} & \textit{0.567 }\\ 
			LO & Near power plant & 0.023 & \textbf{0.933 }& 0.044 \\ 
			MO & Near power plant & \textit{0.115}& \textit{0.294} & \textit{0.590} \\ 
			\hline
		\end{tabular}
	}
	\caption{Membership degrees for the 20 monitoring stations in Galicia by considering the QCD-FCMn model and a 3-cluster partition.}
	\label{fuzzytablestations}
\end{table}

\bibliography{mybibfile}

\end{document}